\documentclass{aa}  

\usepackage{graphicx}
\usepackage{tabularx}
\usepackage{array}
\usepackage{txfonts}
\usepackage{lipsum}
\usepackage[colorlinks=true,linkcolor=blue,citecolor=blue]{hyperref}
\usepackage{url}
\usepackage{natbib}
\usepackage{breqn}
\usepackage{diagbox}
\usepackage{cuted} 
\usepackage[mathscr]{euscript}

\newcommand{\orcid}[1]{\href{https://orcid.org/#1}{\includegraphics[width=10pt]{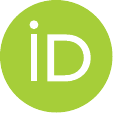}}}

\renewcommand{\bf}{}

\usepackage[belowskip=-13pt,aboveskip=2pt]{caption}
\DeclareMathOperator\erf{erf}
\newcolumntype{C}{>{\centering\arraybackslash}p{0.2\textwidth}}
\usepackage{enumitem}
\usepackage{booktabs}
\bibpunct{(}{)}{;}{a}{}{,} 
\begin{document} 
\title{The Mechanical Alignment of Dust (MAD) I: On the spin-up process of fractal grains by a gas-dust drift}
\titlerunning{The Mechanical Alignment of Dust (MAD)}
\author{Stefan Reissl \orcid{0000-0001-5222-9139} \inst{ \ref{inst1}} \and Paul Meehan \inst{\ref{inst1}} \and Ralf S. Klessen \orcid{0000-0002-0560-3172} \inst{\ref{inst1},\ref{inst2}} }

\institute{
\centering Universit\"{a}t Heidelberg, Zentrum f\"{u}r Astronomie, Institut f\"{u}r Theoretische Astrophysik, Albert-Ueberle-Straße 2, \\  D-69120 Heidelberg, Germany \label{inst1} 
\and
\centering Universit\"{a}t Heidelberg, Interdisziplin\"{a}res Zentrum f\"{u}r Wissenschaftliches Rechnen, Im Neuenheimer Feld 205, \\ D-69120 Heidelberg, Germany  \label{inst2}
}
						
\abstract
   {Observations of polarized light emerging from aligned dust grains are commonly exploited to probe the magnetic field orientation in astrophysical environments. However, the exact physical processes that result in a coherent large-scale grain alignment are still far from being fully constrained.}
{In this work, we aim to investigate the impact of a gas-dust drift on a microscopic level potentially leading to a mechanical alignment of fractal dust grains and subsequently to dust polarization.}
   {We scan a wide range of parameters of fractal dust aggregates in order to statistically analyze the average grain alignment behavior of distinct grain ensembles. In detail, the spin-up efficiencies for individual aggregates are determined utilizing a Monte-Carlo approach to simulate the collision, scattering, sticking, and evaporation processes of gas on the grain surface. Furthermore, the rotational disruption of dust grains is taken into account to estimate the upper limit of possible grain rotation. The spin-up efficiencies are analyzed within a mathematical framework of grain alignment dynamic in order to identify long-term stable grain alignment points in the parameter space. Here, we distinguish between the cases of grain alignment in direction of the gas-dust drift and the alignment along the magnetic field lines. Finally, the net dust polarization is calculated for each collection of stable alignment points per grain ensemble. }
   {We find the purely mechanical spin-up processes within the cold neutral medium to be sufficient enough to drive elongated grains to a stable alignment. The most likely mechanical grain alignment configuration is with a rotation axis parallel to the drift direction. Here, roundish grains require a supersonic drift velocity while rod-like elongated grains can already align for subsonic conditions. We predict a possible dust polarization efficiency in the order of unity resulting from mechanical alignment. Furthermore, a supersonic drift may result in a rapid grain rotation where dust grains may become rotationally disrupted by centrifugal forces. Hence, the net contribution of such a grain ensemble to polarization becomes drastically reduces.\\
   In the presence of a magnetic field the drift velocity required for the most elongated grains to reach a stable alignment is roughly one order of magnitude higher compared to the purely mechanical case. We demonstrate that a considerable fraction of a grain ensemble can stably align with the magnetic field lines and report a possibly dust polarization efficiency of ${0.6-0.9}$ indicating that a gas-dust drift alone can provide the conditions required to observationally probe the magnetic field structure. We predict that the magnetic field alignment is highly inefficient when the direction of the gas-dust drift and the magnetic field lines are perpendicular. }
 {Our results strongly suggest that mechanical alignment has to be be taken into consideration as an alternative driving mechanism where the canonical radiative torque alignment theory fails to account for the full spectrum of available dust polarization observations. }
  
\maketitle
%
\newpage
\section{Introduction}
Magnetic  fields  play  a quintessential role  in  the  physics  of the  interstellar medium (ISM) of galaxies as well as the subsequent star-formation. Observationally, the magnetic field orientation can be probed by polarization measurements of aligned elongated dust grains. The fact that an ensemble of dust grains aligns coherently on large scales within the ISM was already independently proposed decades ago by \cite{Hiltner1949a,Hiltner1949b} and  \cite{Hall1949}, respectively. Originally, this phenomenon was interpreted as consequence of a supersonic velocity difference between the gas and the dust content of the ISM. Such a gas-dust drift would simply align spheroidal grains mechanically by means of minimizing their geometrical cross section \citep[][]{Gold1952a,Gold1952b}. Hence, the dust polarization was expected to be related to the direction of the drift velocity.\\
Alternatively, mechanisms based on ferromagnetic alignment \citep[e.g.][]{Spitzer1949} or the alignment caused by internal paramagnetic dissipation \citep[][]{Davis1949,Davis1951,Mathis1986}, respectively, were proposed. In such a case, the light polarized by dust was expected to be correlated with the magnetic field orientation. Furthermore, charged grains were believed to cause a grain precession around the direction of the magnetic field by means of the Rowland effect \citep[][]{Martin1971}.\\
However, for all of these magnetic field associated mechanisms to work the dust requires a significant amount of grain rotation. Considering, the average galactic field strength of about ${10\ \mu\mathrm{G}}$ it seemed to be questionable that ferromagnetic alignment or paramagnetic dissipation alone can account for the observed dust polarization. For instance, dust grains exposed to gas collisions would easily be kicked out of a long-term stable alignment \citep[][]{JonesSpitzer1967,PurcellSpitzer1971}.\\
One way to remedy the situating was suggested in \cite{JonesSpitzer1967} by considering super-paramagnetic dust where small pallets of iron were baked into the grain material itself. Later, in \cite{Dolginov1976} and \cite{Purcell1979} it was noted that the Barnett effect \citep{Barnett1915} would be much more efficient in coupling paramagnetic grains to the magnetic field lines.\\
Moreover, considering the diffusive processes involved in the grain growth the dust surface is usually expected to be highly irregularly shaped \citep{Mathis1989,Lazarian1995,Greenberg1995,Dominik1997,Beckwith2000,Wada2007,Kim2021}.
It was already pointed out in \cite{Purcell1975,Purcell1979} that regular dust grains may be subject to some long-term torques leading to a rapid grain rotation.  \cite{Purcell1979} identified three processes for the potential spin-up of dust namely: (i)  the conversion of atomic (H) to molecular hydrogen (H$_{\mathrm{2}}$) on the grain surface,  (ii) the rebound of colliding gas particles, and (iii) the absorption and subsequent emission of photo-electrons. However, it was first noticed by \cite{Dolginov1976} that irregular grains may be expedience some additional systematic torques. 
The proposed systematic torques are inevitably related to the grains surface and would subsequently lead to super-thermal rotation compensating for random gas collisions. Hence, from now onward an irregular shape of interstellar grains became an integral parameter for the description of grain alignment.\\
The spin up-process of such irregularly shaped grains in the presence of a radiation field was studied in \cite{DraineWeingartner1996} using the Discrete Dipole Scattering code DDSCAT \citep[see][]{Draine1994}. It was numerically demonstrated that starlight can lead to substantial radiative torques (RAT) acting on an irregular grain. In followup studies further phenomena such as H$_{\mathrm{2}}$ formation \citep[][]{DraineWeingartner1997}  and the thermal flipping of grains \citep[][]{Weingartner2001} were incorporated. Later, an analytical model (AMO) of RAT alignment was provided by \cite{Lazarian2007} based on a microscopic toy-model in combination with numerical calculations using DDSCAT. The latter work initiated a series of studies \citep[e.g.][and references therein]{HoangLazarian2008,LazarianHoang2008ApJ,LazarianHoang2009ApJ,Hoang2016} dealing with the physical implications of RATs. Eventually, the AMO allowed for accurate modelling of synthetic dust polarization observations \citep[e.g.][]{Cho2005ApJ,Bethell2007,HoangLazarian2014MNRAS,HoangLazarianAndersson2015,Reissl2016,Bertrang2017,Valdivia2019,Reissl2020,Seifried2020MNRAS,Kuffmeier2020,Reissl2021MNRAS} assuming a set of free alignment parameters to fine tune the model. More recently, an approach was presented in \cite{Hoang2016} and \cite{Herranen2021}, respectively, based on exact exact light scattering solutions of RAT physics to constrain the remaining free parameters of the AMO for a large ensemble of irregular grain shapes.\\
Actual observations suggest a certain predictive capability of the AMO \citep[][]{Andersson2007,Whittet2008,Andersson2010,Matsumura2011,Vaillancourt2015}. For instance, a hallmark of RAT alignment is that the grain rotation scales with the strength of the radiation field. Hence, the AMO predicts for the grain alignment efficiency and the subsequent dust polarization to drop towards dense starless cores. This prediction appears to be partly backed up by corresponding core observations \citep[see e.g.][]{Alves2014,Jones2015,Jones2016}. However, other observations casts some doubt if the AMO covers the full range of observed grain alignment physics. For instance \cite{LeGouellec2020} presented a statistical analysis of several star-forming cores observations and conclude that the grain alignment efficiency seems to be higher as predicted by the AMO.\\ Moreover, abrupt changes of the polarization direction allows the surmise that the magnetic field lines do not necessarily dictate the grain alignment direction \citep[][]{Rao1998,Cortes2006,Tang2009}. Indeed, it was already noted in \cite{Lazarian2007} that dust grains may align in direction of a strong directed radiation field. However, it is commonly speculated in literature that a sudden change of the polarization direction and the high degree of dust polarization of some cores may better be accounted for by means of mechanical alignment especially in the presence of molecular outflows \citep[see e.g.][and references therein]{Sadavoy2019,Takahashi2019,Kataoka2019,Cortes_2021,Pattle2021}. In particular, observations of a proto-stellar system presented in \cite{Kwon2019} strongly suggest that more than one grain alignment mechanism seems to be simultaneously at work.
{\bf The required gas-dust drift may arise due to by magnetohydrodynamic turbulence \citep[][]{Yan2003} or resonant drag instabilities \citep[][]{Hopkins2018,Squire2021}, respectively.}\\
Up to this date an analytical description of mechanic alignment comparable to the AMO of RAT alignment is still a matter of ongoing research. Attempts to model the impact of mechanical alignment to dust polarization remain mostly qualitatively in nature \citep[see e.g.][]{Kataoka2019}. An early model was provided in \cite{Lazarian1994Gold} based on the original Gold alignment mechanism by incorporating internal dissipation of energy within the grain. This first attempt to account for mechanically driven dust alignment in the presence of a magnetic field, however, did not take the increased spin-up efficiency of irregularly shaped grains into consideration. Hence, this model required a super-sonic gas-dust drift \citep[see also][]{Lazarian1995Gold,Lazarian1997Gold} to work at all. A description of mechanical alignment based on a toy-model similar to that of the AMO was provided in \cite{LazarianHoang2007Mech}. This description allowed also for a subsonic alignment but the grain shape itself remained still a free parameter. About a decade later numerical studies  were provided by \citep{DasWeingartner2016} and \citep{HoangChoLazarian2018}, respectively, describing the of the mechanical torque (MET) acting on irregular grain shapes. Here, the irregular grains were modelled by means of cubical structures and Gaussian random spheres. However, these studies remain inconclusive concerning the exact set of parameters that would allow for an long-term stable mechanical alignment since they only provide limited data for a few distinct grain shapes.\\
The aim of this paper is to examine the effectiveness of mechanical alignment of dust  (MAD) for large ensembles of distinct fractal grain shapes. In our study the grain shapes are modelled as fractal aggregates composed of smaller building blocks. We present a novel method to simulate the spin-up process of such aggregates by means of Monte-Carlo (MC) simulations. The paper is structured as follows: First, in Sect.~\ref{sect:DustGrainModel} we introduce the methods and grain parameters applied to mimic the growth of fractal dust aggregates. Then, in Sect.~\ref{sect:GasDustInt} we present the gas-dust interaction processes considered in our simulations. In Sect.~\ref{sect:MCSimulation} we outline the MC scheme applied to simulate specific mechanical alignment efficiencies. The physical processes and conditions that give raise to METs, drag torques, and paramagnetic torques are discussed in Sect.~\ref{sect:TorqueMech}, Sect.~\ref{sect:TorqueDrag}, and Sect.~\ref{sect:MagTorque}, respectively. The equations governing the grain alignment dynamics are introduced in Sect.~\ref{sect:GrainDyn}. We outline the considered methods and measures to account for dust destruction and polarization in Sect.~\ref{sect:DustPolDest}. Our results are presented Sect.~\ref{sect:Alignemnt}. Finally, Sect.~\ref{sect:Summary} is devoted to the summary and outlook.

\section{Dust grain model}
\label{sect:DustGrainModel}
\begin{figure*}[ht!]
     \begin{center}
     \begin{tabular}{ | c | c | c | c | }
     \hline \diagbox{$a_{\mathrm{eff}}$}{$D_{\mathrm{f}}$}
       & $1.6$ & $2.0$ & $2.4$ \\

			\hline
			\raisebox{-\totalheight}{$200\ \mathrm{nm}$}

			&
      \raisebox{-\totalheight}{\vspace{-10mm}\includegraphics[width=0.28\textwidth]{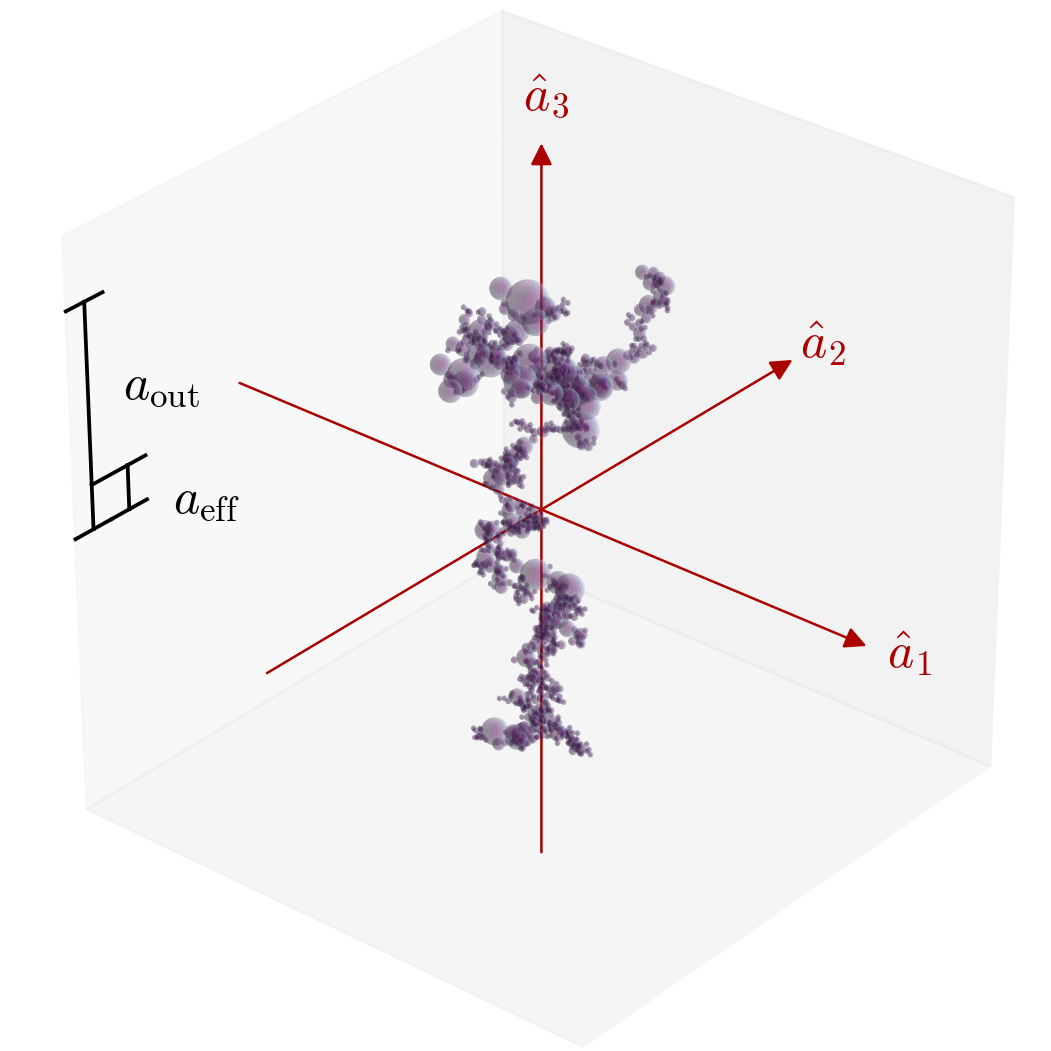}}

      & 
      \raisebox{-\totalheight}{\vspace{-10mm}\includegraphics[width=0.28\textwidth]{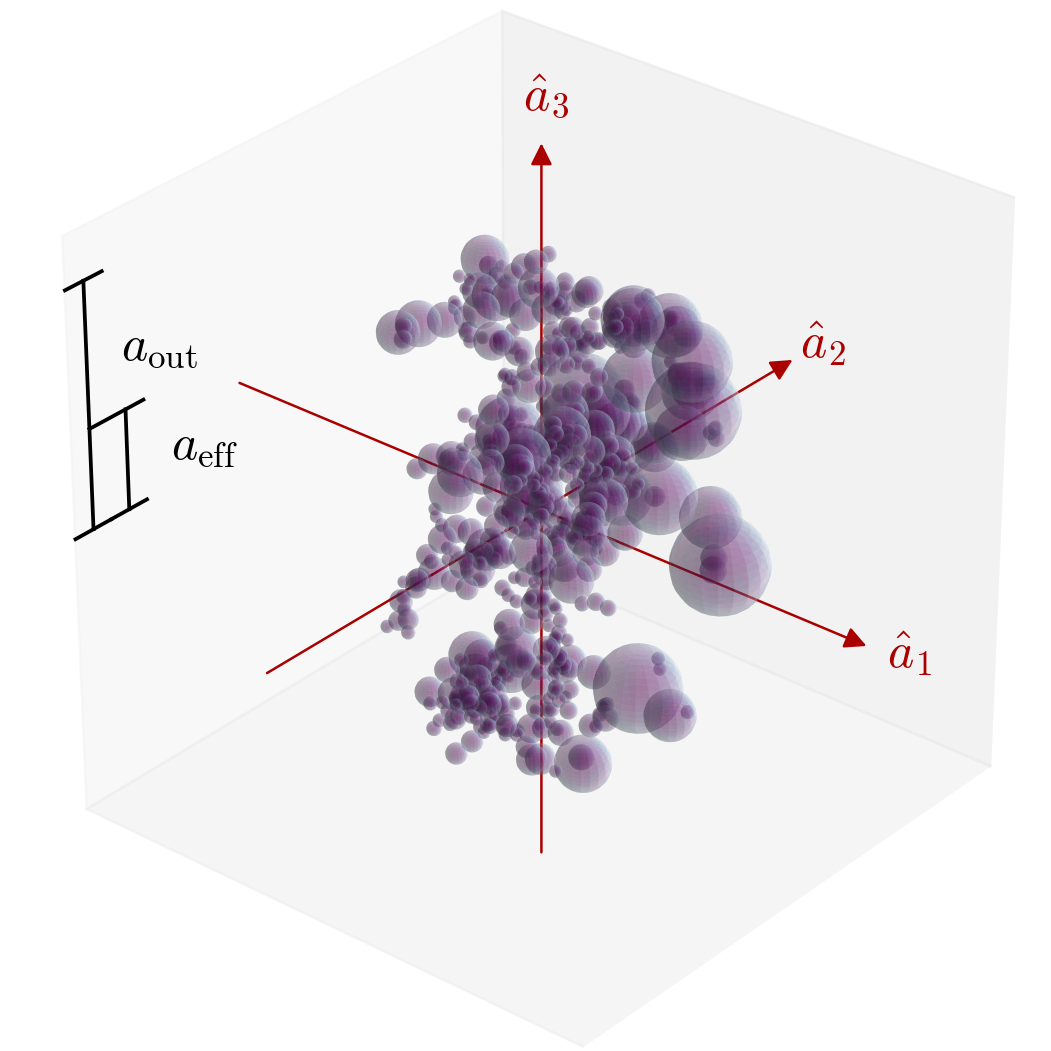}}
      & 
      \raisebox{-\totalheight}{\vspace{-10mm}\includegraphics[width=0.28\textwidth]{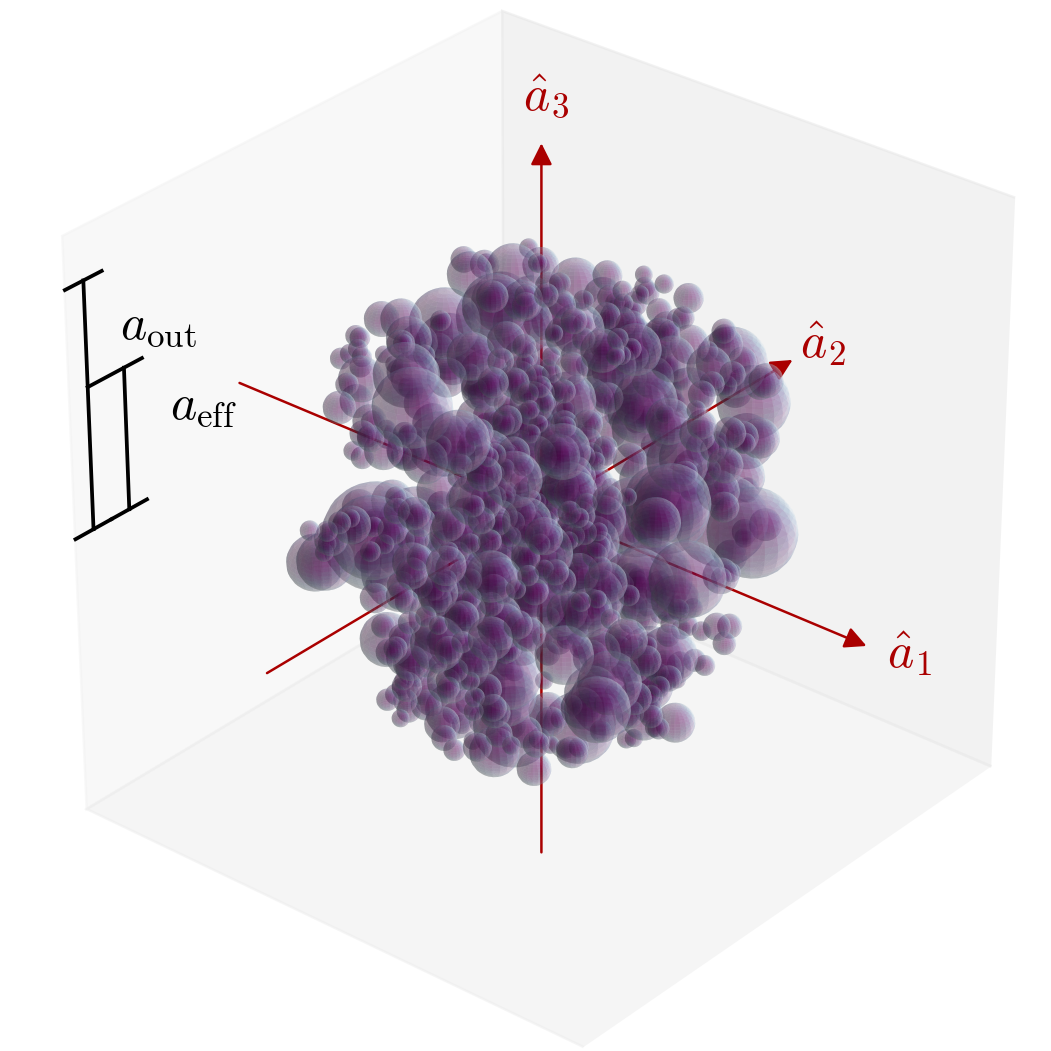}} \vspace{-1mm}
      \\ 
			& $a_{\mathrm{out}}=955\ \mathrm{nm}$, $N_{\mathrm{mon}}=500$ & $a_{\mathrm{out}}=446\ \mathrm{nm}$, $N_{\mathrm{mon}}=420$ & $a_{\mathrm{out}}=370\ \mathrm{nm}$,  $N_{\mathrm{mon}}=870$\\
	\hline
\raisebox{-\totalheight}{$400\ \mathrm{nm}$}
			&
      \raisebox{-\totalheight}{\vspace{-10mm}\includegraphics[width=0.28\textwidth]{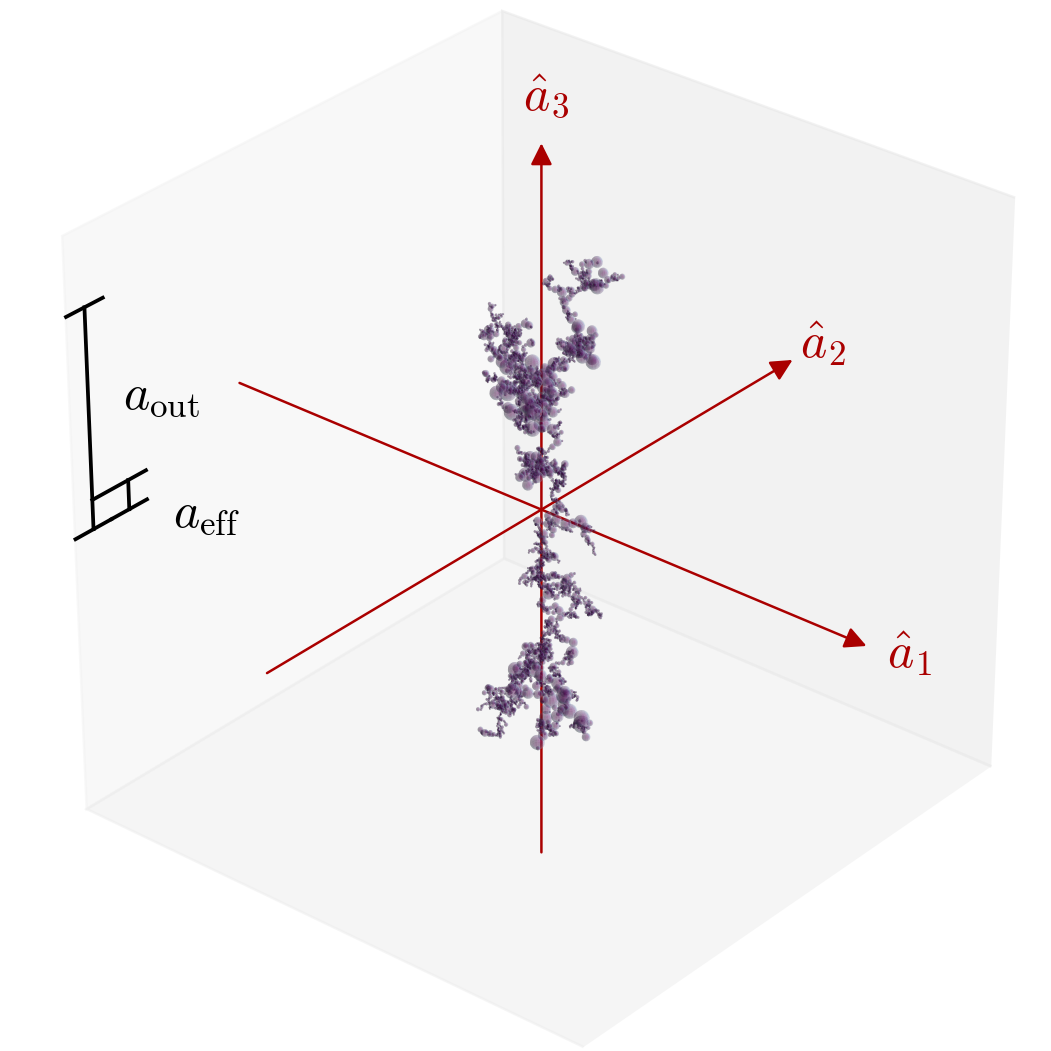}}

      & 
      \raisebox{-\totalheight}{\vspace{-10mm}\includegraphics[width=0.28\textwidth]{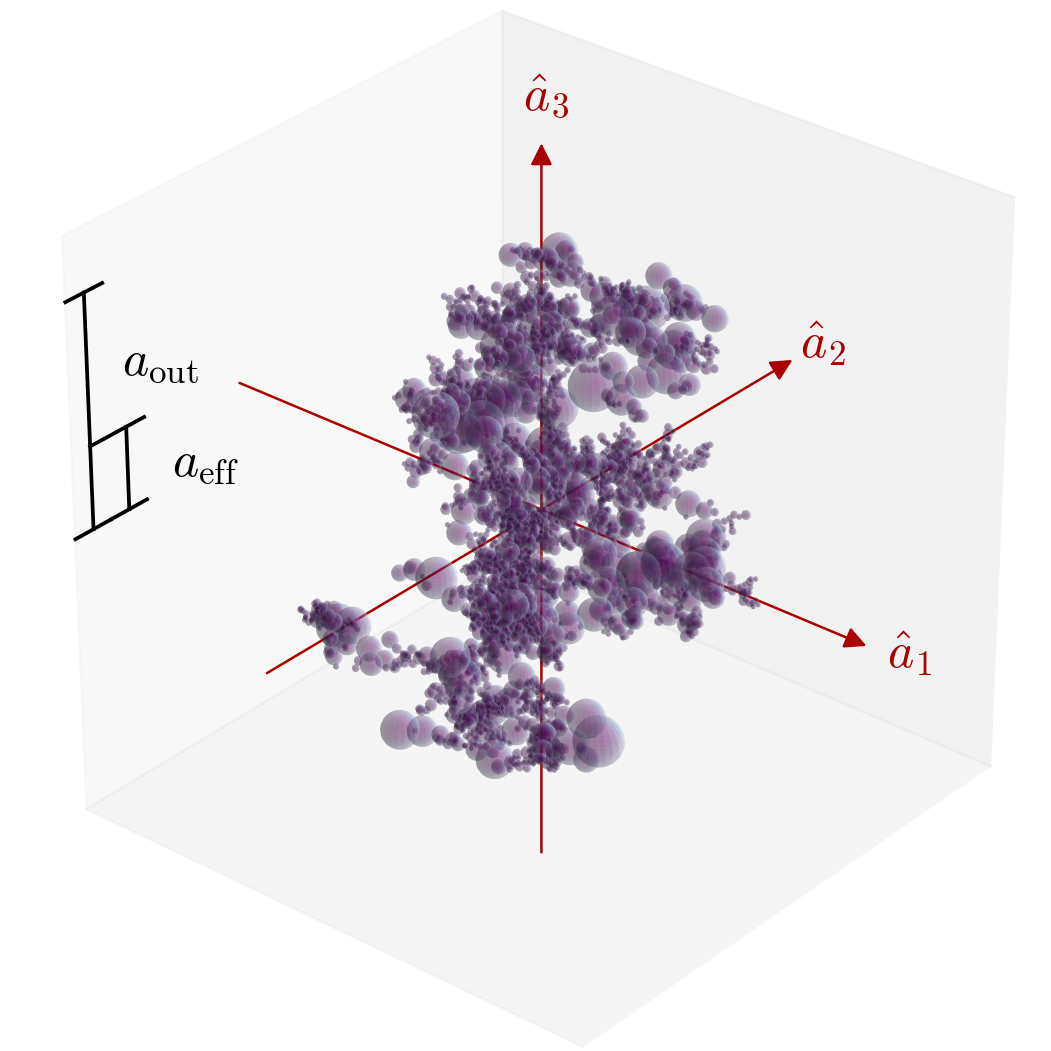}}
      & 
      \raisebox{-\totalheight}{\vspace{-10mm}\includegraphics[width=0.28\textwidth]{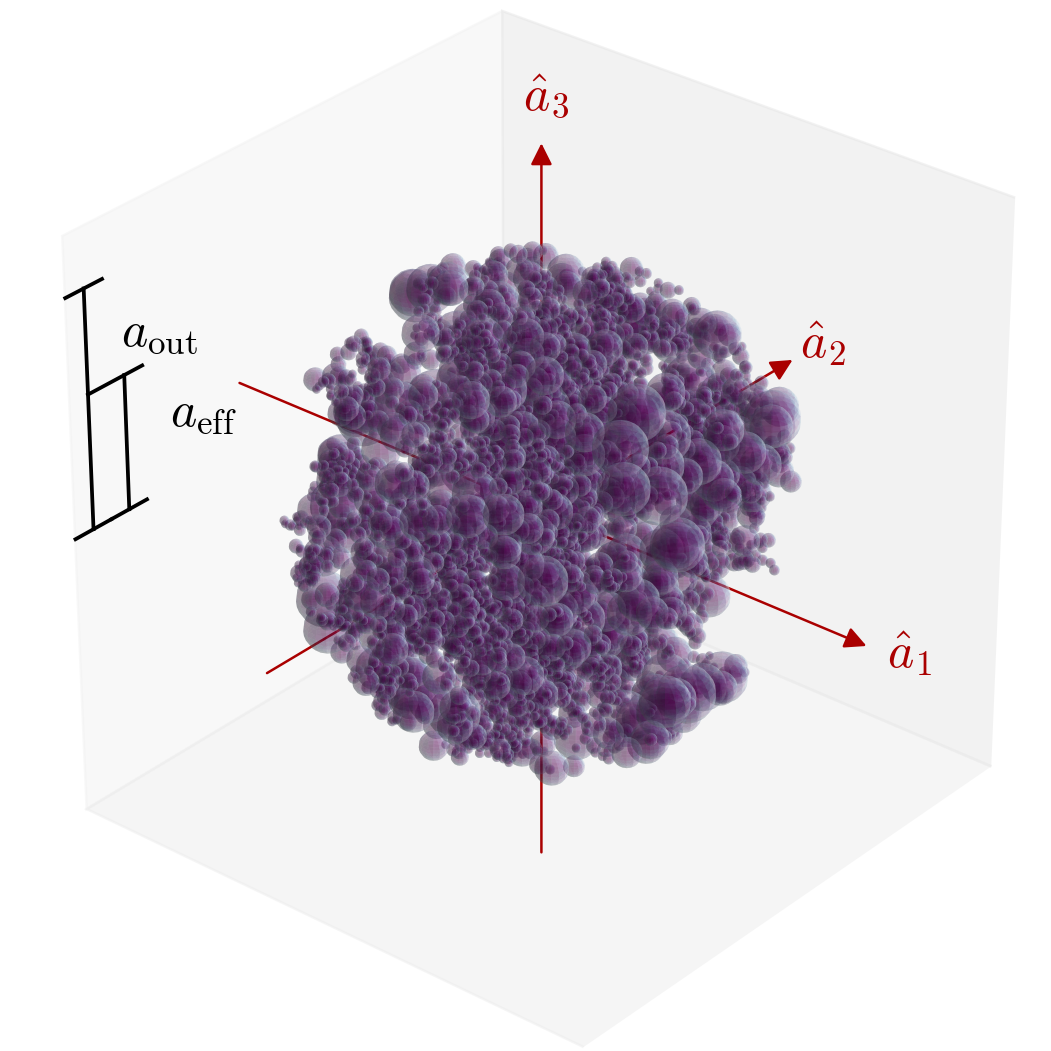}} \vspace{-1mm}
      \\
			& $a_{\mathrm{out}}=2891\ \mathrm{nm}$, $N_{\mathrm{mon}}=3350$ & $a_{\mathrm{out}}=1186\ \mathrm{nm}$, $N_{\mathrm{mon}}=4882$  & $a_{\mathrm{out}}=731\ \mathrm{nm}$, $ N_{\mathrm{mon}}=3280$\\
      \hline
\raisebox{-\totalheight}{$800\ \mathrm{nm}$}
			&
      \raisebox{-\totalheight}{\vspace{-10mm}\includegraphics[width=0.28\textwidth]{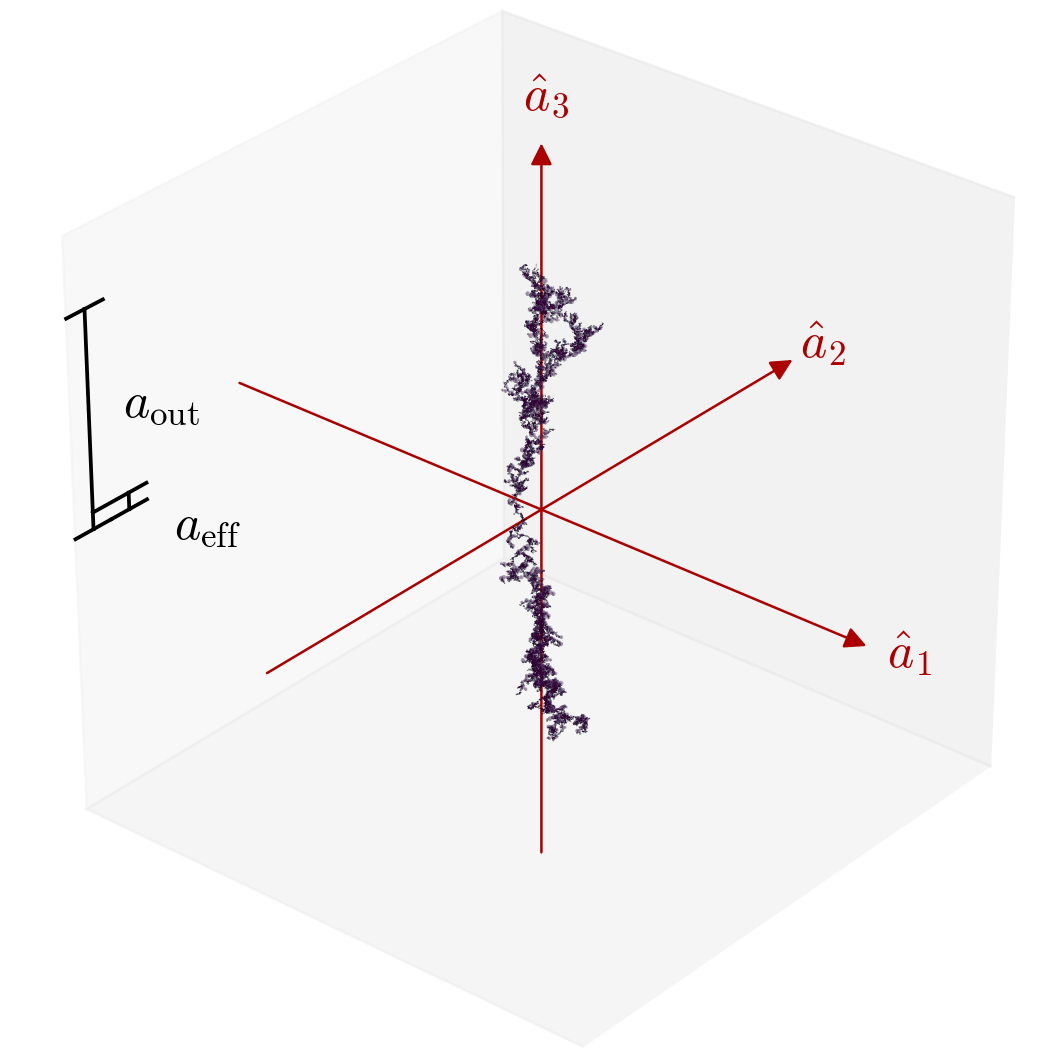}}

      & 
      \raisebox{-\totalheight}{\vspace{-10mm}\includegraphics[width=0.28\textwidth]{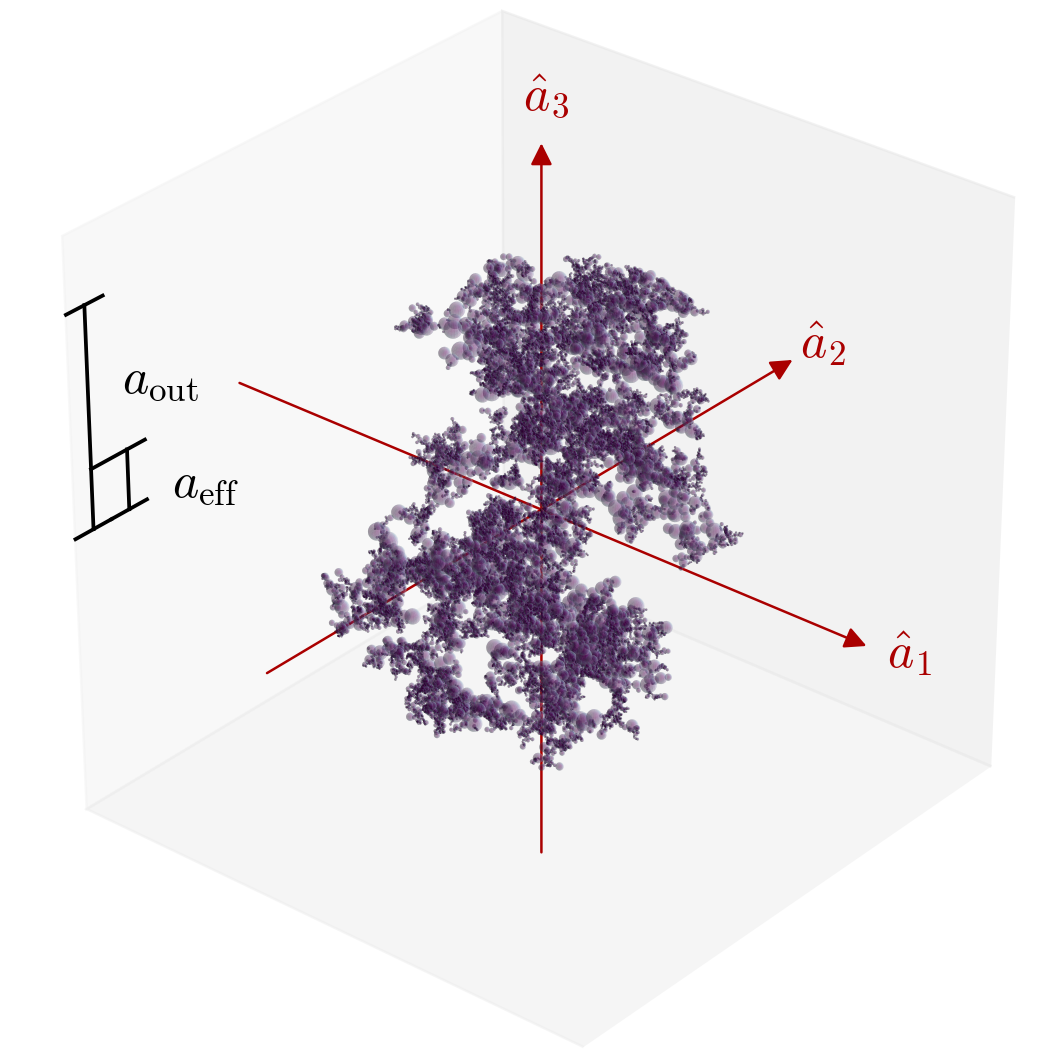}}
      & 
      \raisebox{-\totalheight}{\vspace{-10mm}\includegraphics[width=0.28\textwidth]{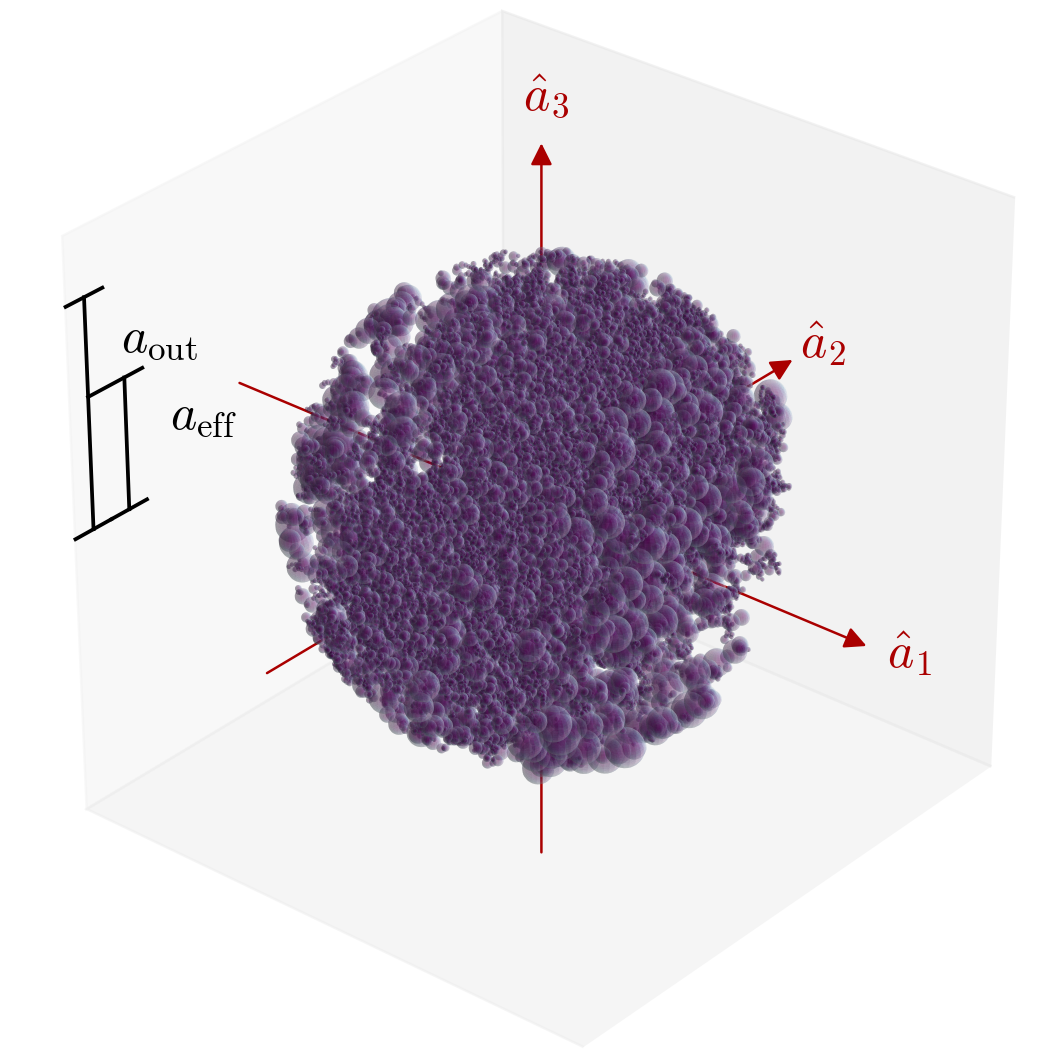}} \vspace{-1mm}
      \\
			& $a_{\mathrm{out}}=10478\ \mathrm{nm}$, $N_{\mathrm{mon}}=22249$  & $a_{\mathrm{out}}=3041\ \mathrm{nm}$, $N_{\mathrm{mon}}=22083$  & $a_{\mathrm{out}}=1463\ \mathrm{nm}$, $N_{\mathrm{mon}}=27645$\\
			\hline
      \end{tabular}
      \caption{Exemplary fractal dust grains with a number of monomers $N_{\mathrm{mon}}$. Monomer sizes and $N_{\mathrm{mon}}$ are selected in such a way to guarantee an exact effective radius $a_{\mathrm{eff}}$. Grains in the same row possess identical  radii $a_{\mathrm{eff}}$ while gains in the same column are constructed with the same fractal dimension $D_{\mathrm{f}}$. The quantity $a_{\mathrm{out}}$ is the radius of the smallest sphere enveloping the entire aggregate. Note that $a_{\mathrm{eff}}$ approaches $a_{\mathrm{out}}$ with increasing $D_{\mathrm{f}}$. Each grain's coordinate system (target-frame)  $\{\hat{a}_{\mathrm{1}},\hat{a}_{\mathrm{2}},\hat{a}_{\mathrm{3}}\}$ is unambiguously defined by its inertia tensor.}
      \label{fig:Grains}
      \end{center}
      \end{figure*}
The process of aggregation of small primary particles (so called monomers) leads to
complex fractal dust grains (see e.g. \cite{Beckwith2000}). The total number of monomers $N_{\mathrm{mon}}$ scales with  
\begin{equation}
N_{\mathrm{mon}} = k_{\mathrm{f}}  \left(  \frac{R_{\mathrm{gyr}}}{{a_{\mathrm{mean}}}}  \right)^{D_{\mathrm{f}}}
\label{eq:Nagg}
\end{equation}
where $k_{\mathrm{f}}$ is a scaling constant in the order of unity, $a_{\mathrm{mean}}$ is the mean monomer radius, $D_{\mathrm{f}}$ is the fractal dimension of the entire aggregate, and $R_{\mathrm{gyr}}$ is the radius of gyration (see below). Note that a grain with $D_{\mathrm{f}}=1$ would be like a rod while a grain with $D_{\mathrm{f}}=3$ would be spherical. The range of $D_{\mathrm{f}}$ depends on the formation history of an individual grain. Diffusive grain growth processes such as aggregate-monomer collisions result in roundish grains with higher fractal dimensions while aggregate-aggregate collisions lead to more elongated aggregates with a lower $D_{\mathrm{f}}$. {\bf A recent study by \cite{Hoang2021X} suggests that the gas-dust drift itself has an impact on the final grain shape.}  \\
In this paper we construct fractal dust grains with the algorithm as outline in \cite{Skorupski2014}. Monomers are consecutively added to an aggregate with $N_{\mathrm{mon}}>2$. Each new monomer position $\vec{X}_{\mathrm N_{\mathrm{mon}}+1}$ is semi-randomly sampled where we demand a connection point to the aggregate with at least one other monomer and the distance from the center of mass needs to follow a scaling law \citep[see][]{Filippov2000} that can be written as:
\begin{equation}
\begin{split}
\left| \vec{X}_{\mathrm{N_{mon}}+1} \right|^2 = \frac{\sigma_{\mathrm{mon}}^2(N_{\mathrm{mon}}+1)^2}{N_{\mathrm{mon}}} \left(\frac{N_{\mathrm{mon}}+1}{k_{\mathrm{f}}}\right)^{2/D_{\mathrm{f}}}  \qquad\qquad\qquad\quad\\-\frac{\sigma_{\mathrm{mon}}^2(N_{\mathrm{mon}}+1)}{N_{\mathrm{mon}}} 
-\sigma_{\mathrm{mon}}(N_{\mathrm{mon}}+1)^2  \left(\frac{N_{\mathrm{mon}}}{k_{\mathrm{f}}}\right)^{2/D_{\mathrm{f}}}\,.
\end{split}
\label{eq:FractalScaling}
\end{equation}
After each iteration we move all monomer positions such that the center of mass of the aggregate coincides with the origin of the coordinate system. \\
While dust aggregates with $D_{\mathrm{f}}<2.0$ are observed in the laboratory \citep[e.g.][]{Bauer2019}  for interstellar conditions such elongated grains are most likely short lived since they are to fragile to withstand super-thermal rotation \citep{Lazarian1995}. However, smaller grains consisting only of a few dozens of monomers may survive \cite{Chakrabarty2007}. Hence, in our study we consider a set of typical fractal dimensions of $D_{\mathrm{f}} \in \{1.6,1.8,2.0,2.2,2.4,2.6\}$. \\
Commonly, fractal dust grain models are constructed utilizing a constant monomer size \citep[see e.g.][]{Kozasa1992,Shen2008}. However, laboratory data suggests that a plethora of different materials form fractal aggregates composed of monomers with variable sizes \citep{Karasev2004,Chakrabarty2007,Slobodrian2010,Kandilian2015,Baric2018,Kelesidis2018,Salameh2017,Wu2020,Zhang2020,Kim2021}. The size distribution of the monomers itself may follow a log-normal distribution \citep{Koylu1994,Lehre2003,Slobodrian2010,Bescond2014,Kandilian2015,ChaoLiu2015,Wu2020,Zhang2020}
\begin{equation}
p\left( a_{\mathrm{mon}} \right) = \frac{1}{\sqrt{2\pi}\ln(\sigma_{\mathrm{mon}})a_{\mathrm{mon}}} \exp\left[ - \left( \frac{\ln(a_{\mathrm{mon}}/a_{\mathrm{mean}})}{\ln(\sigma_{\mathrm{mon}})}\right)^2 \right]\,,
\label{eq:SizeDist}
\end{equation}
where $\sigma_{\mathrm{mon}}$ is the standard deviation of possible monomer sizes. In order to construct our dust aggregates we take typical laboratory values of $\sigma_{\mathrm{mon}} = 1.25\ \mathrm{nm}$, $a_{\mathrm{mean}} = 16\ \mathrm{nm}$, and for the scaling factor we choose $k_{\mathrm{f}}=1.3$ \citep{Salameh2017, Wang2019,Wu2020,Zhang2020,Kim2021}. All monomer sizes are sampled from Eq. \ref{eq:SizeDist} within the limits of 
${ a_{\mathrm{mon}}\in [10\ \mathrm{nm};100\ \mathrm{nm}] }$ until a certain volume of 
\begin{equation}
V_{\mathrm{agg}} = \frac{4\pi}{3}\sum_{\mathrm{i}=1}^{ N_{\mathrm{mon}} } a_{\mathrm{mon,i}}^3
\end{equation}
is reached. This choice is also consistent with the model based on observations of dust extinction as presented in \cite{Mathis1989} with $a_{\mathrm{mon}}=5\ \mathrm{nm}$. whereas in more recent studies aggregate models are utilized with a monomer sizes in the order of $a_{\mathrm{mon}}\approx100\ \mathrm{nm}$ \citep[e.g][]{Seizinger2012,Tazaki2019}. We note that the sizes of the last three monomer are biased such that an exact effective radius of 
\begin{equation}
a_{\mathrm{eff}}=\left(  \frac{3}{4\pi} V_{\mathrm{agg}}  \right)^{1/3}
\end{equation}
is guaranteed for each individual grain.\\
Finally, the inertia tensor of the entire aggregate is calculated. This allows to define an unique coordinate system for each grain where the moments of inertia $I_{\mathrm{a_1}} > I_{\mathrm{a_2}} > I_{\mathrm{a_3}}$ are along the grain's principal axes $\hat{a}_{\mathrm{1}}$, $\hat{a}_{\mathrm{2}}$, and $\hat{a}_{\mathrm{3}}$, respectively (compare Fig.~\ref{fig:Grains} and also Appendix~\ref{app:Interia} for greater details). 
In total we construct individual grains with $50$ random seeds and a set of distinct sizes of ${a_{\mathrm{eff}}\in \{50\ \mathrm{nm},100\ \mathrm{nm},200\ \mathrm{nm},400\ \mathrm{nm},800\ \mathrm{nm}\}\ }$ for each fractal dimension $D_{\mathrm{f}}$ leading to an ensemble of $1500$ unique grains. As material of the dust we assume silicate with a typical material density of ${\rho_{\mathrm{dust}}=3000\ \mathrm{kg\ m}^{-3}}$.

\section{Gas-dust interaction}
\label{sect:GasDustInt}
\subsection{The gas velocity distribution for drifting dust}
\begin{figure}[ht!]
	\begin{center}
	\includegraphics[width=0.5\textwidth]{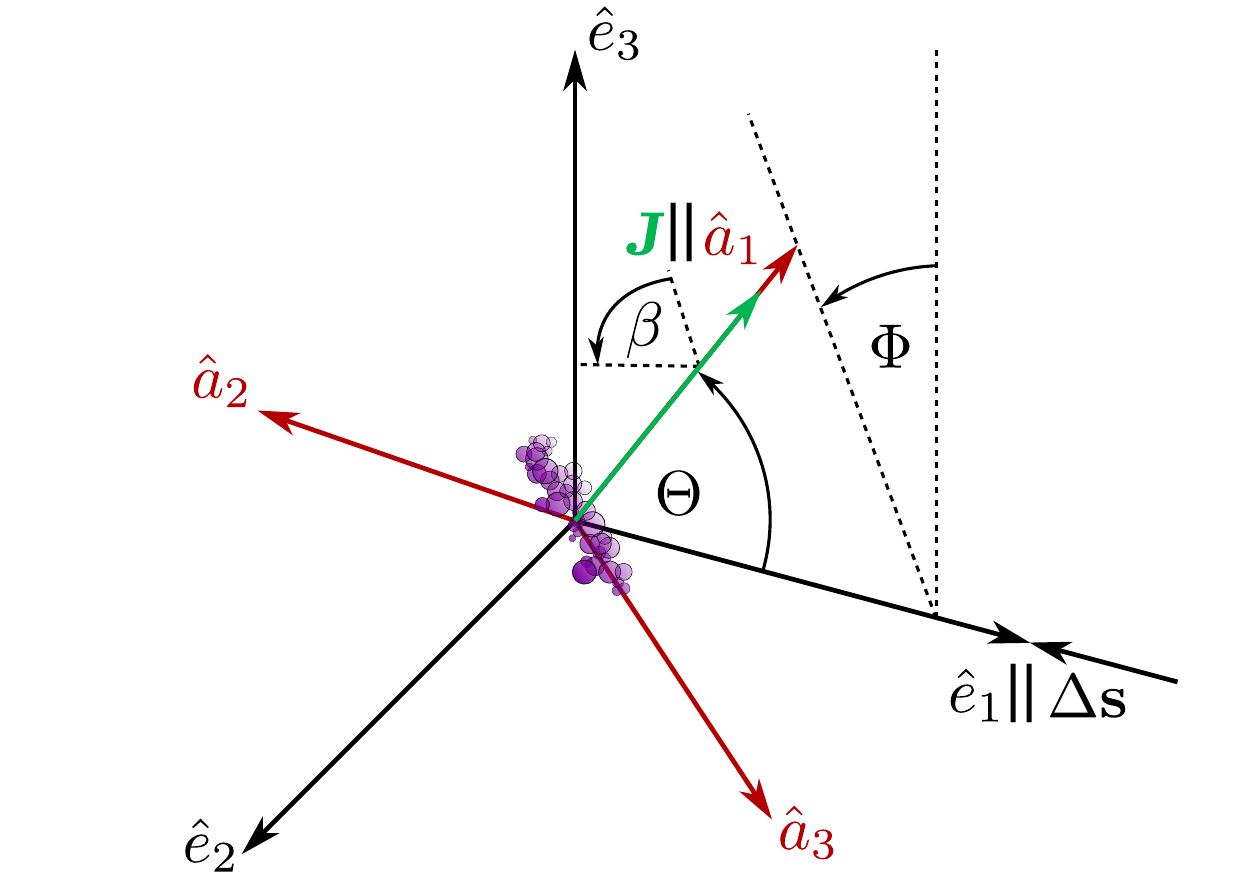}
	\end{center}
\caption{Sketch of the relation of vectors and angles between the lab-frame $\{\hat{e}_{\mathrm{1}},\hat{e}_{\mathrm{2}},\hat{e}_{\mathrm{3}}\}$ and the target-frame $\{\hat{a}_{\mathrm{1}},\hat{a}_{\mathrm{2}},\hat{a}_{\mathrm{3}}\}$. In the applied MC setup the unit vector $\hat{e}_{\mathrm{1}}$ is anti-parallel to the direction of the drift velocity $\Delta \vec{s}$. The vectors  $\hat{e}_{\mathrm{2}}$ and $\hat{e}_{\mathrm{3}}$ are chosen to create an orthonormal basis together with  $\hat{e}_{\mathrm{1}}$ for the gas properties. The target-frame $\{\hat{a}_{\mathrm{1}},\hat{a}_{\mathrm{2}},\hat{a}_{\mathrm{3}}\}$ is defined by the moments of inertia of the dust aggregate. It is assumed that the grain is rapidly rotating around the principal axes $\hat{a}_{\mathrm{1}}$ and that the angular momentum $\vec{J}$ of the grain is parallel to $\hat{a}_{\mathrm{1}}$. The rotating dust may perform a stable precession with respect to $\Delta \vec{s}$. Here, the quantities $\beta$, $\Theta$, and $\Phi$ are the angles of rotation, alignment, and precession, respectively.}
\label{fig:SketchCoordMech}
\end{figure}
\begin{figure}[ht!]
	\begin{center}
	\includegraphics[width=0.5\textwidth]{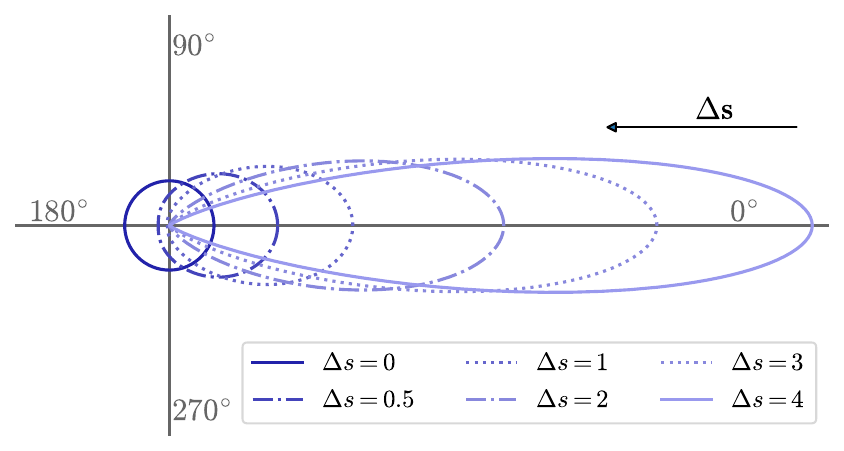}
	\end{center}
\caption{Plot of the distribution function $P_{\Delta s}(\vartheta)$. 
The gas-dust drift of $\Delta \vec{s}$ is anti-parallel with the axis $\hat{e}_{\mathrm{1}}$  while the dust grain is situated in the origin of the lab-frame. For a gas-dust drift of $\Delta s=0$ the gas velocity field is isotropic concerning the trajectories of individual gas particles. With $\Delta s>0$ the gas is more likely to approach the gain with an trajectory parallel to $\Delta \vec{s}$. Note that for a drift with $\Delta s \geq 1$ the probability to find a gas particle approaching the dust under an angle of $\vartheta=180^{\circ}$ is already virtually zero.}
\label{fig:DistPs}
\end{figure}
The momentum transfer from the gas phase onto the dust surface depends on the number of impinging gas particles per unit time and their velocity $\varv_{\rm g}$. All gas quantities are defined with respect to the lab-frame  $\{\hat{e}_{\mathrm{1}},\hat{e}_{\mathrm{2}},\hat{e}_{\mathrm{3}}\}$. The correlation between the lab-frame and the target-frame is depicted in Fig.~\ref{fig:SketchCoordMech}.\\
As for the gas component we assume an ideal gas of hydrogen atoms surrounding the dust grain. Hence, the velocities of individual atoms is governed by the Maxwell-Boltzmann distribution. This distribution predicts for the most likely velocity of all gas atoms to be 
\begin{equation}
\varv_{\rm th}=\sqrt{\frac{2 k_{\rm B}T_{\rm g}}{m_{\rm g}}}\,,
\end{equation}
whereas 
\begin{equation}
\left< \varv_{\rm th} \right>=\sqrt{\frac{8 k_{\rm B}T_{\rm g}}{\pi m_{\rm g}}}
\end{equation}
is the average velocity. Here, the quantities $T_{\rm g}$, $m_{\rm g}$, and $k_{\rm B}$ are the gas temperature,  the gas mass, and the Boltzmann constant, respectively.\\
If the gas and the dust phase decouple the grains move with a velocity $\vec{\varv}_{\mathrm{d}}$ relatively to the gas leading leading to a drift velocity of ${ \Delta \left|\vec{\varv}\right| = \left|\vec{\varv}_{\rm g}-\vec{\varv}_{\rm d}\right| }$. In this case the Maxwell-Boltzmann distribution needs to be modified to account for the gas-dust drift. For this we introduce the dimensionless velocity $s=\varv/\varv_{\rm th}$ and subsequently the gas-dust drift may be represented by ${\left|\Delta \vec{s}\right| = |\vec{s}_{\rm g}-\vec{s}_{\rm d}|}$. We emphasize that in this paper the drift $\Delta \vec{s}$ is always anti-parallel to the $\hat{e}_{\mathrm{1}}$ vector of the lab-frame. Without loosing generality for any element of the solid angle, ${ \mathrm{d}\Omega = \sin\vartheta \mathrm{d} \varphi \mathrm{d} \vartheta  }$, the dimensionless drift velocity may be written as ${ \Delta\vec{s} = (\Delta s \cos \vartheta,\Delta s \sin \vartheta,0)^T }$. Here, the quantity $\vec{s}$ represents an arbitrary gas velocity and $\vartheta$ and $\varphi$ are the polar angle  and the azimuthal angle with respect to the lab-frame. Hence, ${ |\vec{s}-\Delta \vec{s}|^2  = (s-\Delta s \cos \vartheta)^2 +(\Delta s \sin \vartheta)^2  = s^2+\Delta s^2-2s\Delta s\cos \vartheta }$. Consequently, the gas velocity distribution modified by the gas-dust drift within $\mathrm{d} \Omega$ may be evaluated as \citep[see e.g.][for further details]{Shull1978,Guillet2008,DasWeingartner2016}
\begin{equation}
f_{\rm{vel}}(s,\Delta s) \mathrm{d} \Omega= \pi^{-3/2} s^2 \exp\left[-(s^2+\Delta s^2-2s\Delta s\cos \vartheta )^2\right] \mathrm{d} \Omega\,.
\label{eq:MaxwellBoltzmanDistribution}
\end{equation}

\subsection{The angular distribution of impinging gas particles}

For zero drift the distribution of directions of individual gas trajectories within the enveloping sphere with radius $a_{\mathrm{out}}$ are uniformly distributed i.e. isotropic. However, with an upcoming gas-dust drift $\Delta s$ the gas trajectories become more likely to be parallel to the  $\hat{e}_{\mathrm{1}}$ axis increasing the anisotropic component to the gas velocity field. The exact probability to find an certain angle $\vartheta$ between an individual gas direction with respect to $\hat{e}_{\mathrm{1}}$ may be evaluated as 
\begin{strip}
\begin{equation}
P_{\Delta s}(\vartheta)=\int_{0}^{\infty}  f_{\rm{vel}}(s,\Delta s) \mathrm{d} s = \frac{ \left( 2 \Delta s \cos \vartheta + \sqrt{\pi} \exp(\Delta s^2 \cos^2 \vartheta) \left[ 1+2\Delta s^2 \cos^2 \vartheta  \right]  \left[ 1+ \erf(\Delta s \cos \vartheta) \right]           \right)   \exp(  -\Delta s^2 / 2)}{ \pi^{3/2} \left[  \left( 1 + \Delta s^2  \right)  \mathrm{I}_{0}(\Delta s^2 / 2) + \Delta s^2  \mathrm{I}_{1}(\Delta s^2 / 2) \right]}
\label{eq:ThetaDistribution}
\end{equation}
\end{strip}
where $\erf(x)$ is the error function and $\mathrm{I}_{n}(x)$ is the modified Bessel function of the first kind. A plot of the distribution function $P_{\Delta s}(\vartheta)$ is provided in Fig.~\ref{fig:DistPs}. 

\subsection{The gas-dust collision probability}
{\bf As soon as the dust starts to drift with respect to the gas the average gas velocity $\left< v_{\mathrm{th}} \right>$ increases with respect to the reference frame of the grain because of the modified Maxwell-Boltzmann distribution. We mimic this process with an event-queue sampling a number of individual gas velocities $\varv_{\mathrm{i}}$ beforehand from Eq.~\ref{eq:MaxwellBoltzmanDistribution}. Hence, each new gas particle is intersecting  the enveloping sphere with radius $a_{\mathrm{out}}$ with a rate of
\begin{equation}
	\mathcal{R}_{\mathrm{i}}=16\pi a_{\mathrm{out}}^2 n_{\mathrm{gas}}\varv_{\mathrm{i}}\,.
\end{equation}
Here, we assumed a sphere with a radius four times larger than $a_{\mathrm{out}}$ in order to guarantee the correct distribution of gas velocities and surface exposure (see also Sect.~ \ref{sect:MCSimulation}). The next time interval between two gas injection events in our MC experiment is then
\begin{equation}
	\Delta t_{\mathrm{i}}=\frac{1}{\mathcal{R}_{\mathrm{i}}}\,.
\end{equation}
In detail, after each interval $\Delta t_{\mathrm{i}}$, we inject a new gas particle from the surface of the enveloping sphere but with a random direction. Here, magnitude of the velocity follows Eq.~\ref{eq:MaxwellBoltzmanDistribution} whereas the  probability of the intersection direction on the surface of the enveloping sphere is sampled from Eq.~\ref{eq:ThetaDistribution}. Naturally, each gas particle that enters the enveloping sphere does not necessarily collide with the aggregate. The exact gas-dust collision rate depends on the fraction of occupied volume within the enveloping sphere as well as the shape of the grain. 


\subsection{Gas sticking probability and desorption}
\label{sect:StickDesorption}
\begin{figure}[ht!]
	\begin{center}
	\includegraphics[width=0.5\textwidth]{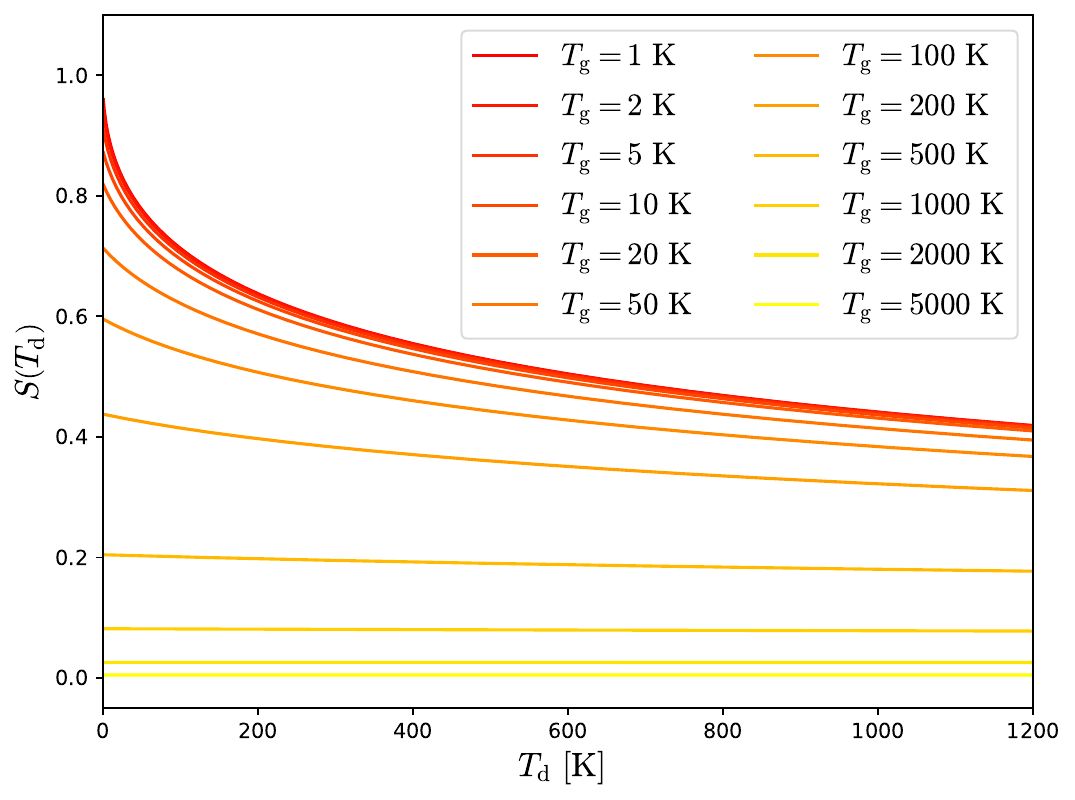} 
	\end{center}
\caption{Plot of the gas sticking probability $S(T_{\mathrm{d}})$ as a function of the dust grain temperature $T_{\mathrm{d}}$ and different gas temperatures $T_{\mathrm{g}}$. Both increasing gas and dust temperatures would decrease the probability for a gas particle to   stick on the surface of a dust grain. We emphasize that $1-S(T_{\mathrm{d}})$ is the probability of gas scattering.}
\label{fig:Sstick}
\end{figure}
Instead of being reflected specularly from the grain surface a fraction of the colliding gas particles may stick on the grain surface. The sticking mechanisms itself is still a matter of debate since the sticking and scattering of hydrogen heavily depends on the ability of the gas and dust phase to form a long lasting bond. This process is governed grain surface properties \citep[][]{Katz1999,Pirronello1999}, dust materials \citep{Katz1999,Cazaux2002}, and the temperatures of the gas and dust phase \citep[][]{Hollenbach1971,Habart2004,LeBourlot2012}.\\
Currently we are lacking these information despite substantial efforts to model the sticking probability. A commonly used sticking function may be written as 
\begin{equation}
S\left( T_{\mathrm{d}} \right) = \left[ 1+0.4\sqrt{\frac{T_{\mathrm{d}}+T_{\mathrm{g}}}{100\ \mathrm{K}}} + 0.2 \left(\frac{T_{\mathrm{g}}}{100\ \mathrm{K}}\right) + 0.08\left(\frac{T_{\mathrm{g}}}{100\ \mathrm{K}}\right)^2  \right]^{-1}
\label{eq:sticking}
\end{equation}
and is derived on the basis of statistical considerations  \citep[see][and references therein]{Hollenbach1979}. We note that this function provides phenomenologically the correct temperature dependency of sticking as presented in Fig.~\ref{fig:Sstick} but the exact coefficients may vastly differ for different grain materials and gas species.\\
We assume that the gas sticks sufficiently long enough on the grain surface to thermalize meaning the sticking gas particle reaches the same temperature as the dust grain. Consequently, the gas leaves the grain surface  with an average velocity of 
\begin{equation}
\left< v_{\mathrm{des}} \right> = \sqrt{\frac{8 k_{\rm B}T_{\rm d}}{\pi m_{\rm g}}}
\label{eq:Vdes}
\end{equation}
by means of desorption.




%

\section{The Monte-Carlo simulation}
\label{sect:MCSimulation}
By reason of the complex topology of the grain surface and the different processes involved in the gas-dust interaction we aim to calculate the resulting mechanical torque $\vec{\Gamma}_{\mathrm{mech}}$ by means of MC simulations. For each dust aggregate we inject a gas particle in intervals of $\Delta t_{\mathrm{int}}$ from a random position on the surface of the enveloping sphere with radius $4 a_{\mathrm{out}}$. Using exactly $a_{\mathrm{out}}$ for the surrounding sphere to inject gas particles may lead to an incorrect exposure of the grain surface due to self-shielding effects. The trajectories through the enveloping sphere are traced until the particle hits the dust or leaves the sphere. If gas hits the dust the particles may stick on the grain surface with a probability of $S\left( T_{\mathrm{d}} \right)$ (see Eq. \ref{eq:sticking}) and may desorb at a later time step. For simplicity we assume that the gas cannot interact with itself.\\
Considering the discrete nature gas-dust interactions the MET as a result of gas impinging on the dust grain surface may be written as ${\vec{\Gamma}_{\mathrm{mech}} =\sum_{\mathrm{i}}^{N_{\mathrm{coll}}} \Delta \vec{L}_{\mathrm{i}} / \Delta t_{\mathrm{i}} = 1/T\sum_{\mathrm{i}}^{N_{\mathrm{coll}}} \Delta \vec{L}_{\mathrm{i}} } $ where $\Delta \vec{L}_{\mathrm{i}}$ is the change in angular momentum and $T=\sum_{\mathrm{i}} {\Delta t_{\mathrm{i}}}$ is the total simulation time. {\bf We note that with an increasing number of collisions $N_{\mathrm{coll}}$ all MC simulation results eventually converge. Hence, we do not control for $T$ but demand a total number of impinging gas particle of $N_{\mathrm{coll}} = 10^3$ for each MC simulation to terminate (see Appendix~\ref{app:MCNoise}).}\\
Each MC simulation run is characterized by the set of dust parameters ${ \{ a_{\mathrm{eff}},D_{\mathrm{f}},T_{\mathrm{d}}, \mathrm{seed} \} }$ as well as the gas parameters ${ \{ n_{\mathrm{g}},T_{\mathrm{g}},\Delta s \} }$. As for the astrophysical environment we assume the typical conditions of the cold neutral medium (CNM) as listed in Tab.~\ref{tab:CNM}. Concerning the dust orientation we use ${\Theta \in [0^\circ;180^\circ]}$ for the alignment and ${\beta \in [0^\circ;360^\circ]}$ for the rotation with a resolution of $2^\circ$. We assume a rapid rotation around $\hat{a}_{\mathrm{1}}$ and average all results over $\beta$ in a final step.


\begin{table}
\centering
\begin{tabularx}{0.49\textwidth} { 
  | >{\centering\arraybackslash}X 
  | >{\centering\arraybackslash}X 
  | >{\centering\arraybackslash}X 
  | >{\centering\arraybackslash}X 
  | >{\centering\arraybackslash}X  | }
 \hline
$n_{\mathrm{g}}\ [\mathrm{m}^{-3}]$  & $T_{\mathrm{g}}\ \mathrm{[K]}$ &  $T_{\mathrm{d}}\ \mathrm{[K]}$ &  $\Delta s$ & $B\ \mathrm{[T]}$ \\
 \hline
$3 \cdot 10^{7}$  & $100$ &  $15$ &  $0.1$ &  $1.5\cdot 10^{-9}$ \\
\hline
\end{tabularx}
\caption{Applied values of gas density $n_{\mathrm{g}}$, gas temperature $T_{\mathrm{g}}$, dust temperature $T_{\mathrm{d}}$, gas-dust drift $\Delta s$ , and magnetic field strength $B$ typical for the cold neutral medium (CNM). }
\label{tab:CNM}
\vspace{2mm}
\end{table}
\begin{figure}[ht!]
	\begin{center}
	\includegraphics[width=.47\textwidth]{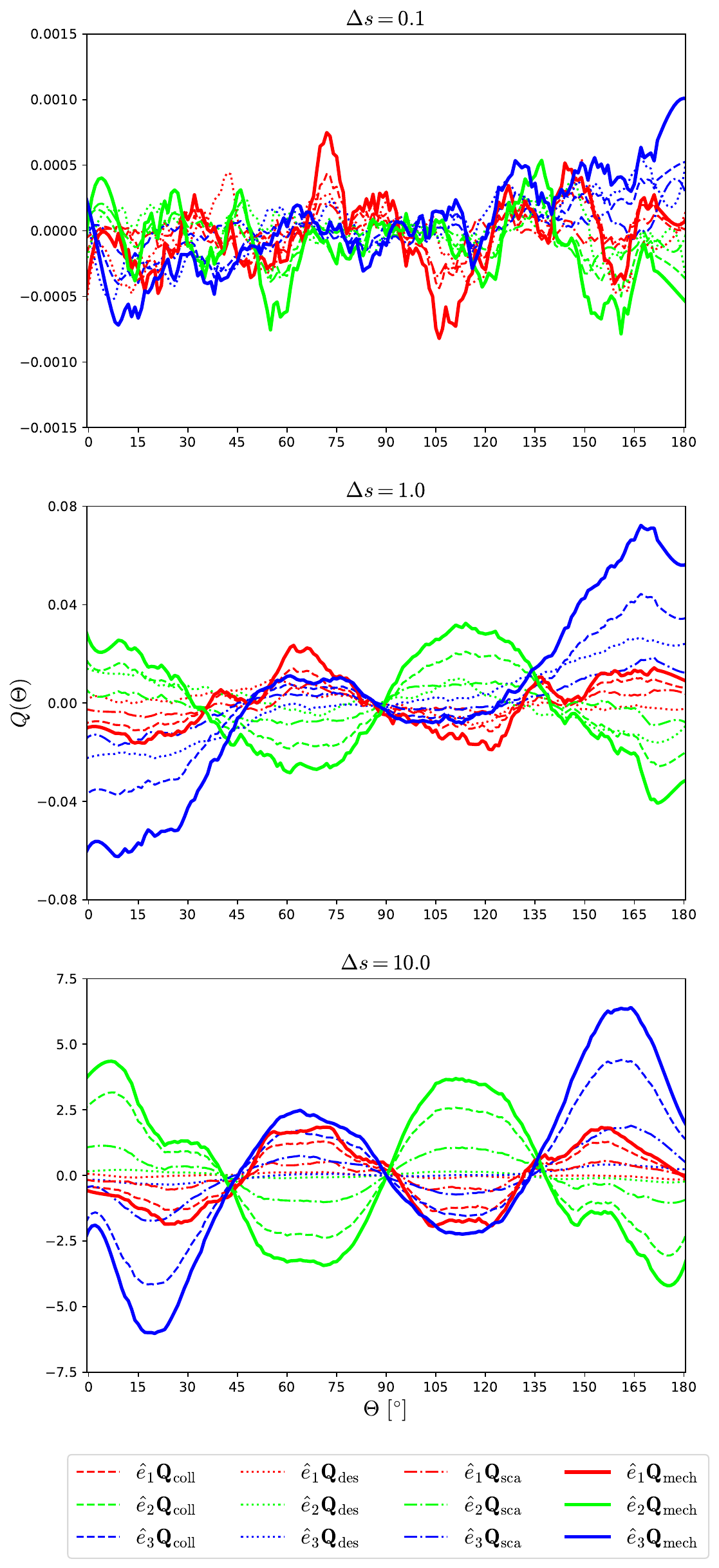}
	\end{center}
\caption{Exemplary efficiencies of gas collision $Q_{\mathrm{coll}}$, desorption $Q_{\mathrm{des}}$, and scattering scattering $Q_{\mathrm{sca}}$ as well as the total mechanical efficiency  $Q_{\mathrm{mech}}$ over alignment angle $\Theta$ and typical CNM conditions. All efficiencies are averaged over the angle $\beta$. The panels are for the different gas-dust drift of $\Delta s=0.1$ (top), $\Delta s=1.0$ (middle), and $\Delta s=10.0$ (bottom), respectively. We emphasize that all efficiencies are simulated for the grain with size ${a_{\mathrm{eff}}=400\ \mathrm{nm}}$ and fractal dimension $D_{\mathrm{f}}=2.0$ depicted in the very center of Fig.~\ref{fig:Grains}.}
\label{fig:Qs}
\end{figure}

\section{Gas induced torques}
\label{sect:TorqueMech}
\subsection{Colliding and scattering of gas particles on dust grains}
{\bf The rate of  gas particles within the velocity range ${ [s, s+\mathrm{d}s] }$ colliding onto the surface of a perfectly spherical grain with radius  $a_{\mathrm{eff}}$ is
${ \mathrm{d} \mathcal{R}_{\mathrm{coll}} = n_{\mathrm{g}} a_{\mathrm{eff}}^2 s \left< \varv_{\rm th} \right>  f_{\rm{vel}}(s,\Delta s) \mathrm{d} s}$. The transfer of angular momentum per collision at a particular position $\vec{r}$ of the surface is ${ \Delta \vec{L}_{\rm coll} =   m_{\mathrm{g}} \left< \varv_{\mathrm{th}} \right> a_{\mathrm{eff}} (\hat{r}\times\vec{s}) \cos \delta  }$. Here, $\hat{N}$ is the normal vector of the dust surface and ${ \hat{r}=\vec{r}/a_{\mathrm{eff}} }$ is the position on the surface where a distinct gas particle collides normalized by $a_{\mathrm{eff}}$. Note that the factor ${ \cos \delta = \hat{N}\vec{s}/(|\hat{N}||\vec{s}|) }$ accounts for the fact that a gas particle impinging under any angle  $\delta>0^\circ$ with respect to the normal $\hat{N}$ of the grain surface cannot fully transfer its momentum. For instance, a gas particle touching the grain perfectly parallel to its surface i.e. $\delta=90^\circ$ would not transfer any momentum at all. The resulting torque over all collision events is then ${ \vec{\Gamma}_{\mathrm{coll}} = \int \Delta \vec{L}_{\rm coll}  \mathrm{d} \mathcal{R}_{\mathrm{coll}}   }$. }\\
For dust where the grain surface can completely analytically be parameterized  by the surface element ${ \mathrm{d}\vec{A} = \hat{N}\mathrm{d}A }$ this yields a net torque of
\begin{equation}
\begin{split}
  \vec{\Gamma}_{\mathrm{coll}}^{\mathrm{AN}} = n_{\mathrm{g}}  m_{\mathrm{g}} \left< \varv_{\mathrm{th}} \right>^2 a_{\mathrm{eff}}^3 \int \int (\hat{r}\times\vec{s}) s  f_{\mathrm{vel}}(s,\Delta s) \cos \delta\  \mathrm{d}s\mathrm{d}\vec{A} = \\  n_{\mathrm{g}}  m_{\mathrm{g}} \left< \varv_{\mathrm{th}} \right>^2 a_{\mathrm{eff}}^3 \vec{Q}_{\mathrm{coll}}
\end{split}
\end{equation}
due to gas-dust collisions. Consequently, the dimensionless quantity $\vec{Q}_{\mathrm{coll}}$ represents the efficiency of collision and encompasses both the grain surface topology as well as the torque amplification by the gas-dust drift. 
For instance ${Q_{\mathrm{coll}} = |\vec{Q}_{\mathrm{coll}}|  =0}$ for a perfectly spherical grain but also for $\Delta s =0$ independent of grain shape.\\
In our MC simulation the collision torque is the sum over all singular collision events. At the i-th gas-dust collision a small force ${ \vec{F}_{\mathrm{i}}=  m_{\mathrm{g}}\vec{\varv}_{\mathrm{i}} /\Delta t_{\mathrm{i}} }$ is exerted onto the grain surface. The collision torque changes then by a discrete amount of $\Delta \vec{\Gamma}_{\mathrm{coll}} = a_{\mathrm{eff}} (\hat{r}_{\mathrm{i}} \times \vec{F}_{\mathrm{i}})$. The resulting net MC torque of collision after a total simulation time $T$ is 
\begin{equation}
\vec{\Gamma}_{\mathrm{coll}}^{\mathrm{MC}} = \gamma \frac{m_{\mathrm{g}} a_{\mathrm{eff}}}{T} \sum_{\mathrm{i}=1}^{N_{\mathrm{coll}}}{(\hat{r}_{\mathrm{i}} \times \vec{\varv}_{\mathrm{i}}) \cos \delta_{\mathrm{i}} }\,.
\end{equation}
The sum of a finite amount of random vectors i.e. the direction of the gas trajectories of the impinging gas would not approximate the zero vector (given a large enough number of random vectors) but describe a random walk. Consequently, the efficiency $\vec{\Gamma}_{\mathrm{coll}}$ cannot exactly reach zero in our MC simulations for a drift of $\Delta s=0$. Hence, we introduce the anisotropy factor of the  gas direction defined as 
\begin{equation}
\gamma=\frac{\sum_{\mathrm{i}} \left| \vec{\varv}_{\mathrm{i}}\right|}{\left| \sum_{\mathrm{i}} \vec{\varv}_{\mathrm{i}}\right|}
\end{equation}
where ${ \gamma=1 }$ represents an unidirectional gas stream and for ${ \gamma=0 }$ the gas collisions are isotropic.\\
Finally, the collision torque efficiency can be evaluated as  
\begin{equation}
  \vec{Q}_{\mathrm{coll}} =  \frac{\vec{\Gamma}_{\mathrm{coll}}^{\mathrm{MC}}}{n_{\mathrm{g}}  m_{\mathrm{g}} \left< \varv_{\mathrm{th}} \right>^2 a_{\mathrm{eff}}^3}\,.
\end{equation}
Exactly the same line of arguments holds for the efficiency $Q_{\mathrm{sca}}$  of scattered gas particles. For a completely specularly reflecting grain surface  $Q_{\mathrm{sca}}=Q_{\mathrm{coll}}$. However,
only a fraction of gas particles scatter because some particles may stick on the grain surface and desorp at a later times step. The scattering rate is ${ \mathcal{R}_{\mathrm{sca}} = \mathcal{R}_{\mathrm{coll}} \left(1-S( T_{\rm{d}} )\right) }$. In general  the relation between the efficiency of scattering and the efficiency of collision is $Q_{\mathrm{sca}} < Q_{\mathrm{coll}}$.

\subsection{Thermal desorption of gas}
The rate of gas particles that leave the surface of a spherical grain by means of desorption is related to ${ \mathcal{R}_{\mathrm{des}} = n_{\mathrm{g}} \left< \varv_{\rm th} \right> a_{\mathrm{eff}}^2 S( T_{\rm{d}} ) }$ (see Sect.~\ref{sect:StickDesorption}). For simplicity we assume that the gas particles evaporate perpendicular to the grain surface i.e. parallel to the normal vector $\hat{N}$. The transfer of angular momentum yields then ${ \Delta \vec{L}_{\rm des} =  m_{\mathrm{g}} \left< \varv_{\mathrm{des}} \right> a_{\mathrm{eff}} (\hat{r}\times\hat{N})}$. Consequently, an analytical expression of the desorption torque reads
\begin{equation}
  \vec{\Gamma}_{\mathrm{des}}^{\mathrm{AN}} =  n_{\mathrm{g}}  m_{\mathrm{g}} \left< \varv_{\mathrm{des}} \right> \left<\varv_{\mathrm{th}} \right> a_{\mathrm{eff}}^3 \vec{Q}_{\mathrm{des}}\,.
\end{equation}
The desorption torque resulting from our MC simulated can be written as
\begin{equation}
 \vec{\Gamma}_{\mathrm{des}}^{\mathrm{MC}} =  \gamma \frac{m_{\mathrm{g}} a_{\mathrm{eff}} \left< \varv_{\mathrm{des}} \right>}{T}    \sum_{\mathrm{i}=1}^{N_{\mathrm{des}}}{(\hat{r}_{\mathrm{i}} \times \hat{N}_{\mathrm{i}})}\,.
\end{equation}
{\bf Hence, the corresponding torque efficiency after $N_{\mathrm{des}}$ desorption events from  the grain surface can be evaluated via }
\begin{equation}
  \vec{Q}_{\mathrm{des}} =  \frac{\vec{\Gamma}_{\mathrm{des}}^{\mathrm{MC}}}{n_{\mathrm{g}}  m_{\mathrm{g}} \left< \varv_{\mathrm{des}} \right> \left< \varv_{\mathrm{th}} \right> a_{\mathrm{eff}}^3}\,.
\end{equation}

\subsection{The total mechanical torque (MET)}
\label{sect:TotalQMech}
From the section above follows that the total MET may be written as 
\begin{equation}
  \vec{\Gamma}_{\mathrm{mech}} = n_{\mathrm{g}}  m_{\mathrm{g}} \left< \varv_{\mathrm{th}} \right>^2 a_{\mathrm{eff}}^3 \vec{Q}_{\mathrm{mech}}
\end{equation}
with a total mechanical efficiency of
\begin{equation}
    \vec{Q}_{\mathrm{mech}}=\vec{Q}_{\mathrm{coll}}+\vec{Q}_{\mathrm{sca}}+\frac{\left< \varv_{\mathrm{des}} \right>}{\left< \varv_{\mathrm{th}}\right>}\vec{Q}_{\mathrm{des}}\,.
\end{equation}
{\bf In our study the efficiency $\vec{Q}_{\mathrm{mech}}$ of each individual dust grain is a result of a MC simulation. Hence, this quantity comes inevitably with a certain level of numerical background noise (see Appendix~\ref{app:MCNoise})}. This adds some additional ambiguity concerning the exact zero points of $\hat{e}_{\mathrm{i}}\vec{Q}_{\mathrm{mech}}$. Here, we utilize a Savitzky-Golay filter \citep[see][]{SavitzkyGolay1964} with a window length of a few degrees to minimize the MC noise while keeping the overall trends intact. \\
In Fig.~\ref{fig:Qs} we present the individual components of 
$\vec{Q}_{\mathrm{mech}}$ as a function of the alignment angle $\Theta$ for an exemplary dust grain. For a gas-dust drift of $\Delta s = 0.1$ the efficiencies are not well defined since the gas-dust interaction is dominated by the random component of the velocity field. With an increasing  $\Delta s$ the characteristics of the efficiencies  such as curve shape and zero points become significant. The exact curve of $\vec{Q}_{\mathrm{mech}}$ is now mostly due to the surface topology. Note that the magnitude of $\vec{Q}_{\mathrm{mech}}$ increases by about three orders of magnitude when the drift jumps from $\Delta s=0.1$ to $\Delta s=1.0$. However, the increase of $\vec{Q}_{\mathrm{mech}}$ is about two orders of magnitude for the jump from $\Delta s=1.0$ to $\Delta s=10.0$. This is because both the anisotropy as well as the average magnitude of the gas velocity affect the efficiency $\vec{Q}_{\mathrm{mech}}$. The isotropic component of the gas velocity field deceases while simultaneously the average gas velocity increases for an $\Delta s=10.0$.\\
We emphasize that the curves of the efficiencies plotted in Fig.~\ref{fig:Qs} are not representative because even grains with an identical fractal dimension $D_{\mathrm{f}}$ and radius $a_{\mathrm{eff}}$ but a different seed may have a vastly different characteristics concerning mechanical alignment.

\section{The drag torques of a rotating grain}
\label{sect:TorqueDrag}
Each gas particle that leaves the grain surface, may it be because of scattering or desorption, carries part of the total angular momentum of the gain away \citep[see e.g.][]{Purcell1979,DraineWeingartner1996}. We assume a rapid rotation of the dust grain with an angular velocity of $\vec{\omega}$ parallel to $\hat{a}_{\mathrm{1}}$. The additional velocity component a gas particle acquires by leaving the grain surface is  ${\vec{\varv}_{\mathrm{out}} = a_{\mathrm{eff}} (\vec{\omega}\times \hat{r}) = a_{\mathrm{eff}} \omega  (\hat{a}_{\mathrm{1}}\times \hat{r}) }$. Hence, the resulting transfer of angular momentum associated with the gas drag is ${ \Delta \vec{L}_{\rm gas} =   m_{\mathrm{g}} a_{\mathrm{eff}}  (\hat{r}\times\vec{\varv}_{\mathrm{out}})}$ \citep[see][]{DasWeingartner2016}. The resulting gas drag torque may be written as the analytical expression  
\begin{equation}
\vec{\Gamma}_{\rm gas}^{\mathrm{AN}} = n_{\mathrm{g}} m_{\mathrm{g}} \left< \varv_{\mathrm{th}} \right> a_{\mathrm{eff}}^4\omega \vec{Q}_{\rm gas}=-\hat{a}_{\mathrm{1}}  \frac{I_{\mathrm{a_1}}\omega}{\tau_{gas}}\,,
\label{eq:GammaGasAn}
\end{equation}
whereas the gas drag efficiency $\vec{Q}_{\rm gas}$ is parallel to $\hat{a}_{\mathrm{1}}$. Consequently, the gas drag acts onto the rotating dust grain with a characteristic time scale of
\begin{equation}
\tau_{\mathrm{gas}} = \frac{I_{\mathrm{a_1}}}{n_{\mathrm{g}} m_{\mathrm{g}} \left< \varv_{\mathrm{th}} \right> a_{\mathrm{eff}}^4 |\vec{Q}_{\rm gas}|}\,.
\label{eq:TotalDumping}
\end{equation}
Note that, in contrast to the mechanical efficiency  $|\vec{Q}_{\rm mech}|$, the gas drag efficiency $|\vec{Q}_{\rm gas}|>0$ even for a perfectly spherical grain.\\
\begin{figure}
\begin{center}
	\includegraphics[width=0.5\textwidth]{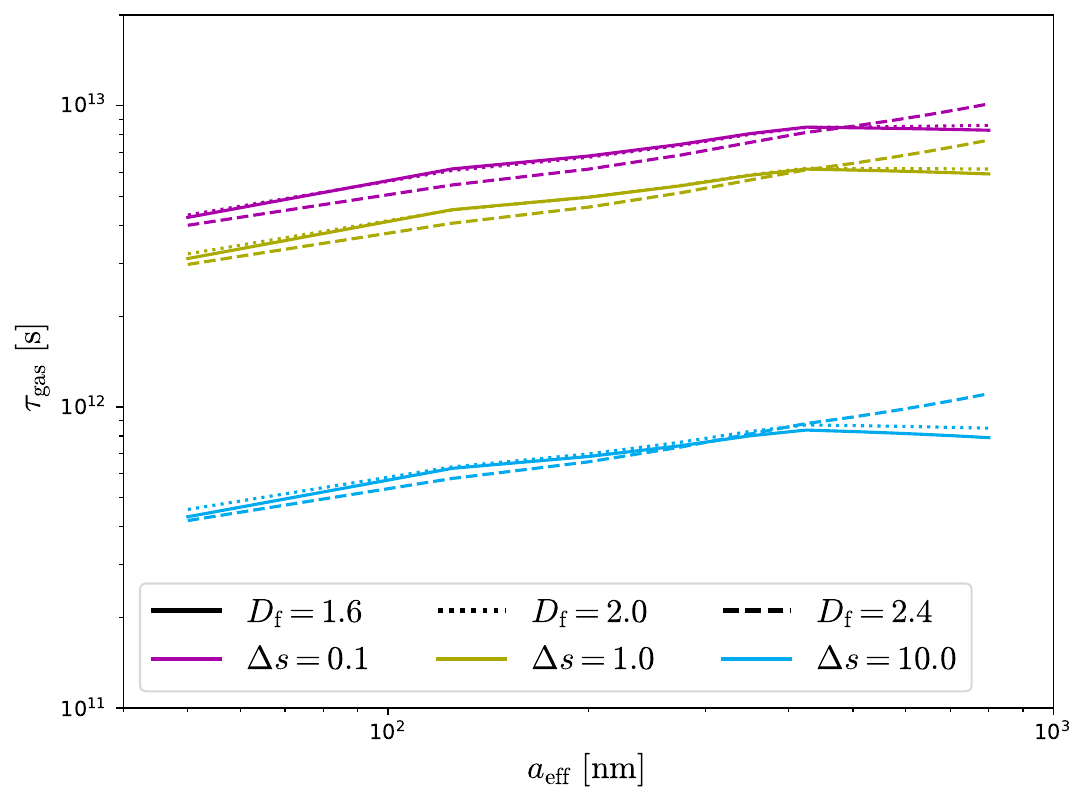}
	\end{center}
\caption{ Characteristic gas drag time scale $\tau_{\mathrm{gas}}$ of rotating grains over the effective radius $a_{\mathrm{eff}}$. The lines represent the average values over the ensemble of dust grains with a distinct fractal dimension $D_{\mathrm{f}}$ and gas-dust drift $\Delta s$ assuming typical CNM conditions.}
\label{fig:TauGas}
\end{figure}
We simulate the total gas drag that results from grain ration in our MC setup via
\begin{equation}
\vec{\Gamma}_{\rm gas}^{\mathrm{MC}} = -\frac{ m_{\mathrm{g}} a_{\mathrm{eff}}^2 \omega }{T} \sum_{\mathrm{i}=1}^{N_{\mathrm{l}}}{ \hat{r}_{\mathrm{i}} \times ( \hat{a}_1 \times \hat{r}_{\mathrm{i}}) }\,,
\label{eq:GammaGasMC}
\end{equation}
{\bf where $N_{\mathrm{l}}$ is the number of gas particles leaving the grain surface.}
\begin{figure}[ht!]
	\begin{center}
	\includegraphics[width=0.5\textwidth]{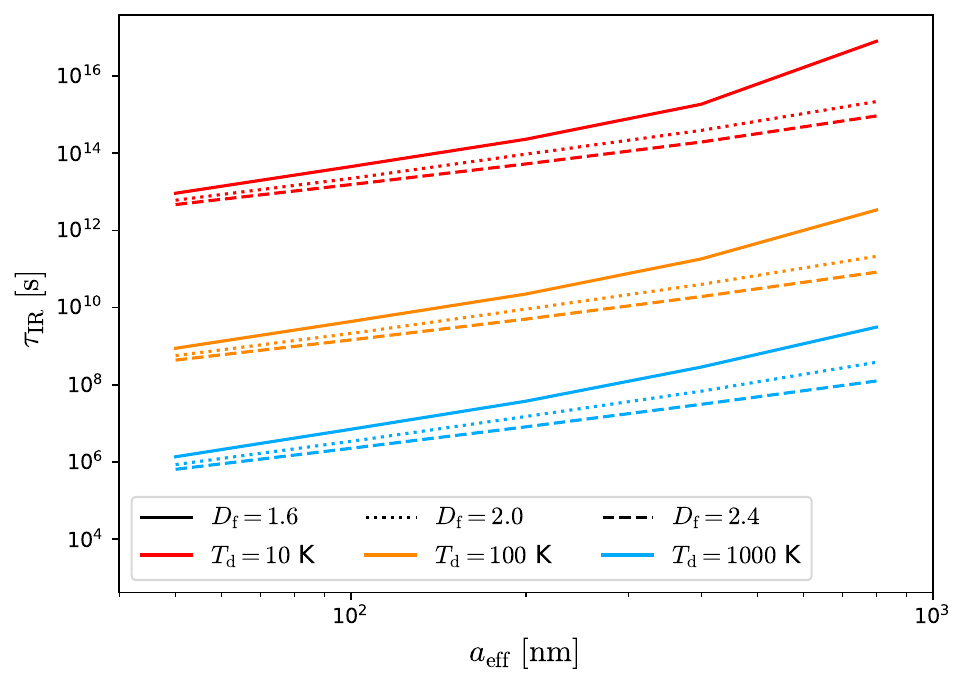}
	\end{center}
\caption{ The same as Fig.~\ref{fig:TauGas} but for the IR drag time scale  $\tau_{\mathrm{IR}}$ and different dust temperatures $T_{\mathrm{d}}$. }
\label{fig:TauIR}
\end{figure}
Hence, the unitless gas drag efficiency may be evaluated as
\begin{equation}
\vec{Q}_{\rm gas} = \frac{\vec{\Gamma}_{\rm gas}^{\mathrm{MC}}}{n_{\mathrm{g}} m_{\mathrm{g}} \left< \varv_{\mathrm{th}} \right> a_{\mathrm{eff}}^4\omega}\,.
\label{eq:dGammaChemDes}
\end{equation}
While the gas drag should act exactly along the axes $\hat{a}_{\mathrm{1}}$ of the target-frame we report some small non-zero values for the components  ${ |\hat{a}_{\mathrm{2}} \vec{Q}_{\mathrm{gas}}| \ll |\hat{a}_{\mathrm{1}} \vec{Q}_{\mathrm{gas}}| }$ and ${ |\hat{a}_{\mathrm{3}} \vec{Q}_{\mathrm{gas}}| \ll |\hat{a}_{\mathrm{1}} \vec{Q}_{\mathrm{gas}}| }$, respectively. {\bf The existence of such component is already noted in \cite{DasWeingartner2016}. However, we cannot find any systematic angular dependencies and the magnitudes are of the same order as the overall MC noise level of about ${\pm 2.5\ \%}$ (see Appendix~\ref{app:MCNoise})}. Thus, in the following sections we assume $|\vec{Q}_{\rm gas}| = |\hat{a}_{\mathrm{1}} \vec{Q}_{\mathrm{gas}}|$.\\
In Fig.~\ref{fig:TauGas} we show the gas drag time scale $\tau_{\mathrm{gas}}$ for different radii $a_{\mathrm{eff}}$.  The gas time scale appears to be only marginally dependent on the fractal dimension $D_{\mathrm{f}}$ but is heavily governed by the gas-dust drift $\Delta s$. As outlined in Sect.~\ref{sect:TotalQMech} this is because $\Delta s$ influence both the anisotropy as well as the average magnitude of the gas velocity field. The overall slope and magnitude of $\tau_{\mathrm{gas}}$ is comparable to that presented in \cite{WeingartnerDraine2003} for grain sizes $a_{\mathrm{eff}}>50\ \mathrm{nm}$.\\
Another torque that may dampen the spin-up process of grains is by means of photon emission \citep[see][]{Purcell1979,DraineLazarian1999}. Since the dust has typically temperatures in the order of $T_{\mathrm{d}}\in[10\ \mathrm{K};1000\ \mathrm{K}]$, the dust emission is for the most part in the infrared (IR) regime of wavelengths. As outlined in \cite{DraineLazarian1999} this IR damping torque may be written as 
\begin{equation}
\vec{\Gamma}_{IR}= - \hat{a}_{\mathrm{1}}\frac{ a_{\mathrm{eff}}^2 f_{\mathrm{IR}}( T_{\mathrm{d}} ) }{c^2} \omega  = -\hat{a}_{\mathrm{1}}\frac{I_{\mathrm{a_1}}\omega}{\tau_{\mathrm{IR}}}\,.
\end{equation}
Here, the constant $c$ is the speed of light and the quantity
\begin{equation}
f_{\mathrm{IR}}( T_{\mathrm{d}} ) = 4\pi\int_{0}^{\infty} \lambda^2  Q_{\mathrm{abs}}(\lambda) B_{\lambda}\left( T_{\mathrm{d}} \right) \mathrm{d}\lambda
\label{eq:fIR}
\end{equation}
represents the unitless efficiency of absorption $Q_{\mathrm{abs}}(\lambda)$ weighted over the wavelength $\lambda$ by the Planck function $B_{\lambda}( T_{\mathrm{d}} )$. Similar to $Q_{\mathrm{mech}}$ and $Q_{\mathrm{gas}}$ the efficiency $Q_{\mathrm{abs}}(\lambda)$ depends on the shape and material of the dust grain. The exact procedure to calculate $Q_{\mathrm{abs}}(\lambda)$ is outlined in Appendix~\ref{app:DDSCAT} in greater detail. Consequently, the characteristic time scale associated with the IR drag yields
\begin{equation}
\tau_{\mathrm{IR}} = \frac{c^2 I_{\mathrm{a_1}}}{  a_{\mathrm{eff}}^2  f_{\mathrm{IR}}( T_{\mathrm{d}} ) }\,.
\end{equation}
In Fig.~\ref{fig:TauIR} we present $\tau_{\mathrm{IR}}$ for grains with different sizes $a_{\mathrm{eff}}$, fractal dimensions $D_{\mathrm{f}}$, and dust temperatures $T_{\mathrm{d}}$. Similar to the gas drag $D_{\mathrm{f}}$ is the least relevant parameter of $\tau_{\mathrm{IR}}$ compared to $a_{\mathrm{eff}}$ and $T_{\mathrm{d}}$, respectively. We note that for more roundish grains i.e. ${ D_{\mathrm{f}}\geq 2.0 }$ the correlation between $\tau_{\mathrm{IR}}$ and $a_{\mathrm{eff}}$ follows a strict power-law while grains with $D_{\mathrm{f}} < 2.0$ seem to have a steeper slope for $a_{\mathrm{eff}}> 200\ \mathrm{nm}$.\\
Finally, the total drag torque acting against the mechanical spin-up process may be written as
\begin{equation}
\vec{\Gamma}_{drag}=  - \hat{a}_{\mathrm{1}}\frac{I_{a_1}\omega}{\tau_{\mathrm{drag}}}\,,
\end{equation}
where 
\begin{equation}
\tau_{\mathrm{drag}} = \left( \tau_{\mathrm{gas}}^{-1} + \tau_{\mathrm{IR}}^{-1} \right)^{-1}
\end{equation}
is the total drag time. Consequently, the net grain drag is dominated by the smaller of the two time scales $\tau_{\mathrm{gas}}$ or $\tau_{\mathrm{IR}}$, respectively.\\
{\bf In Fig.~\ref{fig:TauDrag} we show the interplay of the gas drag and IR drag, respectively, for different gas number densities $n_{\mathrm{g}}$ and dust temperatures $T_{\mathrm{d}}$. For typical CNM conditions but $T_{\mathrm{d}}=10\ \mathrm{K}$ the drag is dominated by the impinging gas. For $T_{\mathrm{d}}=100\ \mathrm{K}$ and $n_{\mathrm{g}} \lessapprox 10^9\ \mathrm{m}^3$ the total drag time is constant because of the smaller IR drag. The gas drag starts to dominate the total drag completely for gas densities $n_{\mathrm{g}} \gtrapprox 10^{11}\ \mathrm{m}^3$, decreasing the drag time by several orders of magnitude. Consequently, the mechanical spin-up process is expected to decrease or stagnate in that density regime.  Assuming hot dust with $T_{\mathrm{d}}=1000\ \mathrm{K}$ the total drag becomes nearly independent of $n_{\mathrm{g}}$ because of the effective loss of angular momentum by means of photon emission.}\\
Additional drag mechanisms such as the collision of ions and charged dust grains or by means of a so called "plasma-drag" may only become of relevance for grains with a typical size of $a_{\mathrm{eff}}<10\ \mathrm{nm}$ \citep[see][]{DraineLazarian1999}. Therefore, such additional drag effects are not considered in this paper.\\
\begin{figure*}[h]
	\begin{center}
	\includegraphics[width=1.0\textwidth]{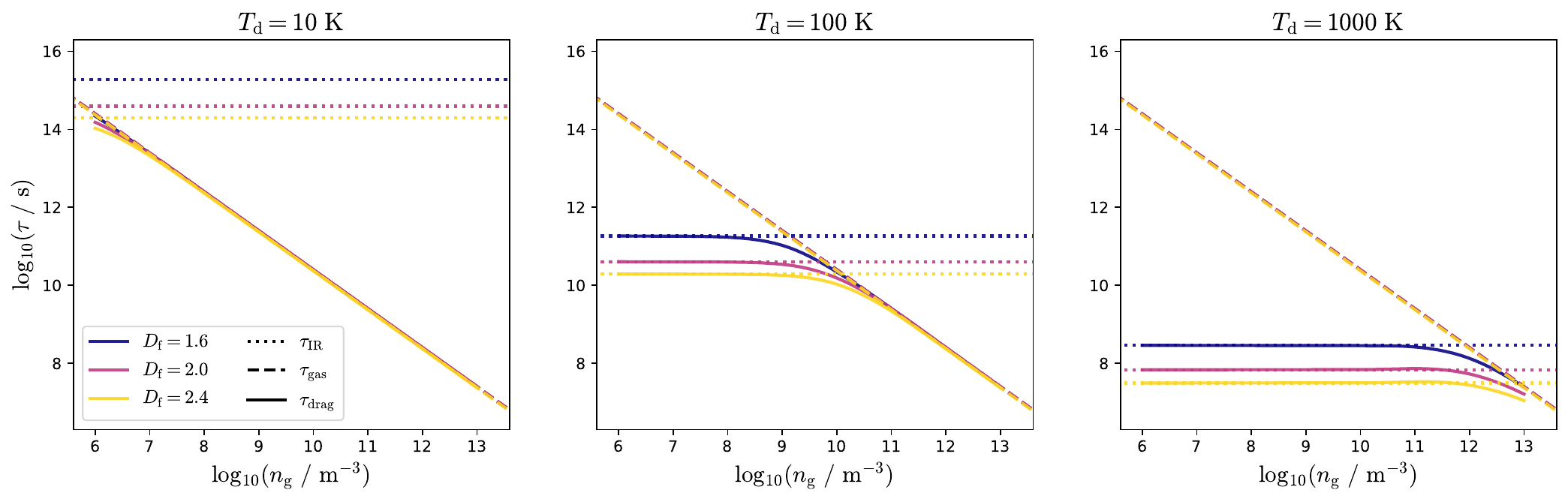}
	\end{center}
\caption{ A comparison of the characteristic time scales of the IR drag $\tau_{\mathrm{IR}}$, the gas drag $\tau_{\mathrm{gas}}$, and the total drag $\tau_{\mathrm{drag}}$, respectively, dependent on gas number densitie $n_{\mathrm{gas}}$ and fractal dimension $D_{\mathrm{f}}$. The lines represent the average values over the entire ensemble of grains with an effective radius of ${ a_{\mathrm{eff}} = 400\ \mathrm{nm} }$. For the gas we assume typical CNM conditions.}
\label{fig:TauDrag}
\end{figure*}

\section{Magnetic field induced torques}
\label{sect:MagTorque}
\begin{figure}[ht!]
	\begin{center}
	\includegraphics[width=0.5\textwidth]{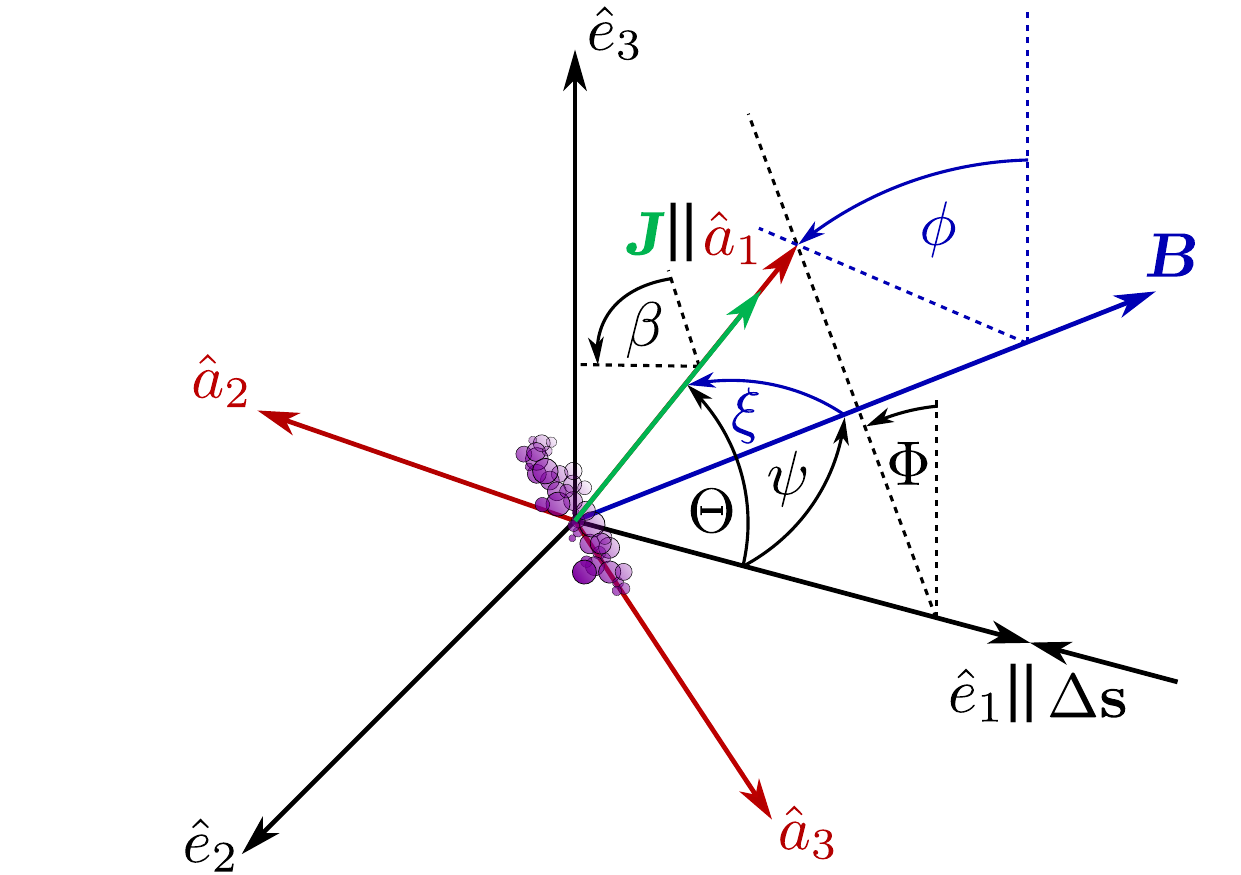}
	\end{center}
\caption{The same as Fig.~\ref{fig:SketchCoordMech} but with an external magnetic field $\vec{B}$. The precession of the grain is now with respect to the orientation of $\vec{B}$ with a precession angel of $\phi$. Note that the angle $\psi$ between the gas-dust drift $\Delta \vec{s}$ and $\vec{B}$ as well as the magnitude of $\vec{B}$ are free parameters in this model. In this configuration the resulting angle of alignment $\xi$ is between $\vec{B}$ and $\hat{a}_{\mathrm{1}}$ whereas the angles $\Theta$ and $\Phi$ may be expressed now as functions of $\xi$ and $\psi$. }
\label{fig:SketchCoordMag}
\end{figure}

\begin{figure}[ht!]
	\begin{center}
	\includegraphics[width=0.5\textwidth]{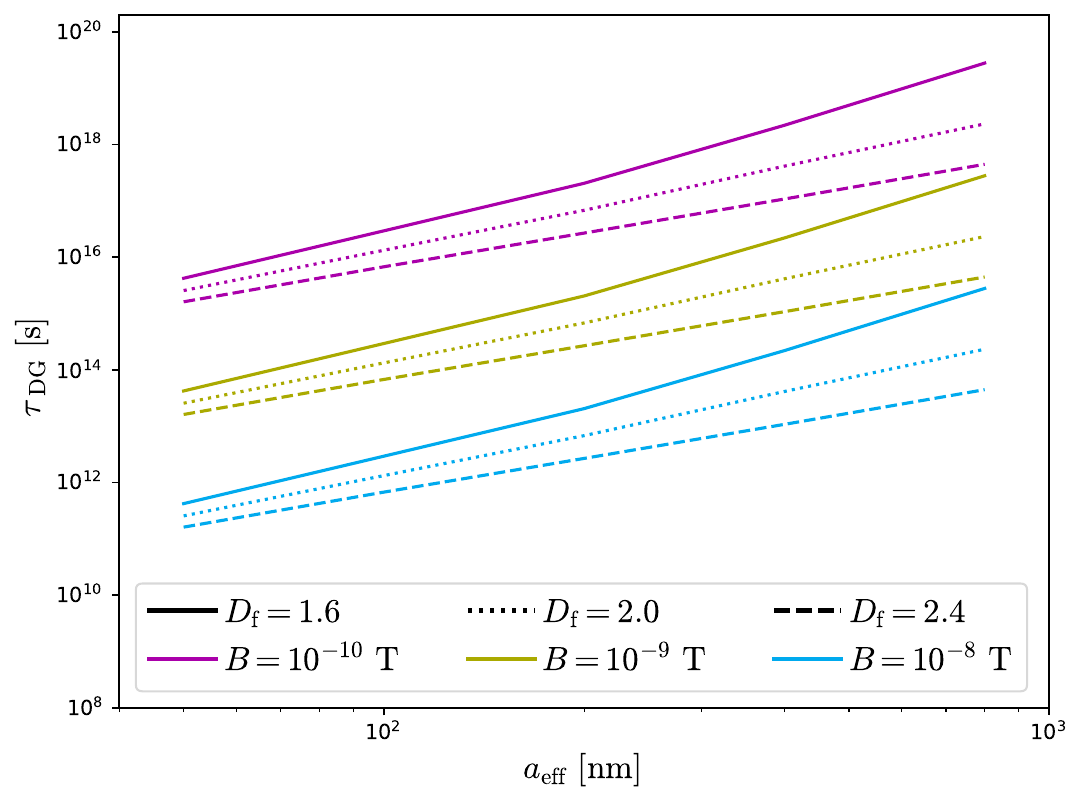}
	\end{center}
\caption{The same as Fig.~\ref{fig:TauGas} but for the DG alignment time scale  $\tau_{\mathrm{DG}}$ for different strengths $B$ of the magnetic field and typical CNM conditions.}
\label{fig:TauDG}
\end{figure}
For a dust grain that performs a precession with $\hat{a}_{\mathrm{1}}$ around the an external magnetic field $\vec{B}$ under an alignment angle of $\xi$, the field orientation in the target-frame would perpetually change. An illustration of the relation between the lab-frame and the target-frame is shown in Fig.~\ref{fig:SketchCoordMag}. In such a case the spins of free electrons within the paramagnetic material cannot follow this change in field orientation instantaneously. This leads to a transfer of the angular velocity component $\omega_{\bot}$ perpendicular to the magnetic field into dust temperature as outlined in \cite{Davis1951}. Hence, this Davis-Greenstein (DG) dissipation process aligns the dust grain with the magnetic field orientation. The characteristic time scale associated with the paramagnetic dissipation of rotational energy is
\begin{equation} 
\tau_{\mathrm{DG}} = \frac{I_{ \mathrm{a_1}}}{K V_{\mathrm{agg}} } \frac{2\mu_{\mathrm{0}}  }{B^2}\,,
\label{eq:time scaleDG}
\end{equation}
where $\mu_{\mathrm{0}}$ is the vacuum permeability, $K=\chi''/\omega_{\bot}$, and $\chi''$ is the imaginary part of the magnetic susceptibility \citep[see][for details]{Davis1951,JonesSpitzer1967}. Following \cite{Draine1996} the quantity $K$ can considered to be a constant for grains rotating with an angular velocity of $\omega \lessapprox 10^{9}s$. For silicate grains we take $K \approx 2.3\cdot 10^{-11}\  \mathrm{K\ s}/T_d$  based on the estimates\footnote{We emphasize that this paper is strictly in SI units. Hence values of K may differ by a factor of $1/(4\pi)$ when compared to other publications.} presented in \cite{JonesSpitzer1967} as well as in \cite{Draine1996}. \\
The DG torque \citep[see e.g.][]{Davis1951,Draine1996} acting on the grain is then defined as \begin{equation}
\vec{\Gamma}_{DG} = \frac{K V_{\mathrm{agg}}}{2\mu_0} \left(\vec{\omega}\times\vec{B}\right)\times\vec{B}=-\frac{I_{ \mathrm{a_1}}\omega}{\tau_{\mathrm{DG}}}\sin \xi \left( \hat{\xi}\cos \xi +\hat{a}_{1}\sin \xi  \right)\,.
\label{eq:TorqueDG}
\end{equation}
Consequently, the torque $\vec{\Gamma}_{DG}$ tends to minimize the
the alignment angle i.e. $\xi \rightarrow 0^\circ$ by means of paramagnetic 
energy dissipation over the time scale of $\tau_{\mathrm{DG}}$.\\
In Fig.~\ref{fig:TauDG} we present a plot of $\tau_{\mathrm{DG}}$ over grain size $a_{\mathrm{eff}}$ for different fractal dimensions $D_{\mathrm{f}}$ and magnetic field strengths $B$ assuming CNM conditions. Using the parametrization presented in \cite{DraineWeingartner1997} we estimate the time scale $\tau_{\mathrm{DG}}$ for roundish grains i.e. $D_{\mathrm{f}}\geq 2.4$, a size of $a_{\mathrm{eff}}=100\ \mathrm{nm}$ and a field strength of $B=5\cdot 10^{9}\ \mathrm{T}$ to be in the order of $\approx 10^{13}\ \mathrm{s}$. This is consistent with our values as shown in Fig.~\ref{fig:TauDG}. However, we note that more elongated grains have a paramagnetic dissipation of at least one order of magnitude larger than that of roundish grains. Similar to the IR drag time $\tau_{\mathrm{IR}}$, the dissipation time scales for different $D_{\mathrm{f}}$ diverge even more for larger grains sizes.\\
Another torque associated with the magnetic field is by means of the Barnett Effect \citep{Barnett1915,Dolginov1976,Purcell1979}. However, the induced Barnett torque is only responsible for the dust precession. Since we apply quantities averaged over grain precision in the following sections, we do not deal with the Barnett torque within the scope of this paper.

\section{Grain alignment dynamics}
\label{sect:GrainDyn}
In this section we outline the equation of motion related to the mechanical spin-up process and the resulting torques. Here, we distinguish between two different cases. In the first case we describe the alignment of dust grains in the mere presence of a gas-dust drift. In a second case, we investigate the grain dynamics by assuming an additional external magnetic field. 


\begin{figure}[ht!]
	\begin{center}
	\includegraphics[width=.48\textwidth]{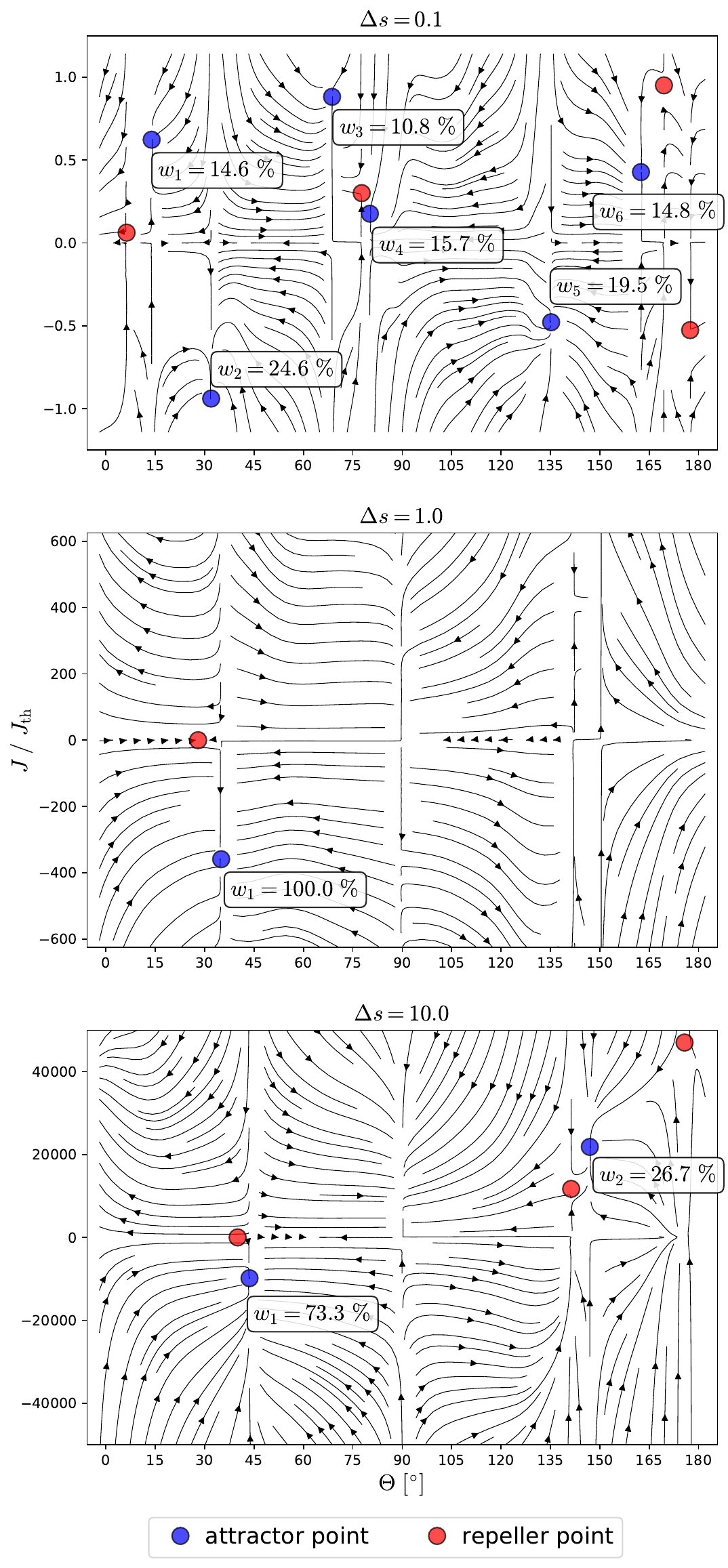}
	\end{center}
\caption{Phase portrait for the correlation between the angular momentum $J/J_{\mathrm{th}}$ and the alignment angle $\Theta$. The plotted grain dynamic is calculated with the parameters of the gas and the considered MET efficiency $\vec{Q}_{\mathrm{mech}}$ being identical to those presented in Fig.~\ref{fig:Qs}. Note that negative values of $J$ represent the case when the vectors of the grain's principal axis $\hat{a}_{\mathrm{1}}$ and $\vec{J}$ are anti-parallel and vice versa. The weights $w_{\mathrm{i}}$ describe the probability to reach a certain attractor point from any arbitrary starting point within the phase portrait.}
\label{fig:PhaseSpaceMech}
\end{figure}
\subsection{Drift velocity alignment}
The grain alignment with respect to the orientation of velocity field follows a set of equations similar to that outlined in \cite{Lazarian2007}. We emphasize that in \cite{Lazarian2007} they use a torque arising from a directed radiation field in order to account for the spin-up of the grains whereas we use the MET $\vec{\Gamma}_{\mathrm{mech}}$ as outlined in the sections above. \\
Given the plethora of internal relaxation processes such as Barnett relaxation \citep{Purcell1979,Lazarian1997}, nuclear relaxation \citep[][]{LazarianDraine1999A} or inelastic relaxation \citep[][]{Purcell1979,LazarianEfroimsky1999} any sufficiently rapidly rotating grain would inevitably align the principal $\hat{a}_{\mathrm{1}}$ to be (anti)parallel with $\vec{J}$. In fact, a stable grain alignment may only be possible under the condition of suprathermal rotation i.e. $3 J_{\mathrm{th}} \leq |\vec{J}|$ as it is claimed in \cite{HoangLazarian2008}. Here, ${  J_{\mathrm{th}}\approx (I_{\mathrm{a_1}} k_{\mathrm{B}} T_{\mathrm{g}} )^{1/2}  }$ is the thermal angular momentum a grain acquires by means of random gas bombardment. Grains with a lower angular momentum than $3 J_{\mathrm{th}}$ would easily be kicked out of alignment by means of gas collision or thermal fluctuations within the grain itself \citep[see e.g.][]{LazarianDraine1999A,LazarianDraine1999B,WeingartnerDraine2003}. 
Since we statistically evaluate only results for grains with a suprathermal rotation, effects associated with slowly rotating grains are neglected within the scope of this paper.\\
Consequently, the mechanical alignment of dust grains is governed by
\begin{equation}
\hat{a}_1\frac{\mathrm{d} J}{\mathrm{d} t}+ J \hat{\Theta}\frac{\mathrm{d} \Theta}{\mathrm{d} t} +J \sin\Theta \hat{\Phi}\frac{\mathrm{d} \Phi}{\mathrm{d} t} = \vec{\Gamma}_{\rm mech}  + \vec{\Gamma}_{\rm drag}\,.
\end{equation}
Note that this system is invariant under a rotation around the normal vector $\hat{e}_{\mathrm{1}}$ i.e. independent of the precession angle $\Phi$ (see e.g. \cite{Lazarian2007} and also Fig.~\ref{fig:SketchCoordMech}).  
We project the MET efficiency $\vec{Q}_{\mathrm{mech}}$ along the direction of the unit vectors $\hat{a}_{\mathrm{1}}=\vec{J}/|\vec{J}|$ and $\hat{\Theta}$ in order to derive the alignment component
\begin{equation}
F\left( \Theta \right) = \hat{e}_{\mathrm{1}}\vec{Q}_{\mathrm{mech}} \cos \Theta - \hat{e}_{\mathrm{2}}\vec{Q}_{\mathrm{mech}} \sin \Theta
\end{equation}
and  the spin-up component 
\begin{equation}
H\left( \Theta \right) =\hat{e}_{\mathrm{1}}\vec{Q}_{\mathrm{mech}} \sin \Theta + \hat{e}_{\mathrm{2}}\vec{Q}_{\mathrm{mech}} \cos \Theta
\end{equation}
of the MAD. By introducing the dimensionless units $\tilde{t}=t/\tau_{\mathrm{drag}}$ and $\tilde{J}=J/J_{\mathrm{th}}$ the time evolution of the angular momentum and the alignment angle may be written as

\begin{equation}
\frac{\mathrm{d} \tilde{J}}{\mathrm{d} \tilde{t}} = MH\left( \Theta \right) - \tilde{J}\,,
\label{eq:dJdt}
\end{equation}
and 
\begin{equation}
\frac{\mathrm{d} \Theta}{\mathrm{d} \tilde{t}} = M\ \frac{  F\left( \Theta \right)}{\tilde{J}}\,,
\label{eq:dJdTheta}
\end{equation}
whereas
\begin{equation}
M =  \frac{n_{\mathrm{g}} m_{\mathrm{H}}\left< \varv_{\mathrm{th}} \right>^2 a_{\mathrm{eff}}^3 \tau_{\mathrm{drag}}}{J_{\mathrm{th}}}    
\end{equation}
simply combines the basic physical parameters into one quantity.\\
A prerequisite for stable grain alignment is the presence of static points i.e. $\mathrm{d} \tilde{J} /  \mathrm{d} \tilde{t} = 0$  and $\mathrm{d} \Theta /  \mathrm{d} \tilde{t} = 0$, respectively. We evaluate the static points as outlined in Appendix~\ref{app:StaticPoints} in order to determine possible attractor and repeller points of the mechanical grain alignment dynamics.\\
Dependent on the dust and gas parameters a distinct grain may have several attractor points in the phase space. Consequently, one grain might contribute multiple times in the ensemble statistic of stable alignment points. In order to deal with this problem we randomly sample a number of points $N_{\mathrm{samp}}$ within the phase space. We trace the trajectories of each sample point on a time scale of $\Delta t \ll \Delta t_{\mathrm{drag}} $ and count the number of sample points $N_{\mathrm{i}}$ that approach the i-th attractor point within a limit of $1\ \%$ of the full range of the phase space. Subsequently, we assign a weight of $w_{\mathrm{i}}=N_{\mathrm{i}}/N_{\mathrm{samp}}$ to each distinct attractor point in our followup analysis.\\
In Fig.~\ref{fig:PhaseSpaceMech} we present the $J/\Theta$ phase space of an exemplary dust grains three distinct values of $\Delta s$. We note that the panels in {\bf Fig.~\ref{fig:PhaseSpaceMech} correspond to the ones shown in Fig.~\ref{fig:Qs}. As shown in Fig.~\ref{fig:Qs} for $\Delta s<0.1$ the grain dynamics is dominated by random gas bombardment. Hence, the grain alignment dynamics is chaotic with multiple attractor points as depicted in the top panel of Fig.~\ref{fig:PhaseSpaceMech}. All attractor points possess a probability of roughly $15\% - 25\%$ for the grain to settle down. However, all of the attractor points have an angular momentum of  $3J_{\mathrm{th}} > |J|$ and cannot be considered to be stable on longer time scales. As the gas-dust drift reaches a value of $\Delta s=1.0$ only the one attractor point at $\Theta = 32^\circ$ remains with ${J\approx -380 J_{\mathrm{th}}}$. In this configuration the dust grain possesses a stable alignment configuration. At a gas-dust drift of $\Delta s=10.0$ the attractor point is slightly shifted from an alignment angle of $\Theta = 32^\circ$  towards $\Theta = 45^\circ$ with an angular velocity of ${J\approx -1.7\cdot 10^4 J_{\mathrm{th}}}$. Simultaneously, a second attractor point starts to appear at $\Theta = 150^\circ$ and ${J\approx 2.4\cdot 10^4 J_{\mathrm{th}}}$. An alignment with an angle of $\Theta = 45^\circ$ remains to be the most likely with a probability of $73.3\ \%$. For typical CNM conditions and $\Delta s\geq 5.0$ most dust grains have one to three stable alignment configurations with $ 3J_{\mathrm{th}} < |J|$.} Note that the phase portraits of Fig.~\ref{fig:PhaseSpaceMech} are only exemplary and not representative for the entire grain ensemble with $a_{\mathrm{eff}}=400\ \mathrm{nm}$ and $D_{\mathrm{f}}=2.0$.\\
For the entire ensemble of considered grains we find that only a small fraction in the order of a few per mille have no attractor points at all. {\bf Hence, it appears to be possible for almost all considered grain shapes to be mechanically aligned as soon as the condition $ 3J_{\mathrm{th}} < |J|$ is given.}

\subsection{Magnetic field alignment}
\label{sect:MagFieldAlignment}
In the presence of an external magnetic field $\vec{B}$ the grain may perform a precession around the direction of $\vec{B}$ instead of the gas-dust drift $\Delta \vec{s}$\footnote{We emphasize that exact conditions of whether $\Delta \vec{s}$ or $\vec{B}$ is the preferential direction of grain alignment is to be dealt with in an upcoming paper. Within the scope of this paper we simply assume an grain alignment with the magnetic field direction for all grains in order to explore the distribution of long-term stable attractor points.}. Hence, the time evolution of the grain alignment dynamics may be evaluated as
\begin{equation}
\hat{a}_{\mathrm{1}}\frac{\mathrm{d} J}{\mathrm{d} t}+J \hat{\xi} \frac{\mathrm{d} \xi}{\mathrm{d} t}+J \sin\xi \hat{\phi}  \frac{\mathrm{d} \phi}{\mathrm{d} t}= \vec{\Gamma}_{\rm mech} + \vec{\Gamma}_{\rm drag} + \vec{\Gamma}_{\rm DG} \,.
\label{eq:MotionBasic}
\end{equation}
Here, the precession angle is $\phi$ and the alignment angle $\xi$ is now defined to be between $\vec{B}$ and the principal axis $\hat{a}_{\mathrm{1}}$ (see Fig.~\ref{fig:SketchCoordMag}). This approach to describe the magnetic field alignment in
terms of the vectors $\hat{a}_{\mathrm{1}}$, $\hat{\xi}$, and $\hat{\phi}$, respectively, is similar to the one presented in \cite{DraineWeingartner1996} \citep[see also e.g.][]{Lazarian2007,DasWeingartner2016}. \\
Note that the MET efficiency $\vec{Q}_{\mathrm{mech}}$ derived by our MC simulations is defined by the alignment angle $\Theta$. The transformation of $\vec{Q}_{\mathrm{mech}}$ into the coordinate system of Eq. \ref{eq:MotionBasic} reads
\begin{equation}
\Theta=\arccos\left( \cos\xi \cos\psi - \sin\xi \sin\psi \cos\phi \right)
\end{equation}
\citep[see e.g.][for furter details]{DraineWeingartner1996,DraineWeingartner1997} while the precession angle is
\begin{equation}
\Phi=2 \arctan \left( \frac{ \sin\Theta-\cos\xi\sin\psi-\cos\psi\sin\xi\cos\phi }{ \sin\xi\sin\phi} \right)\,.
\end{equation}
By assuming an external magnetic field the alignment component $F$ and the spin-up component $H$ of the MET efficiency may now be written as
\begin{equation}
\begin{split}
        F\left(\xi,\psi,\phi\right)=\hat{e}_{\mathrm{1}}\vec{Q}_{\mathrm{mech}} \left( -\sin \psi\cos \xi\cos \phi - \cos \psi\sin \xi \right) + \\
          \hat{e}_{\mathrm{2}}\vec{Q}_{\mathrm{mech}} \left[ \cos \Phi\left(\cos \psi\cos \xi\cos \phi - \sin \psi\sin \xi\right) +\right. \\ \left. \sin \Phi\cos \xi\sin \phi \right] + 
          \hat{e}_{\mathrm{3}}\vec{Q}_{\mathrm{mech}} \left[ \cos \Phi\cos \xi\sin \phi\ + \right. \\ \left. \sin \Phi\left(\sin \psi\sin \xi - \cos \psi\cos \xi\cos \phi\right)\right] 
\end{split}
\end{equation}
and 
\begin{equation}
\begin{split}
        H\left(\xi,\psi,\phi\right)=\hat{e}_{\mathrm{1}}\vec{Q}_{\mathrm{mech}} \left( \sin \psi\sin \phi\right) +\qquad\qquad\qquad\qquad\qquad \\ \qquad
          \hat{e}_{\mathrm{2}}\vec{Q}_{\mathrm{mech}} \left( \sin \Phi\cos \phi - \cos \Phi\cos \psi\sin \phi \right) +\\ 
          \hat{e}_{\mathrm{3}}\vec{Q}_{\mathrm{mech}} \left( \cos \Phi\cos \phi + \sin \Phi\cos \psi\sin \phi \right)   
\end{split}
\end{equation}
\citep[see e.g.][]{WeingartnerDraine2003,Lazarian2007}. The grain precession is expected to be much faster than the grain alignment time scale \citep[][]{DraineWeingartner1997}. {\bf This allows to  average the alignment component and the spin-up component over the precession  angle $\phi$ via}
\begin{equation}
\overline{F}\left(\xi,\psi\right) = \frac{1}{2\pi} \int_0^{2\pi}{F\left(\xi,\psi,\phi\right) \mathrm{d}\phi}
\end{equation}
and
\begin{equation}
\overline{H}\left(\xi,\psi\right) = \frac{1}{2\pi} \int_0^{2\pi}{H\left(\xi,\psi,\phi\right) \mathrm{d}\phi}\,,
\end{equation}
 respectively. Finally, splitting the equation of grain dynamics into its individual variables gives 
\begin{equation}
\frac{\mathrm{d} \xi}{\mathrm{d} \tilde{t}}=M\frac{\overline{F}\left(\xi, \psi \right)}{\tilde{J}}  - \delta_{\mathrm{m}}\sin \xi \cos \xi
\label{eq:dXidt}
\end{equation}
and
\begin{equation}
\frac{\mathrm{d} \tilde{J}}{\mathrm{d} \tilde{t}}=M \overline{H}\left(\xi, \psi  \right)  - \tilde{J}\left(1+ \delta_{\mathrm{m}}\sin^2 \xi\right)\,,
\label{eq:dJdtMag}
\end{equation}
where 
\begin{equation}
\delta_{\mathrm{m}}=\frac{f_{\mathrm{mag}}\tau_{\mathrm{drag}}}{\tau_{\mathrm{DG}}}
\label{eq:DeltaM}
\end{equation}
is a measure of the impact of the applied gas parameters as well as the magnetization of the grain.\\
Note that the characteristic DG alignment time $\tau_{\mathrm{DG}}$ goes with the field strength $B^{-2}$ (see Eq. \ref{eq:time scaleDG} and also Fig.~\ref{fig:TauDG}). Furthermore, silicate grains are believed to harbor small clusters of pure iron \citep[see e.g.][and references therein]{Jones2013} increasing the susceptibility $\chi$ of the dust material and subsequently the quantity $K$ may be larger as assumed in Sect.~\ref{sect:MagTorque}. {\bf Since both magnetic field strength $B$ and the quantity $K$ are free parameters in our alignment models the quantity $\delta_{\mathrm{m}}$ may vary by several orders of magnitude. Hence, we introduce the amplification factor $f_{\mathrm{mag}}$ in Eq. \ref{eq:DeltaM} to explore the impact of the two parameters of grain magnetization, i.e. possible variations in the magnetic field strength $B$ and the susceptibility $\chi$ of the grain material, respectively.}

%


%
\section{Dust destruction and polarization}
\label{sect:DustPolDest}
\subsection{Rotational disruption of dust grains}
\begin{figure}[ht!]
	\begin{center}
	\includegraphics[width=0.5\textwidth]{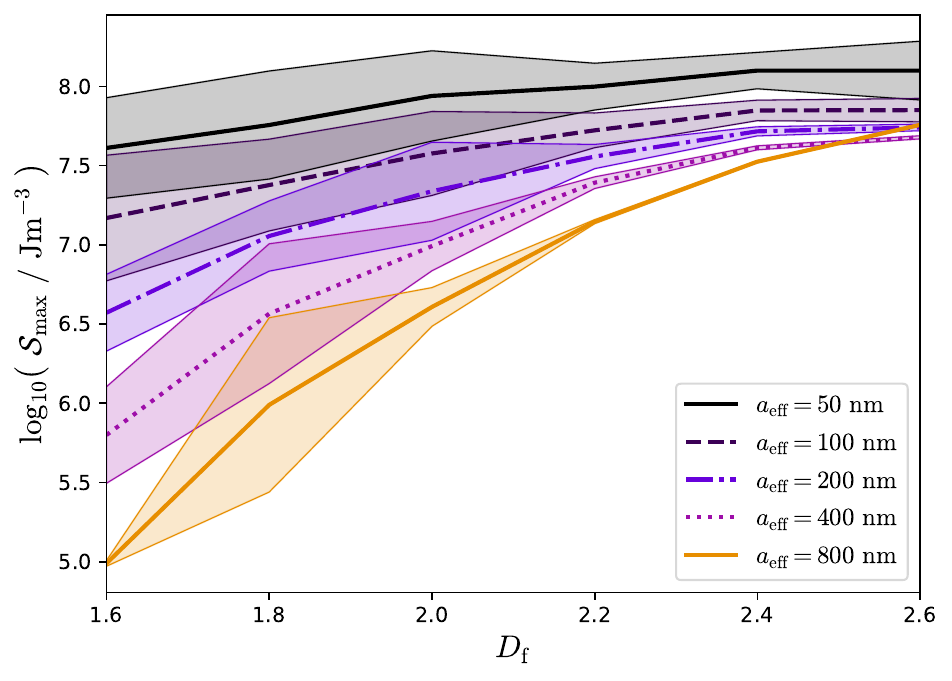}
	\end{center}
\caption{The estimated tensile strength $\mathcal{S}_{\mathrm{max}}$ over fractal dimension $D_{\mathrm{f}}$ for all the considered grain sizes $a_{\mathrm{eff}}$.}
\label{fig:Smax}
\end{figure}

Rapidly rotating grains may be disrupted by means of centrifugal forces. Rotational disruption in the context of RATs was recently studied in \cite{Hoang2019}. A similar process may occur in the presence of a strong MET.  Following \cite{Hoang2019} a dust grain might become rotationally disrupted when the grain exceeds the critical angular momentum of
\begin{equation}
    J_{\mathrm{disr}} = \frac{I_{\mathrm{a_1}}}{a_{\mathrm{eff}}}  \left(  \frac{\mathcal{S}_{\mathrm{max}}  }{ \rho_{\mathrm{dust}}  }  \right)^{1/2}\,.
\end{equation}
Here, the quantity $\mathcal{S}_{\mathrm{max}}$ is the tensile strength related to the dust material. We emphasize that for fractal aggregates the magnitude of $\mathcal{S}_{\mathrm{max}}$ is still a matter of debate. An analytical expression for a solid body is discussed in \cite{Hoang2020galaxies}. However, numerical simulations suggest that the tensile strength $\mathcal{S}_{\mathrm{max}}$ of fractal grains  may vary by several orders of magnitude depending on monomer size and initial grain shape \citep[][]{Seizinger2013,Tatsuuma2019}. We estimate the tensile strength by 
\begin{equation}
    \mathcal{S}_{\mathrm{max}} =  \frac{3}{2} \left< N_{\mathrm{con}} \right>  \frac{\left(1 - \mathcal{P} \right)E_{\mathrm{b}}}{h\ a_{\mathrm{mean}}^2}
\end{equation}
as outlined in \cite{Greenberg1995} where $\left< N_{\mathrm{con}} \right>$ is the average number of connections between all monomers within the dust aggregate and the quantity $h$ is associated with the overlap at each connection point. Here, we assume the overlap to be ${ h\approx10^{-3}a_{\mathrm{mean}} }$ and apply a binding energy of ${ E_{\mathrm{b}}=1.22\cdot 10^{-21}\ \mathrm{J} }$ \citep[][]{Greenberg1995}.\\
The porosity $\mathcal{P} \in [0;1]$ of a grain quantifies the amount of empty space within the grain aggregate where $\mathcal{P}=0$ would represent a solid body. However, $\mathcal{P}$ is not well defined in literature. For instance in \cite{Ossenkopf1993} the porosity is calculated based on the geometric cross section of the dust grain whereas a method based on comparing the moments of inertia is utilized in \cite{Shen2008}. Hence, the values of $\mathcal{P}$ provided in the literature may differ within a few percent. For the fractal grains presented in this paper we apply the expression 
\begin{equation}
    \mathcal{P} = 1- \left( \frac{a_{\mathrm{eff}}}{R_{\mathrm{c}}} \right)^3
\end{equation}
as suggested by \cite{Kozasa1992}, where the critical radius $R_{\mathrm{c}}=\sqrt{5/3}R_{\mathrm{gyr}}$ and $R_{\mathrm{gyr}}$ is the radius of gyration. Since $R_{\mathrm{gyr}}$ is connected to the total number of monomers $N_{\mathrm{mon}}$ and the fractal dimension $D_{\mathrm{f}}$ (see Eq.~\ref{eq:Nagg}) this seems to be the natural way to define the porosity of the fractal aggregates utilized in this paper. We emphasise that all aggregates applied in this study are shifted to have the center of mass coinciding with the origin of the coordinate system (see Sect.~\ref{sect:DustGrainModel}). Consequently, the radius of gyration may be written as 
\begin{equation}
    R_{\mathrm{gyr}} =  \sqrt{ \frac{ \sum_{\mathrm{i}}^{N_{\mathrm{mon}}} a_{\mathrm{mon,i}} ^3  \left|\vec{X}_{\mathrm{i}}  \right|^2    }{ \sum_{\mathrm{i}}^{N_{\mathrm{mon}}}  a_{\mathrm{mon,i}} ^3 }}\,.
\end{equation}
In Fig.~\ref{fig:Smax} we present the resulting tensile strength $\mathcal{S}_{\mathrm{max}}$ over fractal dimension $D_{\mathrm{f}}$. We note that $\mathcal{S}_{\mathrm{max}}$ converges as the fractal dimension $D_{\mathrm{f}}\rightarrow 3$. However, for more elongated grains we report that $\mathcal{S}_{\mathrm{max}}$ differs by three orders of magnitude for different grain sizes $a_{\mathrm{eff}}$ where the largest grains have the smallest $\mathcal{S}_{\mathrm{max}}$.\\
For the average number of connections $\left< N_{\mathrm{con}} \right>$ we find typically values of about $2$ - $4$ independent of $D_{\mathrm{f}}$ but a porosity $\mathcal{P} \gg 0.9$ for $D_{\mathrm{}}\leq 1.8$ and $a_{\mathrm{eff}}\geq 400\ \mathrm{nm}$. Furthermore, for interstellar dust a porosity $\mathcal{P} \approx 0.2$ is usually applied \citep[][]{Guillet2018}. Considering the low tensile strength and subsequent threshold $J_{\mathrm{disr}}$ together with the high porosity it is highly unlikely to find such elongated grains in greater numbers in the CNM.

\subsection{Dust polarization measure}
\label{sect:RRF}
The maximally possible polarization of a grain is determined by its differential cross section parallel and perpendicular to its axis of rotation. However, several factors may reduce the maximal polarization. One factor an imperfect alignment i.e. an alignment angle of $\Theta>0^\circ$ for drift alignment or an angle $\xi>0^\circ$, respectively, in case of magnetic field alignment. A way to quantify imperfect alignment and subsequently the reduction is polarization is by means of the Rayleigh reduction factor (RRF)
\begin{equation}
    R=\frac{3\zeta}{2}\left( \left<\cos^2 \Theta  \right> -\frac{1}{3} \right)
\end{equation}
\citep[][]{Greenberg1968}, where $\left<\cos^2 \Theta  \right>$ is the ensemble average over all alignment angles $\Theta$ of the individual grains. For instance $R=1$ stands for perfect alignment, i.e. $\Theta=0^\circ$, and $R=0$ may represent a completely randomized ensemble of dust grains\footnote{Note that $R=0$ might also represent the case where all grains coincidentally have a stable alignment at an angle of $\Theta=54.7^\circ$.}.\\
Note that we use a slightly modified  RRF by introducing the quantity $\zeta=N_{\mathrm{att}}/N_{\mathrm{tot}}$ where $N_{\mathrm{att}}$ is the number of grains possessing at least one attractor point with an angular momentum $J$ in between ${ 3J_{\mathrm{th}} < |J| <  J_{\mathrm{disr}} }$ and $N_{\mathrm{tot}}$ is the total number considered grains per set of input parameters. For instance $\zeta=0$ for an ensemble of completely randomized grains ($3J_{\mathrm{th}}<|J|$) but also for the case when the entire grain ensemble becomes rotationally disrupted ($|J|>J_{\mathrm{disr}}$).\\
Finally, we evaluate the ensemble average by \begin{equation}
    \left<\cos^2 \Theta  \right> =  \frac{\sum_{\mathrm{i}} w_{\mathrm{i}} \cos^2 \Theta_{\mathrm{i}} }{ \sum_{\mathrm{i}}  w_{\mathrm{i}} }  
\end{equation}
in order to get the RRF of mechanical alignment. Note that we only add up attractor points within the range ${ 3J_{\mathrm{th}} < |J| <  J_{\mathrm{disr}} }$. Furthermore, attractor points at an alignment angle of $\Theta=0^\circ$ or $\Theta=180^\circ$, respectively would contribute equally to the net dust polarization. Hence, we map all attractor points  with $\Theta>90^\circ$ in the following sections to ${ \Theta \rightarrow 180^\circ-\Theta }$ in order to get more data points for the statistics of dust polarization. We follow exactly the same procedure to calculate the RRF for the alignment angle $\xi$ in case of the magnetic field alignment.

\section{The alignment behavior of grain ensembles}
\label{sect:Alignemnt}

\subsection{The mechanical spin-up process}
\label{sect:MechJ}
\begin{figure*}[ht!]
\begin{center}
     \includegraphics[width=0.99\textwidth]{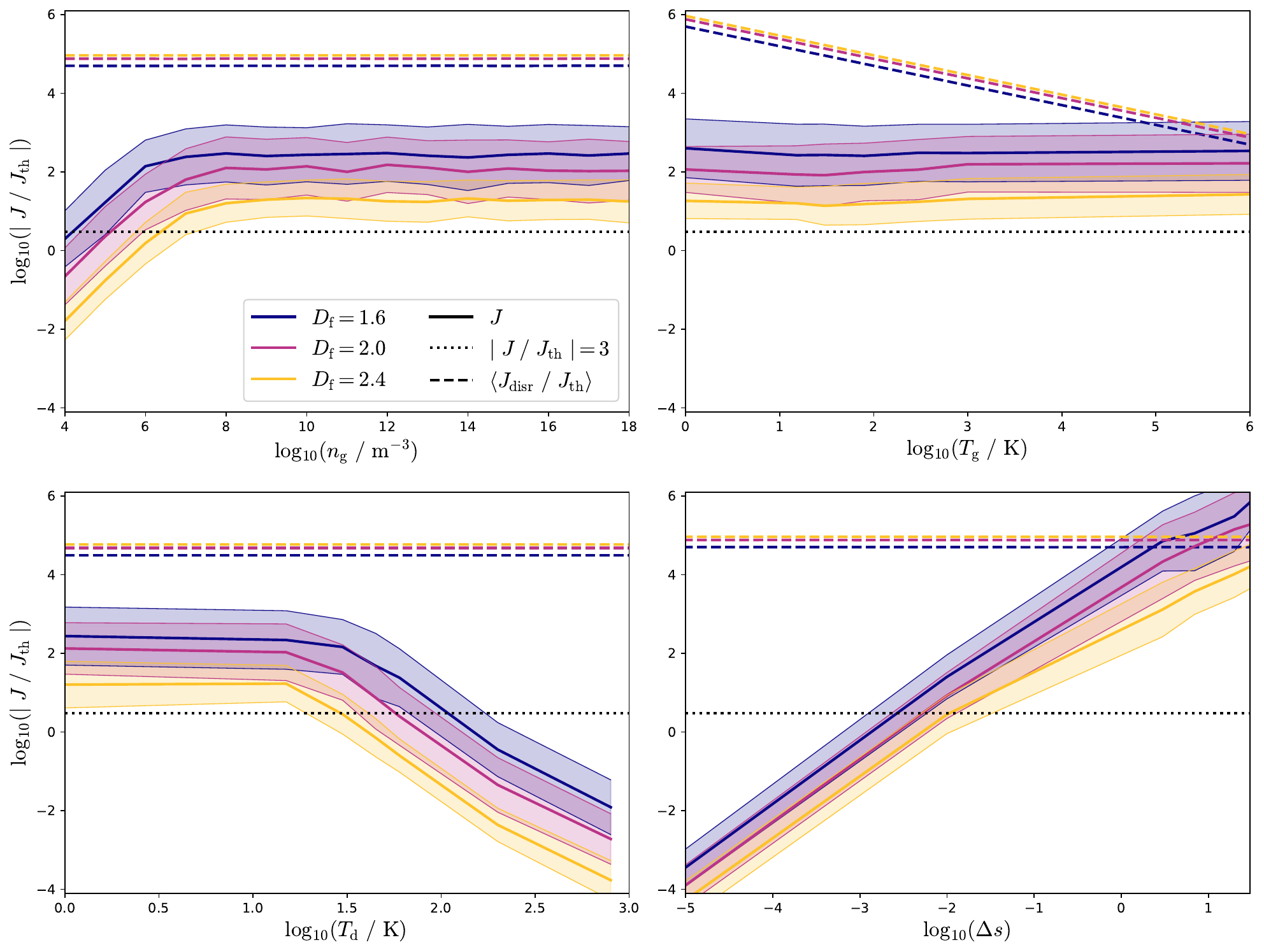}
\end{center}
\caption{ The average angular momentum $|J/J_{\mathrm{th}}|$ (solid lines) over gas number density $n_{\mathrm{g}}$ (top left), gas temperature $T_{\mathrm{g}}$ (top right), dust temperature $T_{\mathrm{d}}$ (bottom left), and gas-dust drift $\Delta s$ (bottom right), respectively, and typical CNM conditions. All results are for the ensemble of dust grains with a size of $a_{\mathrm{eff}} = 400\ \mathrm{nm}$ and a fractal dimension $D_{\mathrm{f}}=1.6$ (purple), $D_{\mathrm{f}}=2.0$ (pink), and $D_{\mathrm{f}}=2.4$ (yellow), respectively. The solid lines represent the ensemble average of the angular momenta while the shaded areas indicate the range of one standard deviation of all data points are situated. Dotted lines are the threshold of $J/J_{\mathrm{th}}=3$ where a stable alignment is assumed while dashed lines is the average of the corresponding critical angular momentum $J=\left<J_{\mathrm{disr}}\right>$ where dust grains are estimated to become rotationally disrupted. We note that we plot only the average threshold $\left<J_{\mathrm{disr}}\right>$ of rotational disruption for clarity while the range of $J_{\mathrm{disr}}$ is about one order of magnitude.}
\label{fig:JattMechDependencyCNM}
\end{figure*}
\begin{figure*}[ht!]
	\begin{center}
	\includegraphics[width=0.99\textwidth]{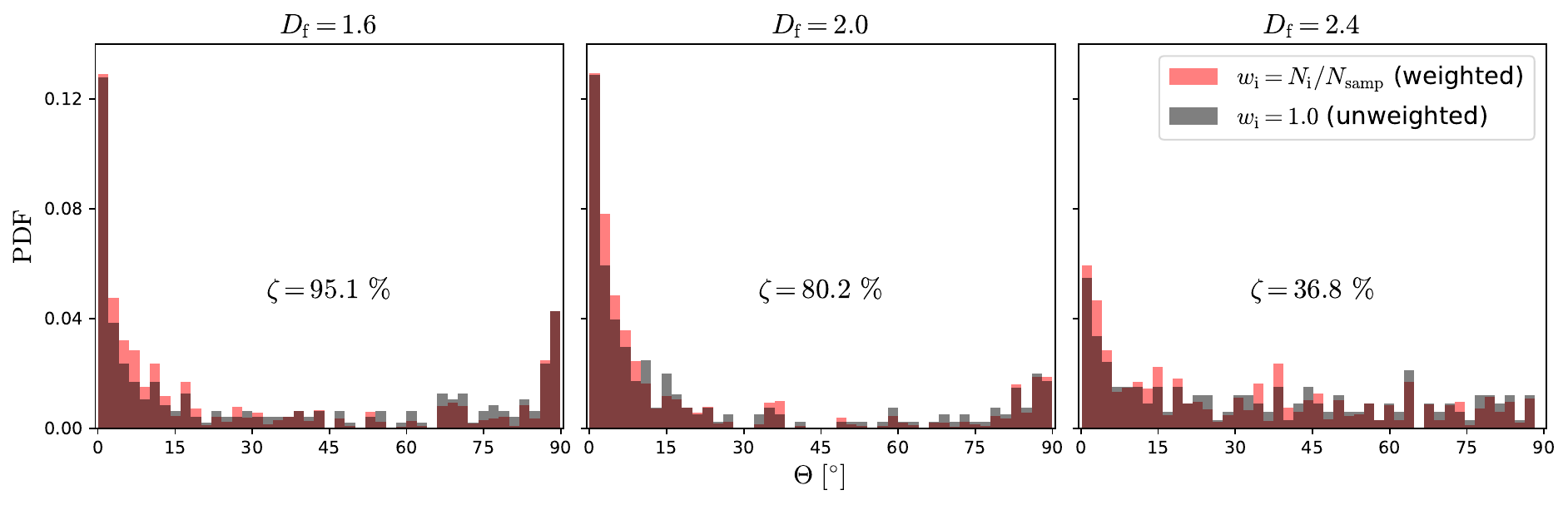}
	\end{center}
\caption{Histograms of the distribution of the alignment angle $\Theta$ of all attractor points with an angular momentum $J$ within the range of ${ 3J_{\mathrm{th}}<  \left|J\right| < J_{\mathrm{disr}} }$.  The parameter $\zeta$ is the ratio of grains with at least one attractor within that range to the total number of considered dust grains. The histograms are for grains with a size of $a_{\mathrm{eff}} = 400\ \mathrm{nm}$ and a fractal dimension of $D_{\mathrm{f}}=1.6$ (left), $D_{\mathrm{f}}=2.0$ (middle), and $D_{\mathrm{f}}=2.4$ (right), respectively. Red bars are weighted attractor points while grey bars are the unweighted ones. We assume CNM conditions but 5with a gas number density of $n_{\mathrm{g}}  =10^{6}\ \mathrm{m}^{-3}$.}
\label{fig:HistMechDistDf}
\end{figure*}
{\bf In Fig.~\ref{fig:JattMechDependencyCNM} we present the angular momentum $|J|$ for typical CNM conditions. In this plot, the quantities of gas number density $n_{\mathrm{g}}$, gas temperature $T_{\mathrm{g}}$, dust temperature $T_{\mathrm{d}}$, and the gas-dust drift $\Delta s$ are varied independently.\\
Increasing $n_{\mathrm{g}}$ allows for the angular momentum $|J|$ to grow continuously because of the higher collision rate $\mathcal{R}_{\mathrm{coll}}$. Hence, the MET $\Gamma_{\mathrm{mech}}$ increases while the drag time scale $\tau_{\mathrm{drag}}$ decreases  simultaneously. Since the CNM dust temperature is assumed to be $T_{\mathrm{d}} = 15\ \mathrm{K}$, the total drag  is identical with the gas drag (compare Fig.~\ref{fig:TauDrag})  . Consequently, the mechanical spin-up process and the total drag reach a balance and the dust grains reach subsequently their terminal angular momentum.\\
Phenomenologically, all fractal dimensions $D_{\mathrm{f}}$ show a similar spin-up behavior. Naturally, elongated grains with $D_{\mathrm{f}} = 1.6$ are more efficient "propellers" than more spherical grains compared e.g. to those with $D_{\mathrm{f}}=2.4$. The difference between the particular fractal dimensions is about one order of magnitude. Hence, grains with $D_{\mathrm{f}} = 1.6$ can already be considered to reach a stable mechanical alignment ($3J_{\mathrm{th}}<|J|$) for $n_{\mathrm{g}} \approx 10^4\ \mathrm{m}^3$ while grains with $D_{\mathrm{f}} = 2.4$ require on average a gas number density of $n_{\mathrm{f}} \approx 10^7\ \mathrm{m}^3$. All grain sizes reach their terminal angular velocity bellow the threshold of rotational disruption ($|J|<J_{\mathrm{disr}}$). Hence, all grains can fully contribute to polarization  for gas number densities of $n_{\mathrm{g}} > 10^7\ \mathrm{m}^3$.\\
For CNM conditions and an increasing gas temperature $T_{\mathrm{g}}$ the angular momentum $|J|$ is increasing for the entire considered range of $T_{\mathrm{g}} \in [1\ \mathrm{K}, 10^6\ \mathrm{K}]$. This is because the gas drag becomes only relevant for $T_{\mathrm{g}} > 10^6\ \mathrm{K}$. We emphasize that $T_{\mathrm{g}}$ appears to be constant in Fig.~\ref{fig:JattMechDependencyCNM} because the thermal angular velocity $J_{\mathrm{th}}$ depends  on the gas temperature  $T_{\mathrm{g}}$ as well. For the same reason $J_{\mathrm{dist}}$ decreases with an increasing $T_{\mathrm{g}}$. Grains with a fractal dimension of $D_{\mathrm{f}} = 2.4$ are in the alignment regime while more elongated grains are rationally disrupted for $T_{\mathrm{g}} > 10^5\ \mathrm{K}$.\\
As shown in Fig.~\ref{fig:JattMechDependencyCNM} for an increase in dust temperature $T_{\mathrm{d}}$  the angular momentum $|J|$ remains constant for lower dust temperatures $T_{\mathrm{d}}<30\ \mathrm{K}$ and starts then to decline. Since, $|J|<J_{\mathrm{disr}}$ grains cannot be rationally disrupted by an increase in $T_{\mathrm{d}}$ for the applied set of parameters. This is a result of the infrared drag acting on the dust grain damping the mechanical spin-up process most efficiently for higher dust temperatures. Hence, a stable mechanical alignment with ${ 3J_{\mathrm{th}}<|J|<J_{\mathrm{disr}} }$ can only be reported within $T_{\mathrm{d}} \in [1\ \mathrm{K}, 30\ \mathrm{K}]$ for a fractal dimension of $D_{\mathrm{f}}=2.4$ and $T_{\mathrm{d}} \in [1\ \mathrm{K}, 100\ \mathrm{K}]$ for $D_{\mathrm{f}}=1.6$.\\
The angular momentum $|J|$ increases with an increasing gas-dust drift $\Delta s$. Here, the mechanical spin-up process of grains with a fractal dimension of $D_{\mathrm{f}} = 1.6$ is the most efficient where the condition $3J_{\mathrm{th}}<|J|$ is already given for $\Delta s \approx 10^{-3}$. In turn grains with $D_{\mathrm{f}} = 2.4$ require a slightly higher drift of $\Delta s \approx 10^{-2}$. Grains with $D_{\mathrm{f}} = 1.6$ become already destroyed at $\Delta s=1.0$ while grains with $D_{\mathrm{f}} = 2.4$ can still contribute to polarization for a drift up to $\Delta s=30.0$.}\\
We emphasize that the mechanical spin-up processes presented in  Fig.~\ref{fig:JattMechDependencyCNM} may be influenced by additional effects of grain destruction. For instance, by sputtering when an impinging gas particle has a sufficiently large energy to separate individual monomers from the aggregate may provide an relevant process to destroy dust grains 
\citep[see e.g.][]{Shull1978,Draine1979,Dwek1992}. Consequently, our results may be modified in the higher end of the regimes of gas number density $n_{\mathrm{g}}$, gas temperature $T_{\mathrm{g}}$, and the gas-dust drift $\Delta s$, respectively. Especially, in the range of gas parameters where $\Delta s > 10$ or $T_{\mathrm{g}} > 10^{5}\ \mathrm{K}$, grains may be efficiently destroyed by sputtering. Such grains would not contribute to the net dust polarization either. However, considering the effects related to sputtering goes beyond the scope of this paper.

\subsection{Distribution of the mechanical alignment directions}
{\bf In Fig.~\ref{fig:HistMechDistDf} we present exemplary  distributions of the alignment angle $\Theta$ for grains with different fractal dimensions $D_{\mathrm{f}}$ and typical CNM conditions but a lower gas number density of ${ n_{\mathrm{g}}  =10^{6}\ \mathrm{m}^{-3} }$ (compare Fig.~\ref{fig:JattMechDependencyCNM}). Alignment angles of $\Theta = 0^\circ$ and $\Theta = 90^\circ$, respectively, are the most likely configurations whereas $\Theta \approx 45^\circ$ is generally a less favourable alignment. In Fig.~\ref{fig:HistMechDistDf} we also compare weighted as well as unweighted attractor points (see Sect.~ \ref{sect:RRF}). We report that the weighting broadens  slightly the peaks at $\Theta = 0^\circ$ and $\Theta = 90^\circ$, respectively. This result holds for all gas and dust parameters considered in this paper. We note that for the particular set of input parameters applied in Fig.~\ref{fig:HistMechDistDf} the preferential alignment direction is $\Theta = 0^\circ$ meaning the rotation axis $\hat{a}_{\mathrm{1}}$ of the grains is (anti)parallel to $\Delta \vec{s}$. The exception are grains with $D_{\mathrm{f}} \geq 2.4$ where an alignment with $\Theta = 90^\circ$ i.e. $\hat{a}_{\mathrm{1}}$ is perpendicular to $\Delta \vec{s}$ is the least likely configuration. However, this result cannot be generalized. For instance for a gas number densities $n_{\mathrm{g}}  > 10^{10}\ \mathrm{m}^{-3}$ the peak at $\Theta = 90^\circ$ would vanish as well for grains with $D_{\mathrm{f}} \leq 2.0$.\\
As shown in Fig.~\ref{fig:HistMechDistDf} for a gas number density of  $n_{\mathrm{g}}  = 10^{6}\ \mathrm{m}^{-3}$ the ensemble of grains with a fractal dimension of $D_{\mathrm{f}} = 1.6$ has the most aligned grains within the range ${ 3J_{\mathrm{th}}<  \left|J\right| < J_{\mathrm{disr}} }$. Here, the exact ratio of grains that possess at least one attractor is $\zeta = 95.1\ \%$. In turn, the grains with $D_{\mathrm{f}} = 2.0$ and $D_{\mathrm{f}} = 2.4$ have a ratio of $\zeta = 80.2\ \%$ and $\zeta = 36.8\ \%$, respectively, because a fraction of all these grains are already below the limit of $|J| < 3J_{\mathrm{th}}$ (see also Fig.~\ref{fig:JattMechDependencyCNM}).}

\subsection{Mechanically induced dust polarization}
\begin{figure*}[ht!]
\begin{center}
    \includegraphics[width=0.99\textwidth]{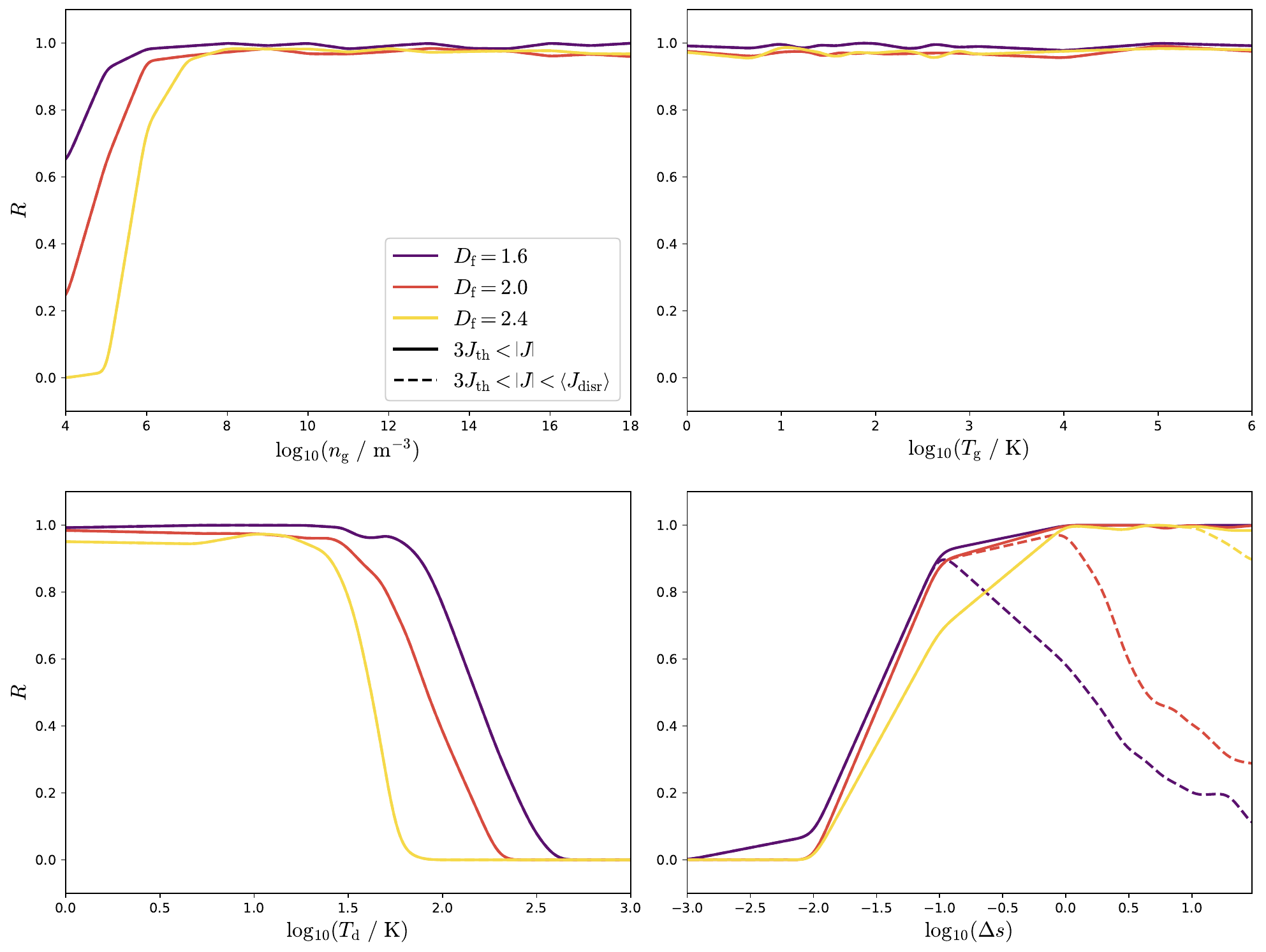}
\end{center}
\caption{ The RRF $R$ corresponding to the conditions presented in Fig.~\ref{fig:JattMechDependencyCNM}. Solid lines represent the ensemble of grains fulfilling the alignment condition $3J_{\mathrm{th}}<|J|$ while for dashed lines rotational disruption is also taken into account i.e. all grains rotating with $|J|>J_{\mathrm{disr}}$ cannot contribute to polarization.}
\label{fig:RRF_Mech}
\end{figure*}
As presented in the sections above various parameter sets allow for a mechanically induced grain rotation with an angular momentum $J$ within the range ${ 3J_{\mathrm{th}} < |J| < J_{\mathrm{disr}} }$. The  most likely alignment directions are at at $\Theta = 0^\circ$ and $\Theta = 90^\circ$, respectively. Consequently, mechanical alignment may result in a high degree of dust polarization.\\
In Fig.~\ref{fig:RRF_Mech} we present the resulting RRF to quantifying the dust polarization of grain ensembles and for different fractal dimensions $D_{\mathrm{f}}$. We emphasize that the panels in Fig.~\ref{fig:RRF_Mech} show exactly the same range of input parameters as the panels in  Fig.~\ref{fig:JattMechDependencyCNM}. Increasing the gas number density $n_{\mathrm{g}}$ in a CNM environment leads to an almost perfect alignment of grains i.e. a RRF $R$ close to unity. For grains with a fractal dimension of $D_{\mathrm{f}}=1.6$ at $n_{\mathrm{g}}>10^4\ \mathrm{m}^3$ the alignment is with a $R \gtrapprox 0.97$ while grains with a $D_{\mathrm{f}}=2.4$ become alignment with a $R \approx 0.95$ for $n_{\mathrm{g}}>10^{7}\ \mathrm{m}^3$. Since all grains reach their terminal angular  momentum before they become rotationally disrupted the high net polarization efficiency retains for higher densities. \\
We report that the RRF barely depends on gas temperature $T_{\mathrm{g}}$ for typical CNM conditions. Hence, almost all of the grains remain within ${ 3J_{\mathrm{th}} < |J| <  J_{\mathrm{disr}} }$ and subsequently the RRF is close to unity independent of $T_{\mathrm{g}}$ and $D_{\mathrm{f}}$. 
Because of the IR emission of photons the grain rotation is heavily governed by the dust temperature for $T_{\mathrm{d}} < 30\ \mathrm{K}$. For $T_{\mathrm{d}} = 30\ \mathrm{K}$ the ensemble of grains with $D_{\mathrm{f}}=2.4$ is already completely randomized while elongated grains with $D_{\mathrm{f}}=1.6$ do no longer contribute to polarization up to a dust temperature of about $T_{\mathrm{d}} = 300\ \mathrm{K}$.\\
Concerning the gas-dust drift $\Delta s$ grains with a fractal dimension of $D_{\mathrm{f}}=1.6$ are most efficiently spun-up. Hence, an ensemble of such $D_{\mathrm{f}}=1.6$ grains start to have a net polarization for $\Delta s=10^{-3}$ while more roundish shaped grains  with $D_{\mathrm{f}}=2.4$ require a $\Delta s = 10^{-2}$. With an increasing $\Delta s$ all grains shapes may eventually become  effectively aligned. Considering rotational disruption grains with a fractal dimension of   $D_{\mathrm{f}}=1.6$ reach their peak polarization at $\Delta s = 0.1$ while grains with  $D_{\mathrm{f}}=2.4$ require a much higher drift of about $\Delta s = 10.0$ to become rotationally disrupted. At a gas-dust drift of $\Delta s = 30$ elongated grains are almost completely destroyed while the ensembles with $D_{\mathrm{f}}\leq 2.4$ would still contribute to the net polarization. In contrast to the other input parameters  for the increasing gas-dust drift it most important to take rotational disruption into account. \\
In compression to the classical Gold alignment mechanism \citep[][]{Gold1952a,Gold1952b} a super-sonic drift ($\Delta s > 1$) is not required to account for dust polarization. A grain alignment for a subsonic-drift is  consistent with the analytical model presented in \cite{LazarianHoang2007Mech} as well as the numerical results of \cite{DasWeingartner2016} and \cite{HoangChoLazarian2018}, respectively. However, the latter studies consider only a limited number of individual grains and lack the information about the distribution of the alignment angle $\Theta$ required to evaluate any net polarization of an ensemble of dust grains.
\subsection{Grain size dependency}
\begin{figure*}[ht!]
	\begin{center}
	\includegraphics[width=1.02\textwidth]{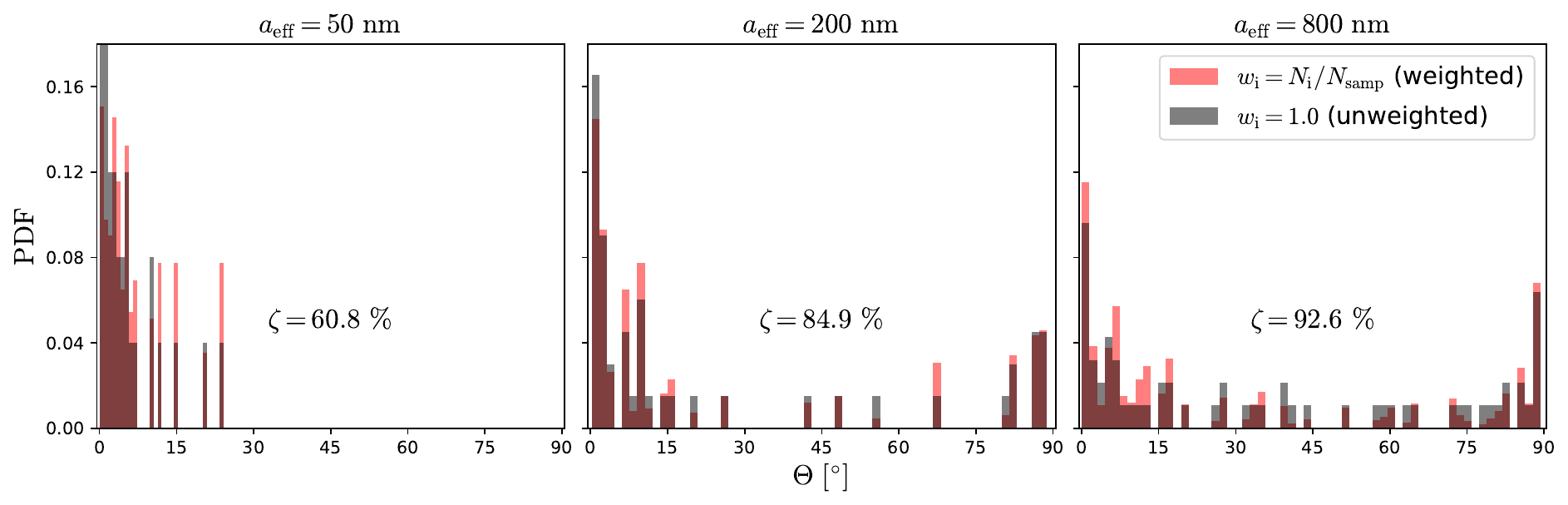}
	\end{center}
\caption{ The same as Fig.~\ref{fig:HistMechDistDf} but for grains with a fractal dimension of $D_{\mathrm{f}}=2.0$ and CNM conditions for the different grain sizes of $a_{\mathrm{eff}}=50\ \mathrm{nm}$ (left), $a_{\mathrm{eff}}=200\ \mathrm{nm}$ (middle), and $a_{\mathrm{eff}}=800\ \mathrm{nm}$ (right), respectively.}
\label{fig:HistQDis}
\end{figure*}
\begin{figure*}[ht!]
\begin{center}
     \includegraphics[width=0.99\textwidth]{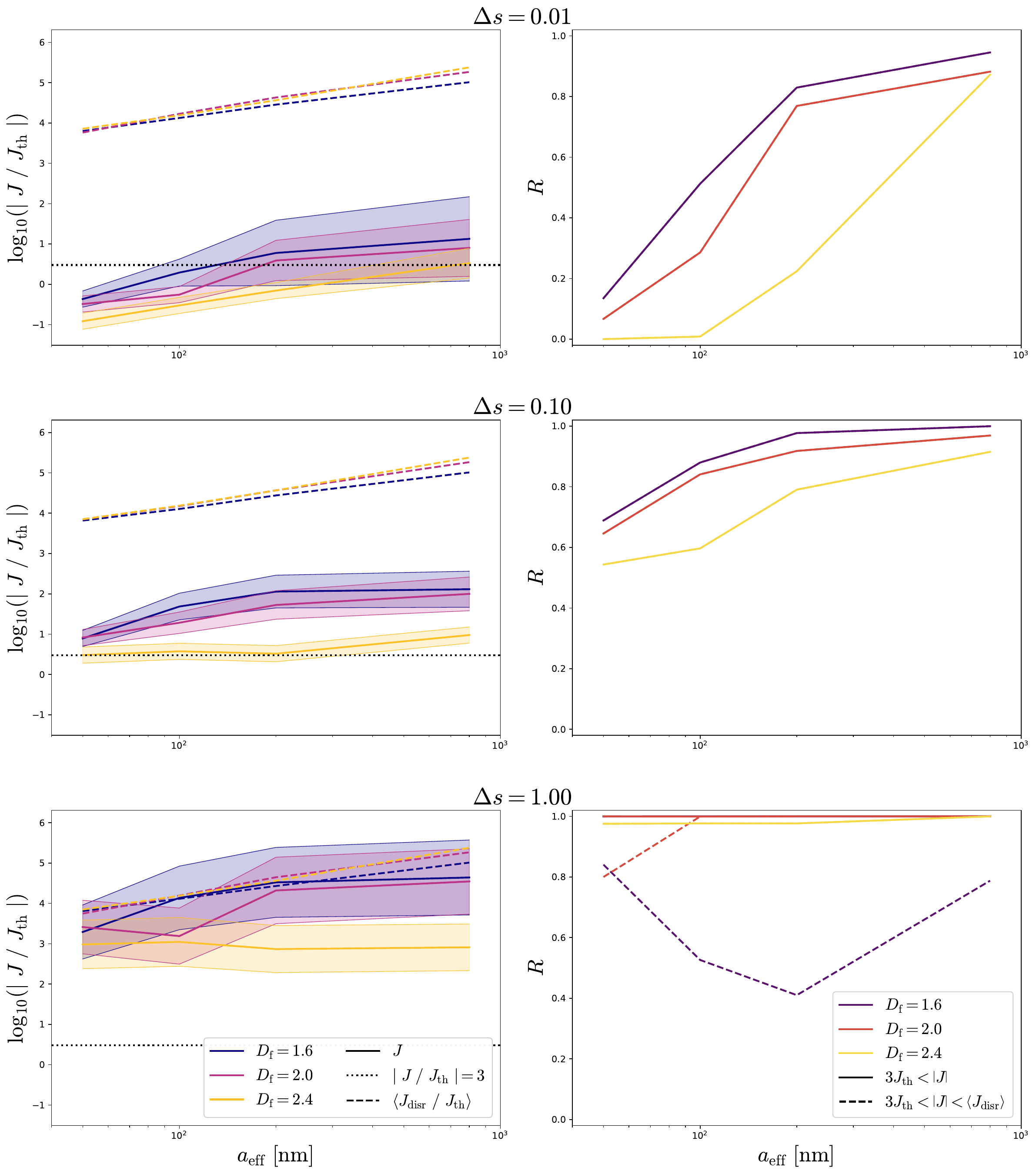}
\end{center}
\caption{ Left panel: The same as Fig.~\ref{fig:JattMechDependencyCNM} but over all considered grain sizes $a_{\mathrm{eff}}$. Right panel: The RRF $R$ corresponding to the conditions presented on the left panel.}
\label{fig:JattDependencyAeff}
\end{figure*}
\begin{figure}[h!]
\begin{center}
     \includegraphics[width=0.49\textwidth]{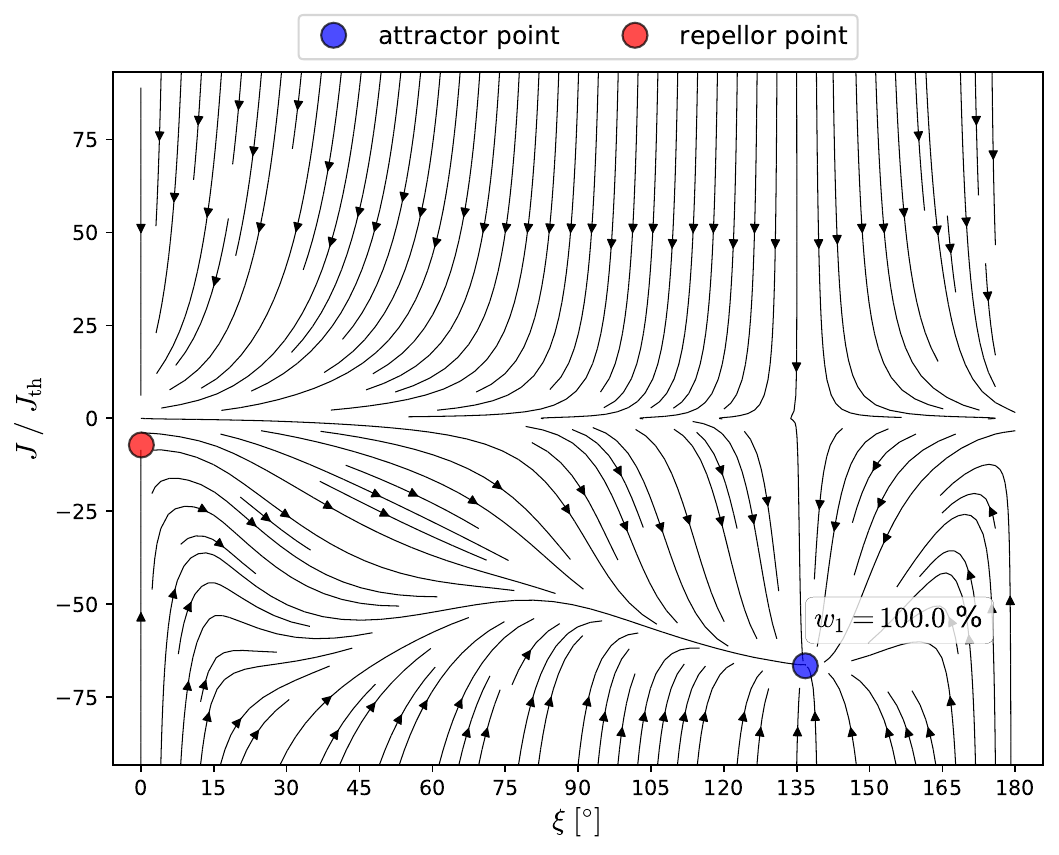}
\end{center}
\caption{ The same as Fig.~\ref{fig:PhaseSpaceMech} but for magnetic alignment with a gas-dust drift of $\Delta s=0.1$, an amplification factor of $f_{\mathrm{mag}}=1$ and an angle $\Psi=15^{\circ}$.}
\label{fig:PhaseSpaceMag}
\end{figure}
So far, we focused on grains with an effective radius of ${ a_{\mathrm{eff}} = 400\ \mathrm{nm} }$. In Fig.~\ref{fig:HistQDis} we present the distribution of alignment angles $\Theta$ for grains with a fractal dimension of $D_{\mathrm{f}}=2.0$ but different effective radii $a_{\mathrm{eff}}$. Similar to the distribution shown in Fig.~\ref{fig:HistMechDistDf} the predominate direction of mechanical grain alignment is $\Theta=0^\circ$ followed by $\Theta=90^\circ$ whereas an alignment with $\Theta\approx 45^\circ$ is the least likely. Here, we note a clear trend concerning the ratio $\zeta$ of dust grains that have a stable alignment within $3J_{\mathrm{th}}<|J|<J_{\mathrm{disr}}$ where larger dust grains are more efficiently spun up. Only  a fraction of $\zeta=60.8\ \%$ of gains with an effective radius of $a_{\mathrm{eff}} = 50\ \mathrm{nm}$ are effectively mechanically aligned while larger grains with $a_{\mathrm{eff}} = 200\ \mathrm{nm}$ and $a_{\mathrm{eff}} = 800\ \mathrm{nm}$ have a higher ratio of $\zeta=84.9\ \%$ and $\zeta=92.6\ \%$, respectively. We attribute this finding to the fact that larger grains are more efficiently spun-up because of the increased surface area of the larger irregular grains.\\
\begin{figure}[h!]
\begin{center}
     \includegraphics[width=0.49\textwidth]{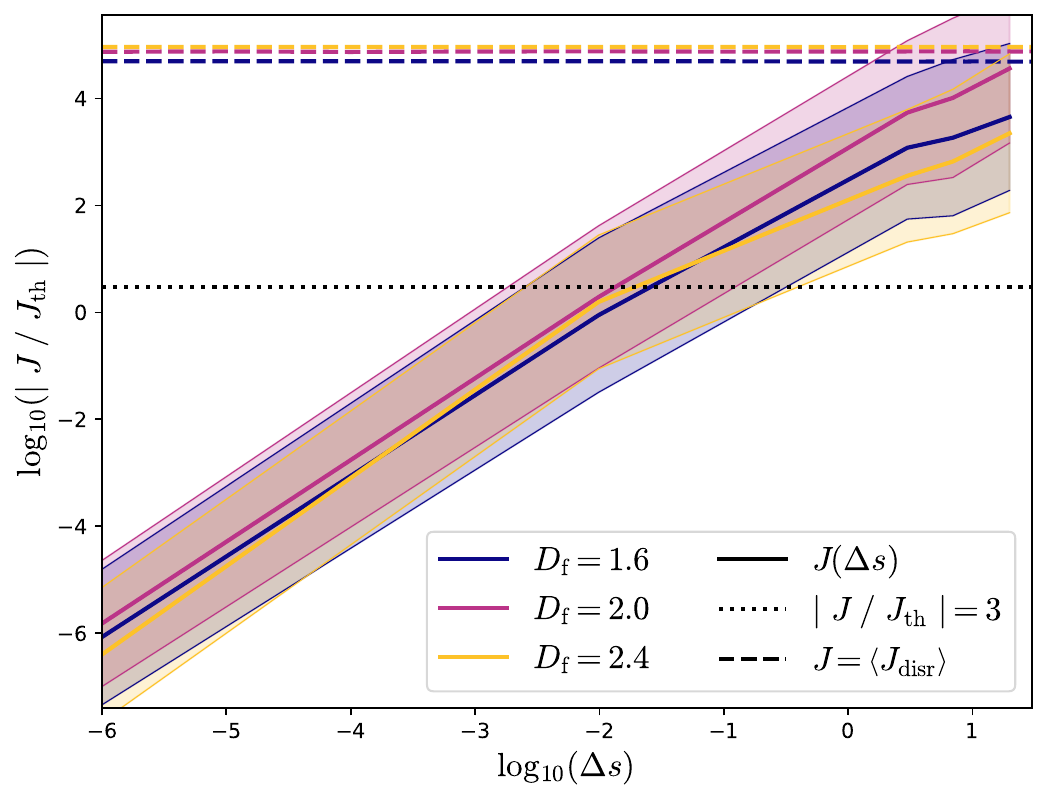}
\end{center}
\caption{ The same as the angular 
momentum $|J/J_{\mathrm{th}}|$ over gas-dust drift $\Delta s$ presented in Fig.~\ref{fig:JattMechDependencyCNM} but for all grains aligning in direction of an external magnetic field instead of the direction of the gas-dist drift $\Delta \vec{s}$. In this plot an amplification factor of $f_{\mathrm{mag}}=1$ is applied. We emphasize that all  of the angles $\xi$ and $\psi=15^\circ$, respectively,  are considered.}
\label{fig:JattDependencyMagds}
\end{figure}
In Fig.~\ref{fig:JattDependencyAeff} we present the angular momentum $|J|$ dependent on the effective grain radius $a_{\mathrm{eff}}$ in comparison with the corresponding RRF $R$ dependent on fractal dimensions as well as gas-dust drift. As noted above larger dust grains are more efficiently spun-up by mechanical alignment. The trend is in general in agreement with the grain size dependency of the angular momentum $|J|$ presented in \cite{HoangChoLazarian2018}. However, they predict a power-law relation between $|J|$ and $a_{\mathrm{eff}}$. We emphasize that our dust grains do not follow strictly a power-law since larger grains are disproportionately spun-up. We attribute this mismatch to the fact that \cite{HoangChoLazarian2018} simply scaled their distinct cubical shapes to get grains of different $a_{\mathrm{eff}}$ whereas we mimic grain growth by pre-calculating grains for different fractal dimensions for each grains size bin individually.\\
In detail, for a gas-dust drift of $\Delta s=0.01$ only grains with $a_{\mathrm{eff}}>400\ \mathrm{nm}$ and a fractal dimension of $D_{\mathrm{f}}=1.6$ and $D_{\mathrm{f}}=2.0$, respectively, surpass the limit $ 3 J_{\mathrm{th}}<|J| $ while more roundish grains with $D_{\mathrm{f}}=2.0$ require a radius of $a_{\mathrm{eff}}=800\ \mathrm{nm}$ to align. Consequently, smaller grains cannot contribute the polarization while only the largest elongated grains reach a RRF close to unity.\\
For $\Delta s=0.1$ larger grains with $a_{\mathrm{eff}} \geq 200\ \mathrm{nm}$ and a fractal dimension of $D_{\mathrm{f}} \leq 2.0$, most grains are within the range 
of $3 J_{\mathrm{th}} < |J| < J_{\mathrm{disr}}$. Hence, the corresponding RRF is $R>0.85$ for such grains.  The magnitude of the angular momenta of grains with a fractal dimension of $D_{\mathrm{f}}=2.4$ are only slightly above the $3 J_{\mathrm{th}}$ limit. Hence, such grains can only reach an RRF in the range $\mathrm{R}\in [0.5,0.85]$.\\
For the a thermal drift of $\Delta s=1.0$ all the more roundish grains with a fractal dimension of $D_{\mathrm{f}}=2.0$ and $D_{\mathrm{f}}=2.4$, respectively, are almost completely within the range $3 J_{\mathrm{th}} < |J| < J_{\mathrm{disr}}$ of stable alignment. Here, the RRF is close to unity independent grain size. The exception are the smallest grains with $D_{\mathrm{f}} = 2.0$ and most of the elongated grains with $D_{\mathrm{f}} = 1.6$ where the grain ensemble becomes partly rotationally disrupted. For the latter ensemble the characteristic interplay of the spin-up process and the rotational disruption limit leads to a dip in the RRF of $\mathrm{R} = 0.4$ for $a_{\mathrm{eff}}=200\ \mathrm{nm}$ whereas grains at the opposite side of the size distribution reach a RRF close to $\mathrm{R} = 0.8$.\\
\begin{figure}[h!]
\begin{center}
     \includegraphics[width=0.49\textwidth]{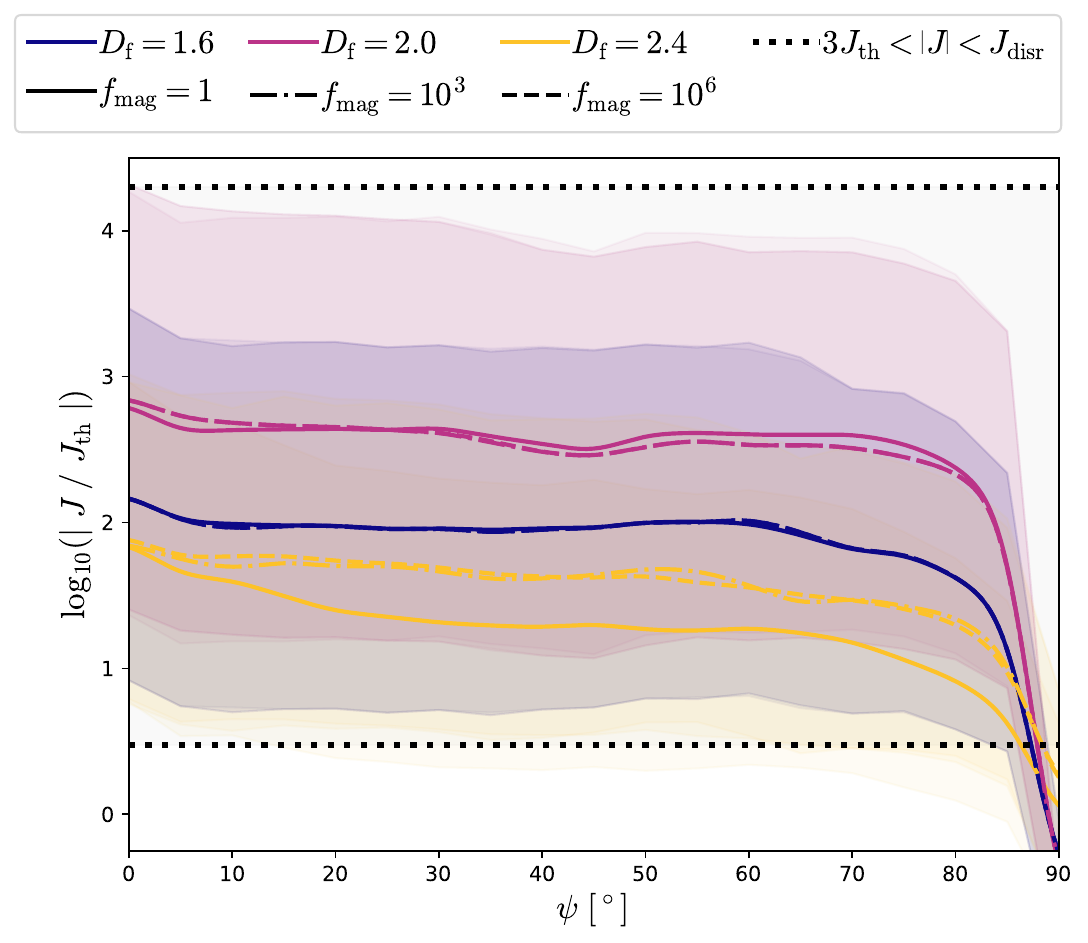}
\end{center}
\caption{ The angular momentum $|J/J_{\mathrm{th}}|$  over the angle $\psi$ for dust grains with a size of $a_{\mathrm{eff}} = 400\ \mathrm{nm}$ and a fractal dimension of $D_{\mathrm{f}}=1.6$ (blue), $D_{\mathrm{f}}=2.0$ (purple), and $D_{\mathrm{f}}=2.4$ (yellow), respectively, for typical CNM conditions but with a gas-dust drift $\Delta s=1.0$.  The results are calculated for a constant alignment angle of $\xi=15^\circ$ and for an amplification factor of $f_{\mathrm{mag}}=1$ (solid), $f_{\mathrm{mag}}=10^3$ (dotted), and $f_{\mathrm{mag}}=10^6$ (dash dotted), respectively. for clarity we plot only the average values of $|J/J_{\mathrm{th}}|$. The range of angular momenta is generally of the same order as in Fig.~\ref{fig:JattDependencyMagds}.}
\label{fig:JattDependencyMagPhi}
\end{figure}
In contrast to that for elongated grains we see the opposite trend where the range of one standard deviation seems to become larger with an increasing grain size. Here, grains with $a_{\mathrm{eff}} = 50\ \mathrm{nm}$ are roundish even for a fractal dimension $D_{\mathrm{f}}=1.6$ because of the low number of monomers whereas grains with $a_{\mathrm{eff}} = 800\ \mathrm{nm}$ are almost a rod with much larger angular momenta. Hence, small variations of the grain shape such as a forking structures especially in the outskirts of the grain may lead to vastly different alignment behavior.\\
The RRF plotted in Fig.~\ref{fig:JattDependencyAeff} reveals that grains with a fractal dimension of $D_{\mathrm{f}}>2.4$ cannot contribute to the polarization for a typical CNM environment. In contrast to that a grain ensemble with $D_{\mathrm{f}}=1.6$ starts to polarize light for sizes of $a_{\mathrm{eff}} \geq 100\ \mathrm{nm}$.\\
This plot demonstrates once more that the parameter of grains size  $a_{\mathrm{eff}}$ alone is not sufficient to quantify the mechanical alignment of dust grains since the net-polarization is highly dependent on the grain shape as well. 

\subsection{The spin-up process of (super)paramagnetic grains}
{\bf Grain alignment dependent on the magnetization of paradigmatic grains is extensively studied in \cite{Hoang2014} and \cite{Hoang2016}, respectively, in the context of RATs. In our study we only consider silicate grains and model the magnetic field strengths as well as  the impact of possible iron inclusions within the dust grains itself by the amplification factor $f_{\mathrm{mag}}$ introduced in Sect.~\ref{sect:MagFieldAlignment}. We also emphasize that we do not scrutinize the exact conditions required for the alignment direction to switch from the mechanical alignment to magnetic field alignment within the scope of this paper.\\
In Fig.~\ref{fig:PhaseSpaceMag} we show a $J/\xi$ phase portrait  exemplary for the magnetic field alignment with an amplification factor of $f_{\mathrm{mag}}=1$ and an angle $\Psi=15^{\circ}$ between $\Delta \vec{s}$ and $\vec{B}$. The phase portrait is to be compared with that presented in Fig.~\ref{fig:PhaseSpaceMech}. For this particular grain we report an alignment with an angle $\xi=135^{\circ}$ assuming typical CNM conditions, but with a gas-dust drift of $\Delta s=0.1$. This result is typical for magnetic alignment in so far as most grains have exactly one attractor in contrast to the "purely MAD" where most grains may have multiple at tractors even for a supra-sonic drift i.e. $\Delta s>1.0$. The magnitude of |J| for this attractor is comparable to the results presented in \cite{DasWeingartner2016}.\\
In Fig.~\ref{fig:JattDependencyMagds} we present the angular momentum $|J|$ over gas-dust drift $\Delta s$ for grains aligned with the magnetic field direction assuming typical CNM conditions, and an angle of $\psi=15^\circ$. The processes involved are outlined in Sect.~\ref{sect:MagFieldAlignment} in detail. In contrast to a "purely MAD" (see Fig.~\ref{fig:JattMechDependencyCNM}) the alignment in the magnetic field direction of the most elongated grains require the gas-dust drift $\Delta s$ to be about one order of magnitude higher in order to surpass the limit $3J_{\mathrm{th}}<|J|$. The spin-up process for different fractal dimensions is nearly identical for a sub-sonic drift. Concerning the rotational disruption of magnetically aligned grains only a small fraction of all grains may be destroyed for the drift of $\Delta s > 1.0$.\\
The dependency of magnetic field alignment of grains on the angle $\xi$ is depicted in Fig.~\ref{fig:JattDependencyMagPhi}.  The resulting angular momentum $J$ remains roughly constant for $\psi<80^\circ$. Note that the entire ensembles of grains cannot rotationally be disrupted for this particular set of parameters. For an angle $\psi\approx 90^\circ$ the spin-up process appears to be most inefficient. Provided that MAD is the only driver for the magnetic alignment of grains in astrophysical environments, a high degree of dust polarization cannot be expected in regions where the predominant directions of the magnetic field $\vec{B}$ and gas-dust drift $\Delta\vec{s}$ are perpendicular. This trend is very similar to grain alignment by RATs as presented e.g. in \cite{LazarianHoang2019ApJ}. However, we note that the magnitude of the angular momentum  $|J|$  depends marginally on $f_{\mathrm{mag}}$ and is also not correlated with the fractal dimension $D_{\mathrm{f}}$ since he angular momentum $|J|$ is  higher for grains with $D_{\mathrm{f}}=2.0$ than for grains with $D_{\mathrm{f}}=2.4$. The exact conditions of this reversal need to be dealt with in an upcoming study.}\\
In \cite{HoangLazarian2007} it is noted that the grain alignment considering RATs correlates with the parameter ${ q_{\mathrm{max}}=\max( \hat{e}_{\mathrm{1}} \vec{Q}_{\mathrm{RAT}}) / \max(\hat{e}_{\mathrm{2}} \vec{Q}_{\mathrm{RAT}}) }$ where $\hat{e}_{\mathrm{i}} \vec{Q}_{\mathrm{RAT}}$ is the RAT efficiency in the i-th direction of the lab-frame. According to the AMO of RAT alignment within the range of ${ q_{\mathrm{max}} \in [1,2] }$ attractor points with supra-thermal rotation become most likely.
More recent studies report \citep[][]{Hoang2016,Herranen2021} about islands in the parameter space of ${ \{ q_{\mathrm{max}}, \psi, \delta_{\mathrm{m}} \} }$  where grain alignment is possible i.e. $J_{\mathrm{th}} \ll |J|$. 
However, as noted in \cite{DasWeingartner2016}, the analogous parameter ${ q_{\mathrm{max}}=\max( \hat{e}_{\mathrm{1}} \vec{Q}_{\mathrm{mech}}) / \max(\hat{e}_{\mathrm{2}} \vec{Q}_{\mathrm{mech}})}$ of mechanical alignment seems to be inconclusive in predicting the MAD. This is consistent with our modeling. We cannot report any trend between the gas and dust input parameters, the resulting quantity $q_{\mathrm{max}}$, and the subsequent grain alignment behavior. 
\begin{figure*}[h!]
    \begin{center}
        \begin{minipage}[c]{1.0\linewidth}
            \begin{center}
                \includegraphics[width=0.99\textwidth]{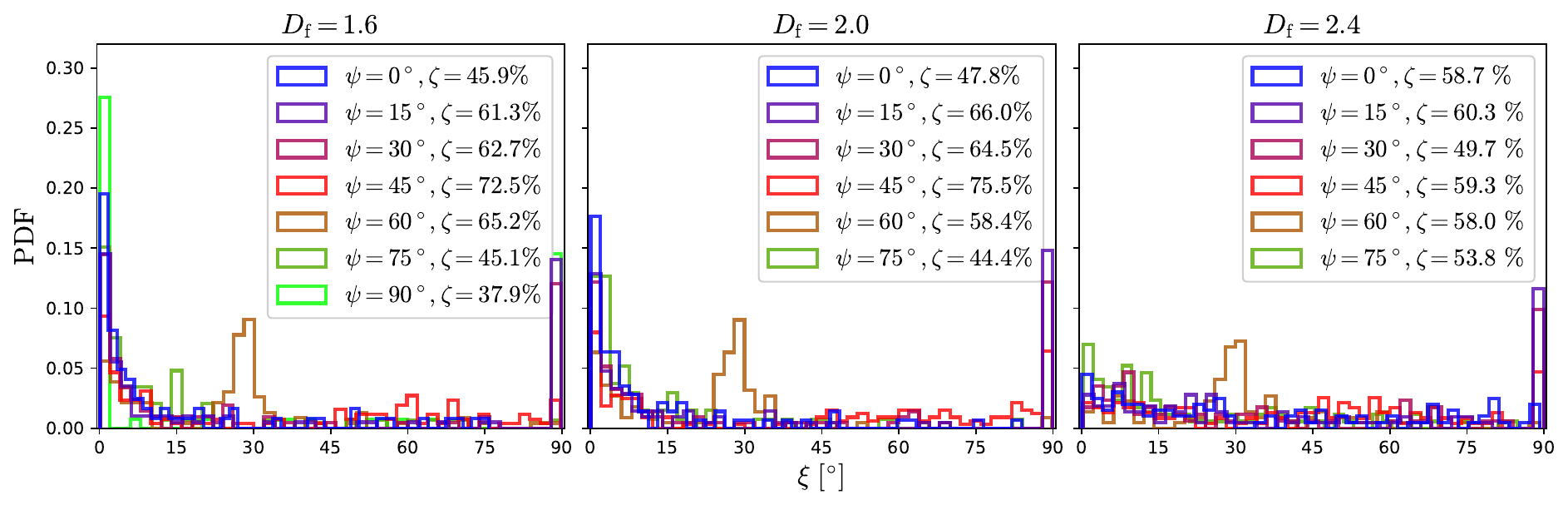}
                \caption{Normalized histogram of the distribution of the alignment angle $\xi$  for dust grains with a size of $a_{\mathrm{eff}} = 400\ \mathrm{nm}$ and a fractal dimension of $D_{\mathrm{f}}=1.6$ (left), $D_{\mathrm{f}}=2.0$ (middles), and $D_{\mathrm{f}}=2.4$ (right), respectively, for typical CNM conditions but with a gas-dust drift of $\Delta s=1.0$. Here, the amplification factor is $f_{\mathrm{mag}}=1.0$ for all panels. Color-coded is the angle $\psi$ while  $\zeta$ is the corresponding ratio of attractor points with an angular momentum $J$ in between ${ 3J_{\mathrm{th}} < |J| <  J_{\mathrm{disr}} }$ to the total number of considered dust grains. We note that an alignment angle of $\xi=0^\circ$ followed by $\xi=90^\circ$, respectively, are the preferential configurations for the magnetic field alignment (compare also Fig.~\ref{fig:HistMechDistDf}). For clarity we only plot configurations with a ratio $\zeta>2.0\ \%$ since distributions with a smaller $\zeta$ would only contribute a few but high peaks to the histogram.}
                \label{fig:HistMagDf}
            \end{center}
        \end{minipage} 
        \vspace{4mm}
        \begin{minipage}[c]{1.0\linewidth}
            \begin{center}
                \includegraphics[width=0.99\textwidth]{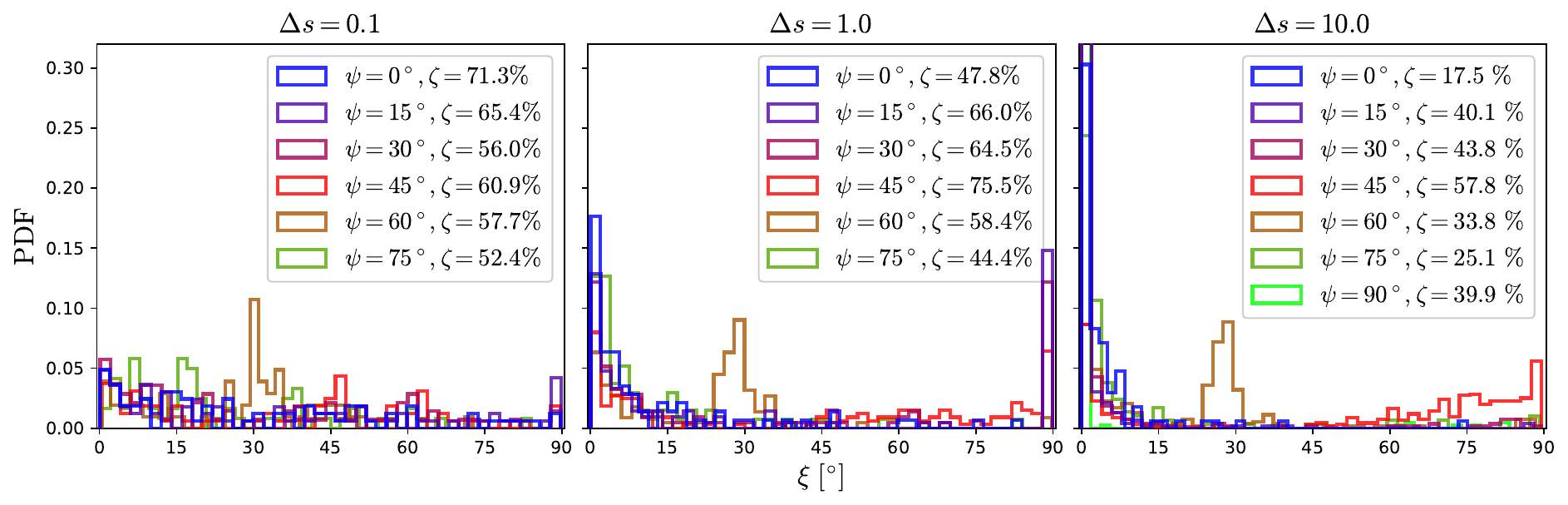}
                \caption{The same as Fig.~\ref{fig:HistMagDf} but for grains with a fractal dimension of $D_{\mathrm{f}}=2.0$ and a gas-dust drift of  $\Delta  s=0.1$ (left), $\Delta  s=1.0$ (middle), and $\Delta  s=10.0$ (right), respectively.}
                \label{fig:HistMagds}
            \end{center}
        \end{minipage} 
         \vspace{4mm}
        \begin{minipage}[c]{1.0\linewidth}
            \begin{center}
                \includegraphics[width=0.99\textwidth]{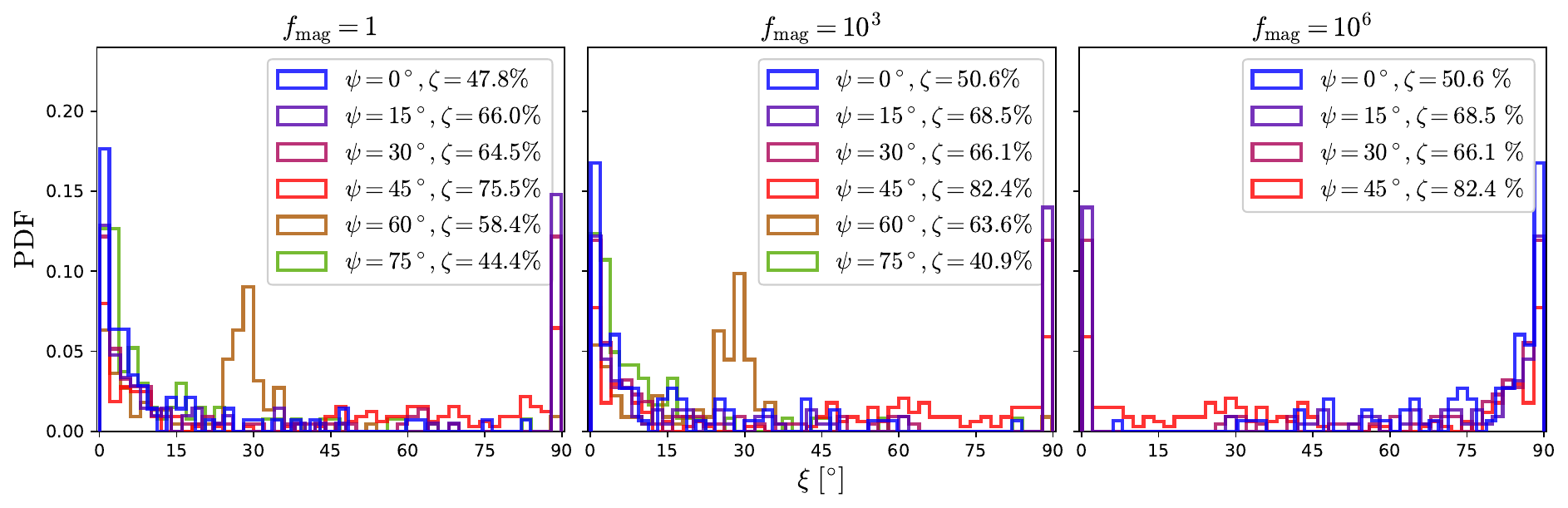}
                \caption{The same as Fig.~\ref{fig:HistMagds} but with a gas-dust drift of $\Delta  s=1.0$ and an amplification factor of $f_{\mathrm{mag}}=1$ (left), $f_{\mathrm{mag}}=10^3$ (middle), and $f_{\mathrm{mag}}=10^6$ (right), respectively.}
                \label{fig:HistMagfmag}
            \end{center}
        \end{minipage} 
\end{center} 
\end{figure*}

\subsection{The angular distribution of magnetic field alignment}
\begin{figure*}[ht!]
\begin{center}
    \includegraphics[width=1.02\textwidth]{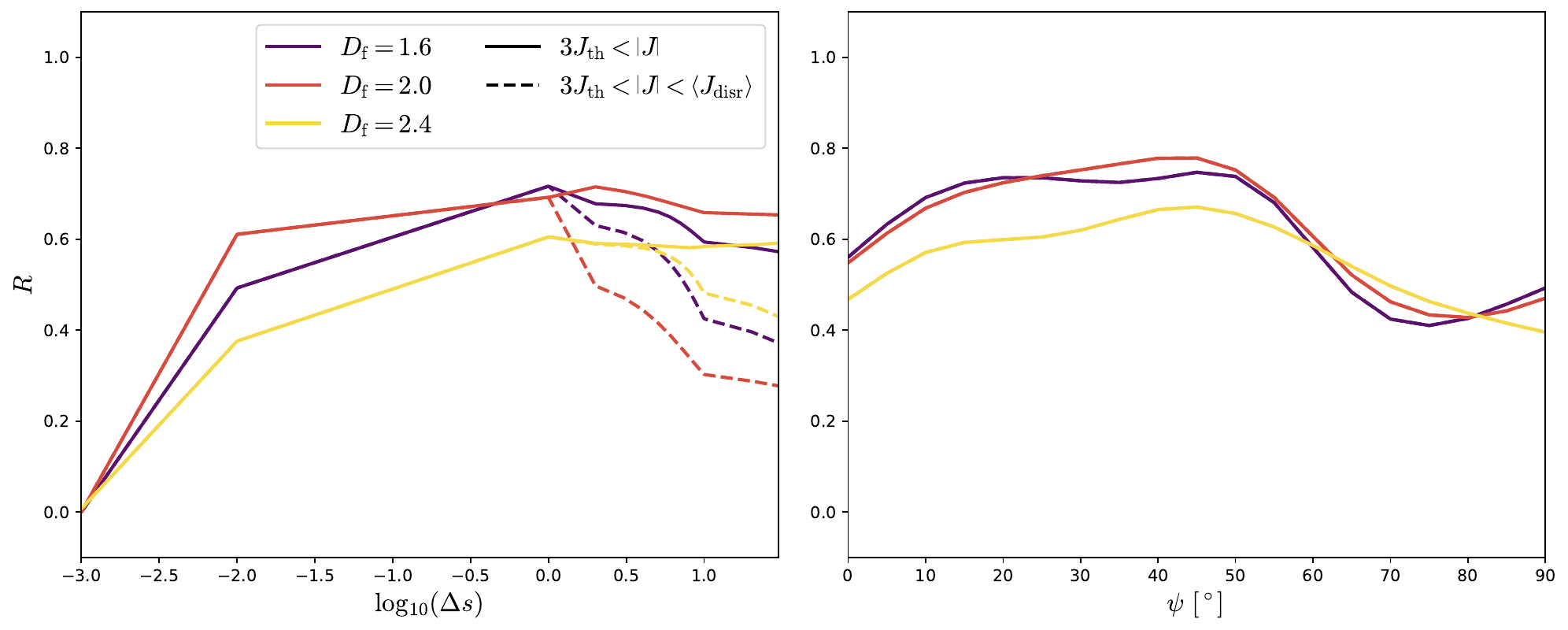}
\end{center}
\caption{Left panel: The RRF $R$ over the gas-dust drift $\Delta s$ for the conditions presented in Fig.~\ref{fig:JattDependencyMagds} but for an angle $\psi=15^\circ$. Right panel: The same as the left panel but for the RRF over the angle $\psi$ but with a gas-dust drift of $\Delta s=1.0$ (compare  Fig.~\ref{fig:JattDependencyMagPhi}).}
\label{fig:RRF_Mag_ds_phi}
\end{figure*} 
{\bf In Fig.~\ref{fig:HistMagDf} we present exemplary cases for the correlation of the alignment angle $\xi$ and the angle $\psi$ between $\vec{B}$ and $\Delta \vec{s}$. For typical CNM conditions and a $\Delta s=1.0$ an alignment angle of $\xi=0^\circ$ is the most likely for all $\psi$ followed by an alignment at $\xi=90^\circ$. The only exception is close to an angle of $\psi\approx 60^\circ$ where the distinct shape of the spin-up component $\overline{H}\left(\xi,\psi\right)$ leads to a most likely alignment at $\xi=30^\circ$. The histogram is the most pronounced for a $D_{\mathrm{f}}=1.6$ while the distribution flattens towards higher fractal dimensions. For $D_{\mathrm{f}}\geq 2.4$ the peak at $\xi=0^\circ$ disappears almost completely. In contrast to a "purely MAD" the fraction $\zeta$ generally increases with an increasing $D_{\mathrm{f}}$. To our understanding this is due to the dependence of paramagnetic alignment timescale $\tau_{\mathrm{DG}}$ on the moment of inertia $I_{ \mathrm{a_1}}$. Note that grains with the same effective radius $a_{\mathrm{eff}}$ have exactly the same volume independent of fractal dimension $D_{\mathrm{f}}$. Hence, the moment of inertia $I_{ \mathrm{a_1}}$ and subsequently $\tau_{\mathrm{DG}}$ decreases toward larger values of $D_{\mathrm{f}}$ and the magnetic alignment becomes more efficient.\\
In Fig.~\ref{fig:HistMagds} we consider CNM conditions and a grain ensemble with $D_{\mathrm{f}}=2.0$ while increasing the gas-dust drift $\Delta s$. For $\Delta s=0.1$ the distribution of the alignment angle is mostly flat where  alignment angles of $\xi=0^\circ$ and $\xi=90^\circ$, respectively, being slightly more likely than other angles. For a gas-dust drift of $\Delta s=1.0$ the peaks at $\xi=0^\circ$ as well as $\xi=90^\circ$ are more pronounced while an alignment within the range $\xi\in [30^\circ,90^\circ[$ become less likely. As the gas-dust drift approaches $\Delta s=10.0$ most of the grains align at $\xi=0^\circ$  almost independent of $\psi$. The only exceptions are for $\psi=45^\circ$ leading to a most likely alignment at $\psi=90^\circ$ and for $\psi=60^\circ$ with a $\xi=90^\circ\ $\footnote{The exact inter-dependencies resulting in these exceptions need to be dealt with in an upcoming paper.}. For the applied fractal dimension of $D_{\mathrm{f}}=2.0$ the tendency of the ratio $\zeta$ of aligned dust grains shows that most of the grains are aligned for $\Delta s=0.1$. This is because for a higher gas-dust drift the some grains become already rationally disrupted.\\
In Fig.~\ref{fig:HistMagfmag} we investigate the impact of grain magnetization by varying the amplification factor $f_{\mathrm{mag}}$. For the default value of $f_{\mathrm{mag}}=1$ an alignment with $\xi=0^\circ$ has the highest probability followed by a very narrow peak at $\xi=90^\circ$. An amplification factor of $f_{\mathrm{mag}}=10^3$ would lead to an  alignment angle distribution comparable to that with $f_{\mathrm{mag}}=1$. Applying the most extreme case of $f_{\mathrm{mag}}=10^6$ virtually all aligned grains have an angular momentum $\vec{J}$ parallel (perpendicular) to $\vec{B}$, i.e. $\xi=0^\circ$ ($\xi=90^\circ$). Here, the grain alignment suddenly stops for angles $\Psi>45^\circ$ . Note that the maximal possible angular momentum ${ J_{\mathrm{max}} \propto \overline{H}\left(\xi, \psi  \right) /(1+\delta_{\mathrm{m}}\sin^2 \xi) }$ (see Eq. \ref{eq:dJdtMag}) considering the grain alignment in the magnetic field direction. Hence, a higher amplification factor and subsequently a higher $\delta_{\mathrm{m}}$ does not necessarily enhance the resulting ratio $\xi$. Rather, an increase of $\delta_{\mathrm{m}}$ leads to a much lower possible values of $J_{\mathrm{max}}$ at angles close to $\xi=45^\circ$. Consequently, the only remaining alignment configurations with $3J_{\mathrm{th}}<J_{\mathrm{max}}$ are at $\xi=0^\circ$ or $\xi=90^\circ$, respectively. For the set of parameters applied in Fig.~\ref{fig:HistMagfmag} this may result in an increase in the net dust polarization. However, this trend cannot be generalized because for grain ensembles where attractor points at $\xi=45^\circ$ are very rare in the first place an increase in $f_{\mathrm{mag}}$ would have less of an impact. The alignment behavior presented in Fig.~\ref{fig:JattDependencyMagds} as well as the variations of input parameters shown in the  Figs.~\ref{fig:HistMagDf} - \ref{fig:HistMagfmag} agree in so far as the as the spin-up process for grains aligned along the magnetic field is most inefficient for as the angle between $\Delta \vec{s}$ and $\vec{B}$ approaches  $\psi \approx 90^\circ$.\\
However, the RRF dependencies cannot easily be generalized. For instance, a slightly higher gas-dust drift $\Delta s$ would push more grains with a fractal dimension of $D_{\mathrm{f}}=2.0$ in the regime of rotational disruption and the RRF depicted in the left panel of Fig.~\ref{fig:RRF_Mag_ds_phi} would behave a rather different curve. This emphasizes once again the importance to describe the grain alignment process and the subsequent dust polarization statistically over a large ensemble of grains instead of investigating the alignment behavior of individual grains. 


\section{Summary and Outlook}
\label{sect:Summary}
This paper explores systematically the impact of the grain shape and gas-dust drift on the mechanical alignment of dust (MAD). Large ensembles of grains aggregates characterized by the fractal dimension $D_{\mathrm{f}}$
and the effective radius $a_{\mathrm{eff}}$ are constructed. 
A novel Monte-Carlo based approach to model the physics of  gas-dust interactions on a microscopic level is introduced. This allows for a statistical description of the grain spin-up process for an environment with an existing gas-dust drift. Finally, stable grain alignment configurations are identified in order to quantify the net polarization efficiency for each grain ensemble. Concerning the net polarization we explore both cases separately, namely (i) the {"purely MAD"} case along the gas-dist drift velocity and (ii) the case of grain alignment with respect to the magnetic field lines. We summarize principal results of this paper as follows:
\begin{itemize}
    \item We demonstrate that the mechanical spin-up process is most efficient for elongated grains with a fractal dimension of $D_{\mathrm{f}}=1.6$. Such grains require only a subsonic gas-dust drift of about $\Delta s=10^{-3}$ to reach a stable grain alignment. In contrast, more roundish grains with  $D_{\mathrm{f}}\geq 2.4$ require a supersonic drift with $\Delta s = 10^{-2}$. 
    
    \item The tensile strength of large elongated grains ($a_{\mathrm{eff}}>400\ \mathrm{nm}$, $D_{\mathrm{f}}< 2.0$) may be about $1-3\ $ dex lower compared to the more roundish considered grain shapes ($D_{\mathrm{f}}\geq 2.0$). Hence, such elongated grains shapes appear to be the most fragile and are rather unlikely to subsist on longer time scales in the ISM once they start to rotate.

    \item Considering non-spherical dust, the characteristic time scales governing the alignment dynamics become closely connected to the grain shape i.e. the fractal dimension $D_{\mathrm{f}}$. However, we find that the gas drag time scale $\tau_{\mathrm{gas}}$ is most impacted by the gas-dust drift $\Delta s$, whereas $D_{\mathrm{f}}$  appears to be only of minor importance. Concerning the IR drag time scale the variation of $\tau_{\mathrm{IR}}$ is about two order of magnitude at most for large grains with different fractal dimensions. The same for the  Davis-Greenstein alignment time scale $\tau_{\mathrm{DG}}$ where we report for the largest grain sizes a variation of $\tau_{\mathrm{DG}}$ up to two orders of magnitude between ensembles with different $D_{\mathrm{f}}$. 
    
    \item Simulating the mechanical spin-up processes of fractal grains reveals that the acceleration rate of the  spin-up process is roughly identical for different grain shapes. Concerning the magnitude of the ensemble average of the angular momentum $J$, a difference in the fractal dimension of $D_{\mathrm{f}}=0.4$ would result in a difference in the magnitude $J$ of about ${0.5 - 1\ \mathrm{dex}}$. 
    
    \item For a "purely MAD" we find that a stable alignment with an angle of $\Theta=0^\circ$ to be the most likely followed by an alignment close to $\Theta=90^\circ$. Hence, most mechanically spun-up grains have a rotation axis $\hat{a}_{\mathrm{1}}$ almost (anti)parallel to the direction of the gas-dust drift $\Delta \vec{s}$. 
    
    \item We report that a mechanically driven magnetic field alignment of fractal dust grains is indeed possible.  The spin-up process for grains aligned with the magnetic field lines is found to be slightly less efficient as the "purely MAD" because a gas-dust drift of $\Delta s=10^{-2}$  is required for a stable grain alignment independent of fractal dimension $D_{\mathrm{f}}$.
    
    
    \item We find that the spin-up process for the magnetic alignment is rather inefficient for an angle $\psi\approx 90^\circ$ between $\Delta \vec{s}$ and the magnetic field $\vec{B}$. This $\psi$-dependency of the spin-up is comparable to that of RAT alignment.
    
    \item The alignment of the "pure MAD" and the alignment case considering a magnetic field are rather comparable. The magnetic field $\vec{B}$ is most likely parallel to the rotation axis $\hat{a}_{\mathrm{1}}$. Hence, a dust polarization efficiency with a ${R=0.8-0.9}$ may be observed given a sufficiently high gas-dust drift $\Delta s$. 
    
\end{itemize}
We emphasize that all the trends discussed above are highly dependent on the applied physical properties of the gas and dust component. Our study remains agnostic concerning the exact conditions that may lead to a grain alignment in direction of the gas-dust drift $\Delta \vec{s}$ or an alignment with the orientation of the magnetic field $\vec{B}$. Moreover, phenomena such as the H$_{\mathrm{2}}$ formation on the surface of the grains and charged dust that may heavily impact the MAD are not taken into consideration within the scope of this paper. Therefore, investigations of those phenomena will be dealt with in a forthcoming paper in tandem with quantifying the likelihood of the occurrence of drift velocity alignment versus magnetic field alignment.

\appendix

\section{The moments of inertia of a dust aggregate}
\label{app:Interia}
The individual monomers of the presented aggregates are considered to be a  perfect sphere. Hence, the k-th monomer with mass $m_{\mathrm{mon,k}} = \frac{4}{3}\pi a_{\mathrm{mon,k}}^3 \rho_{dust}$ has a strictly diagonal inertia tensor where the elements may be written as 
\begin{equation}
	I_{\mathrm{sph,ii,k}}=\frac{2}{5} m_{\mathrm{mon,k}} a_{\mathrm{mon,k}}^2 = \frac{8}{15} \pi \rho_{\mathrm{dust}} a_{\mathrm{mon,k}}^5\,.
\end{equation}
Applying the parallel axis theorem (Steiner's theorem) the diagonal elements for the entire aggregate are 
\begin{equation}
	I_{\mathrm{11}}=\frac{4}{3}\pi \rho_{dust} \sum_{\mathrm{k}=1}^{N_{\mathrm{mon}}} a_{\mathrm{mon,k}}^3  (X_{\mathrm{2,k}}^2+X_{\mathrm{3,k}}^2) + \frac{2}{5}a_{mon,k}^5\,,
\end{equation}

\begin{equation}
	I_{\mathrm{22}}=\frac{4}{3}\pi \rho_{dust} \sum_{\mathrm{k}=1}^{N_{\mathrm{mon}}} a_{\mathrm{mon,k}}^3  (X_{\mathrm{1,k}}^2+X_{\mathrm{3,k}}^2) + \frac{2}{5}a_{mon,k}^5\,,
\end{equation}
and 
\begin{equation}
	I_{\mathrm{33}}=\frac{4}{3}\pi \rho_{dust} \sum_{\mathrm{k}=1}^{N_{\mathrm{mon}}} a_{\mathrm{mon,k}}^3  (X_{\mathrm{1,k}}^2+X_{\mathrm{2,k}}^2) + \frac{2}{5}a_{mon,k}^5\,, 
\end{equation}
respectively, whereas the off-diagonal terms in the inertia tensor are
\begin{equation}
	I_{\mathrm{12}}=I_{\mathrm{21}}=-\frac{4}{3}\pi \rho_{dust} \sum_{\mathrm{k}=1}^{N_{\mathrm{mon}}} a_{\mathrm{mon,k}}^3(X_{\mathrm{1,k}} X_{\mathrm{2,k}})\,,
\end{equation}

\begin{equation}
	I_{\mathrm{13}}=I_{\mathrm{31}}=-\frac{4}{3}\pi \rho_{dust} \sum_{\mathrm{k}=1}^{N_{\mathrm{mon}}} a_{\mathrm{mon,k}}^3(X_{\mathrm{1,k}} X_{\mathrm{3,k}})\,,
\end{equation}
and
\begin{equation}
	I_{\mathrm{23}}=I_{\mathrm{32}}=-\frac{4}{3}\pi \rho_{dust} \sum_{\mathrm{k}=1}^{N_{\mathrm{mon}}} a_{\mathrm{mon,k}}^3(X_{\mathrm{2,k}} X_{\mathrm{3,k}})\,.
\end{equation}
Here, ${ \vec{X}_{\mathrm{k}}=(X_{\mathrm{1,k}},X_{\mathrm{2,k}},X_{\mathrm{3,k}})^T }$ represents the position of the k-th monomer. By calculating the eigenvalues, the tensor may be written in an orthogonal basis with principal axis ${ \{ \hat{a}_{\mathrm{1}}, \hat{a}_{\mathrm{2}}, \hat{a}_{\mathrm{3}} \}  }$. In accordance with previous publications we refer to this basis as the target-frame \citep[see e.g.][]{Draine1996,Lazarian2007,Draine2013}. Finally, we rotate each aggregate such that the moments of inertia ${ I_{\mathrm{a_1}}>I_{\mathrm{a_2}}>I_{\mathrm{a_3}} }$ run along the principal axis. We emphasize that each dust grain's target-frame is unique and well defined.

\section{Monte Carlo noise estimation}
\label{app:MCNoise}
A certain level of noise is an inevitable drawback of any MC simulation. In this section we estimate the noise of our particular MC setup. For this, a subset of all dust aggregates consisting of $20$ distinct grains per fractal dimension $D_{\mathrm{f}}$ and size $a_{\mathrm{eff}}$ is selected. Furthermore, $100$ input parameter sets of gas and dust are randomly sampled within the range applied in Sect.~\ref{sect:MechJ}. For each individual grain and input parameter set MC simulations are repeatedly performed for different numbers of collisions within ${ N_{\mathrm{coll}}\in[10^2,10^4] }$. We quantify the noise of the resulting angular momentum $J_{\mathrm{att}}$ and the alignment angle $\Theta_{\mathrm{att}}$ at each attractor point via 
\begin{equation}
	\mathrm{err}\left(  J_{\mathrm{att}} \right)   = \frac{ J_{\mathrm{att}} - \left. J_{\mathrm{att}} \right|_{N_{\mathrm{coll}}=10^4}          }{ \left. J_{\mathrm{att}} \right|_{N_{\mathrm{coll}}=10^4} } 
\end{equation}
and
\begin{equation}
	\mathrm{err}\left(  \Theta_{\mathrm{att}} \right)   = \frac{ \Theta_{\mathrm{att}} - \left. \Theta_{\mathrm{att}} \right|_{N_{\mathrm{coll}}=10^4}          }{ \left. \Theta_{\mathrm{att}} \right|_{N_{\mathrm{coll}}=10^4} } \,,
\end{equation}
respectively, where all differences are normalized with respect to the corresponding results at $N_{\mathrm{coll}}=10^4$.\\
In Fig.~\ref{fig:ErrJ} we present the noise level of $J_{\mathrm{att}}$ for the exemplary grains with a size of ${ a_{\mathrm{eff}} = 400\ \mathrm{nm} }$ dependent on fractal dimension $D_{\mathrm{f}}$ and $N_{\mathrm{coll}}$. For $N_{\mathrm{coll}}<10^3$ we report a noise level up to $\pm 25\ \%$. With an increasing number of collisions $N_{\mathrm{coll}}$ the noise range declines. Approaching $N_{\mathrm{coll}} \approx 10^3$ the noise reaches a range below $\pm 2.6\ \%$ and remains within that limit for $N_{\mathrm{coll}} \geq 10^3$ independent of  fractal dimensions $D_{\mathrm{f}}$. The overall trend is similar for the noise of the alignment angle $\Theta_{\mathrm{att}}$ at each attractor point as shown in Fig.~\ref{fig:ErrTh}. However, the variation of the noise of $\Theta_{\mathrm{att}}$ is slightly larger for $N_{\mathrm{coll}} \geq 10^3$ than that of $J_{\mathrm{att}}$ but remains within the limit of $\pm 2.6\ \%$ as well. We emphasize that this limit is independent of grain size.\\
We note that for an increasing $N_{\mathrm{coll}}$ the run-time increases linearly for each individual run. For instance, by increasing the number of collisions from $N_{\mathrm{coll}} = 10^3$ to $N_{\mathrm{coll}} = 10^4$ the average run-time increases by a factor of about $8.5$ while there is no further benefit by means of noise reduction. Running all MC simulations with $N_{\mathrm{coll}} = 10^3$, therefore, is the optimal compromise between run-time and noise. Hence, we estimate for our MAD MC setup to operate within an accuracy of $\pm 2.6\ \%$.

\begin{figure}[ht!]
	\begin{center}
	\includegraphics[width=0.5\textwidth]{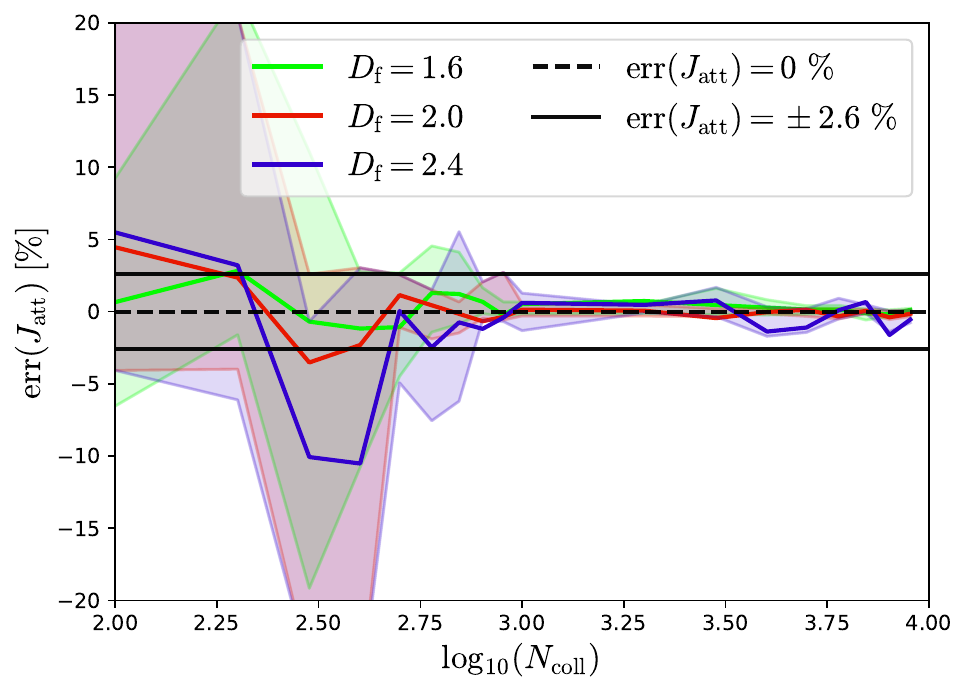}
	\end{center}
\caption{Distribution of the noise level of the angular momentum $J_{\mathrm{att}}$ at distinct attractor points calculated for grains with an effective radius $a_{\mathrm{eff}} = 400\ \mathrm{nm}$ and the different fractal dimensions of $D_{\mathrm{f}}=1.6$ (solid green), $D_{\mathrm{f}}=2.0$ (solid red), and $D_{\mathrm{f}}=2.4$ (solid blue), respectively, dependent on the number of collisions  $N_{\mathrm{coll}}$. All data points are normalized by the corresponding results at $N_{\mathrm{coll}}=10^4$. Solid lines are the average of all $J_{\mathrm{att}}$ while the shaded areas represent the minima and maxima of the noise level per $N_{\mathrm{coll}}$. Horizontal black lines indicate the  range of $\pm 2.6\ \%$.}
\label{fig:ErrJ}
\end{figure}
\begin{figure}[ht!]
	\begin{center}
	\includegraphics[width=0.5\textwidth]{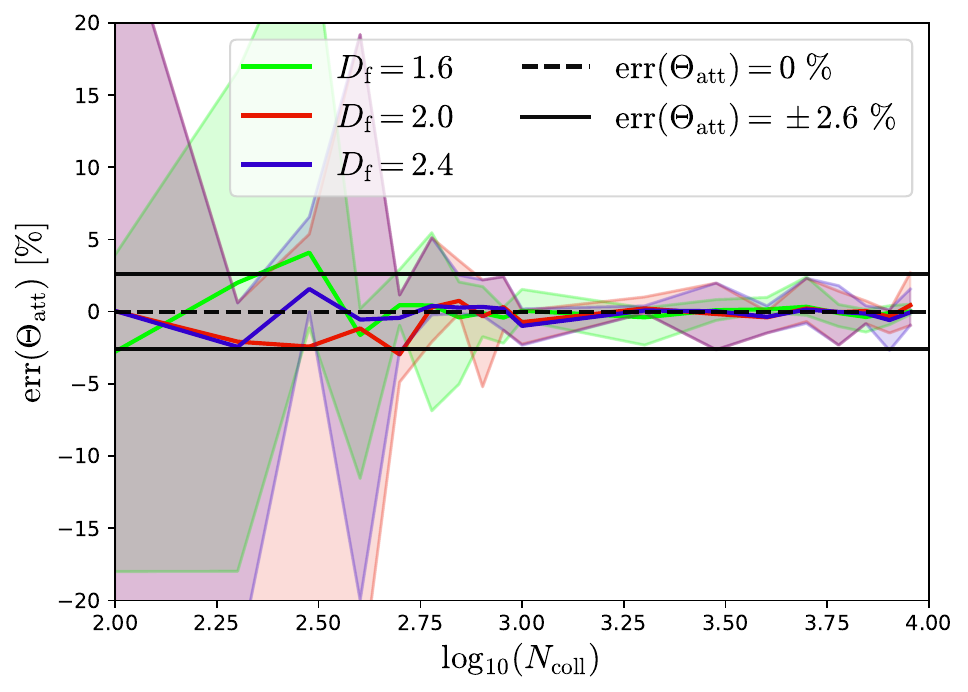}
	\end{center}
\caption{The same as Fig.~\ref{fig:ErrJ} but for the distribution of the alignment angle $\Theta$.}
\label{fig:ErrTh}
\end{figure}

\section{Optical properties of dust aggregates}
\label{app:DDSCAT}
\begin{figure*}[ht!]
	\begin{center}
	\includegraphics[width=1.0\textwidth]{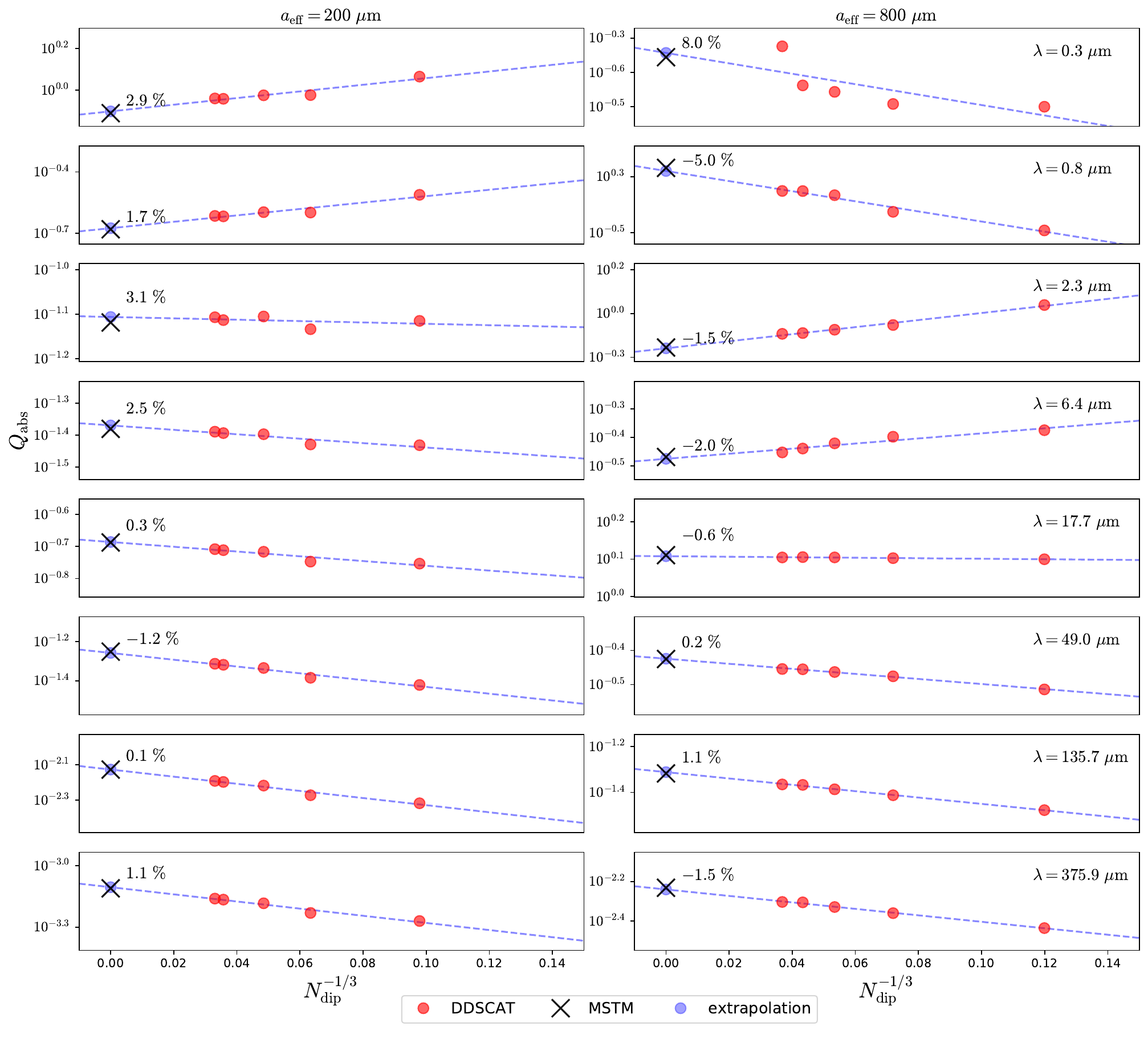}
	\end{center}
\caption{Plots of the efficiency of absorption $Q_{\mathrm{abs}}(\lambda)$ for two exemplary grain shapes with a fractal dimension of $D_{\mathrm{f}}=2.0$, different wavelengths $\lambda$ (rows), and grain sizes $a_{\mathrm{eff}}$ (columns). Solutions are calculated with the DDSCAT code for different values  of the number of dipoles $N_{\mathrm{dip}}$ (red dot). The DDSCAT solutions are extrapolated assuming $N_{\mathrm{dip}}\rightarrow \infty$ (blue dots). The percentage values represent the error between the extrapolated solutions and the exact solutions of the MSTM code (black cross).}
\label{fig:DDSCAT}
\end{figure*}
In order to determine the contribution of the IR drag to the grain alignment dynamics we need to calculate the efficiency of light absorption $Q_{\mathrm{abs}}(\lambda)$ per aggregate. Usually, this quantity may be calculated for spherical grains as a series of Bessel functions and Legendre functions, respectively \citep[e.g.][]{Wolf2004} on the basis of refractive indices of distinct materials. \\
Calculating $Q_{\mathrm{abs}}(\lambda)$ for an entire aggregate adds considerable complexity to the problem. An exact solution for an aggregate may be achieved  by the Multiple Sphere T-Matrix (MSTM) code \citep[][]{Mackowski2011,Egel2017}. However, we find that calculating the optical properties with the help MSTM for a larger ensemble of dust aggregates and several grain orientations cannot be achieved within a reasonable time frame. Alternatively, an approximate solution may be calculated with the dipole approximation code DDSCAT \cite[][]{Draine2013}. Here, an arbitrary grain shape can approximated by a number of discrete dipoles $N_{\mathrm{dip}}$ \citep[see e.g.][for details]{DeVoe1965,Draine1994}. The numerical limitations of DDSCAT per wavelength $\lambda$ are given by 
\begin{equation}
a_{\mathrm{eff}} < 9.9\frac{\lambda}{|m|} \left( \frac{N_{\mathrm{dip}}}{10^6}  \right)^{1/3}\,,
\label{eq:DDSCATLimit}
\end{equation}
where $m$ is the imaginary refractive index of the grain material. Note that because of this limitation the run-time increases with $N_{\mathrm{dip}}^3$. Consequently, calculating the optical properties of an ensemble of large grain aggregates is still not feasible. In order to overcome these limitations we apply the extrapolation method suggested in \cite{Shen2008}. The efficiencies $Q_{\mathrm{abs}}(\lambda)$ are calculated for $45$ different grain orientations around $\hat{a}_{\mathrm{1}}$ and $200$ wavelengths logarithmically distributed within the interval ${ \lambda\in[0.2\ \mu\mathrm{m},2000\ \mu\mathrm{m}] }$. Here, we apply the refractive indices of silicate presented in \cite{Weingartner2001}. However, each solution is calculated with four different values of $N_{\mathrm{dip}}$ even though the condition given in Eq. \ref{eq:DDSCATLimit} may be violated. Finally, the efficiency $Q_{\mathrm{abs}}(\lambda)$ is extrapolated by assuming ${ N_{\mathrm{dip}}\rightarrow \infty }$. For about $1\ \%$ of all runs we repeat the calculations utilizing the MSTM code in order to estimate the error of this procedure. This way the $Q_{\mathrm{abs}}(\lambda)$ for all aggregates may be calculated in a reasonable time frame. \\
In Fig.~\ref{fig:DDSCAT} we present the result for to exemplary grains with a fractal dimension $D_{\mathrm{f}}=2.0$ size of $a_{\mathrm{eff}}=200\ \mathrm{nm }$ and $a_{\mathrm{eff}}=800\ \mathrm{nm }$, respectively. We find typical fractional errors of a few percent between the solutions of the MSTM code and the extrapolated solutions of DDSCAT for wavelengths $\lambda > 1\ \mu\mathrm{m}$. We note no systematic trend for different wavelength. The errors are generally larger with values up to $\pm 10\ \%$ for $\lambda < 1\ \mu\mathrm{m}$. However, we consider only a maximal dust temperatures of $1000\ \mathrm{K}$ in our alignment models corresponding to a peak wavelength of the Planck function of $\lambda \approx 2.9\ \mu\mathrm{m}$. Hence, the impact to the integral in Eq. \ref{eq:fIR} should only be marginal when using the extrapolation method of \cite{Shen2008} instead of more the more precise but time consuming MSTM calculations.

\section{Stationary points}
\label{app:StaticPoints}
In this section we briefly outline general criteria to characterize the stationary points of the time evolution of the grain alignment dynamics. For convenience we write the first time derivatives of Eq. \ref{eq:dJdt} and Eq. \ref{eq:dJdTheta} as 
${\dot{J}=\mathrm{d}\tilde{J}/\mathrm{d}\tilde{t}}$ and ${\dot{\Theta}=\mathrm{d} \Theta/\mathrm{d}\tilde{t}}$, respectively. Each stationary point $(\tilde{J}_{\mathrm{s}},\Theta_{\mathrm{s}})$ of this system of differential equations is defined by the sufficient conditions
\begin{equation}
\left.\dot{J}\right\vert_{\tilde{J}_{\mathrm{s}},\Theta_{\mathrm{s}}}=0\, ,
\end{equation}
and
\begin{equation}
\left.\dot{\Theta}\right\vert_{\tilde{J}_{\mathrm{s}},\Theta_{\mathrm{s}}}=0\, .
\end{equation}
Consequently, $\dot{J}$ and $\dot{\Theta}$ may be approximated around $(\tilde{J}_{\mathrm{s}},\Theta_{\mathrm{s}})$ as a Taylor series up to the first order as
\begin{equation}
\dot{J}\approx\left.\dot{J}\right\vert_{\tilde{J}_{\mathrm{s}},\Theta_{\mathrm{s}}}+\left(\tilde{J}-\tilde{J}_{\mathrm{s}}\right)\left.\frac{\mathrm{d}\dot{J}}{\mathrm{d}\tilde{J}}\right\vert_{\tilde{J}_{\mathrm{s}},\Theta_{\mathrm{s}}}+\left(\Theta-\Theta_{\mathrm{s}}\right)\left.\frac{\mathrm{d}\dot{J}}{\mathrm{d}\Theta}\right\vert_{\tilde{J}_{\mathrm{s}},\Theta_{\mathrm{s}}}
\end{equation}
and
\begin{equation}
\dot{\Theta}\approx\left.\dot{\Theta}\right\vert_{\tilde{J}_{\mathrm{s}},\Theta_{\mathrm{s}}}+\left(\tilde{J}-\tilde{J}_{\mathrm{s}}\right)\left.\frac{\mathrm{d}\dot{\Theta}}{\mathrm{d}\tilde{J}}\right\vert_{\tilde{J}_{\mathrm{s}},\Theta_{\mathrm{s}}}+\left(\Theta-\Theta_{\mathrm{s}}\right)\left.\frac{\mathrm{d}\dot{\Theta}}{\mathrm{d}\Theta}\right\vert_{\tilde{J}_{\mathrm{s}},\Theta_{\mathrm{s}}}
\end{equation}
, respectively. This linearization defines a equation system of the form:
\begin{equation}
\begin{pmatrix} \dot{J}\\ \dot{\Theta} \end{pmatrix} = \begin{pmatrix} \mathrm{d}\dot{J}/\mathrm{d}\tilde{J} &  \mathrm{d}\dot{J}/\mathrm{d}\Theta \\  \mathrm{d}\dot{\Theta}/\mathrm{d}\tilde{J} & \mathrm{d}\dot{\Theta}/\mathrm{d}\Theta \end{pmatrix}\begin{pmatrix} \tilde{J}-\tilde{J}_{\mathrm{s}}\\ \Theta-\Theta_{\mathrm{s}}  \end{pmatrix}\, .
\end{equation}
Here, the Jacobian matrix may be evaluated at the corresponding static points with the imaginary eigenvalues $\lambda_{\rm 1}$ and $\lambda_{\rm 2}$. Under the condition $\Re(\lambda_{\rm 1})\cdot\Re(\lambda_{\rm 2})\neq0$ for the real parts eigenvalues the nature of the static points can be quantified as listed in Tab. \ref{tab:Stability}.\\
\begin{table}[ht!]
\centering
\begin{tabular}{ l | c | c  | c }
\hline
  attractor & stable  & $\Re(\lambda_{\rm 1})<0$  & $\Re(\lambda_{\rm 2})<0$\\
  repeller & unstable  &  $\Re(\lambda_{\rm 1})>0$  & $\Re(\lambda_{\rm 2})>0$\\
\hline
\end{tabular}
\caption{Necessary conditions to identify a stationary point as an attractor point or repeller point, respectively. }
\label{tab:Stability}
\end{table}\\
Exactly, the same procedure is applied to characterize the stationary points $(\tilde{J}_{\mathrm{s}},\xi_{\mathrm{s}})$ of Eq. \ref{eq:dXidt} and Eq. \ref{eq:dJdtMag}.\\
We note that criteria are presented in \cite{Draine1996} and \cite{Lazarian2007} to characterize stationary points for the particular scenarios of magnetic field alignment and RAT alignment, respectively.

\section{Size distribution test}
Dust aggregates are usually constructed using a constant monomer size $a_{\mathrm{mon}}$. However, in our study the ensemble of dust grain follows the scaling law presented in \cite{Filippov2000} to guarantee a certain fractal dimension (see Eq. \ref{eq:FractalScaling}) and simultaneously a size distribution law to account for different monomer sizes (see Eq. \ref{eq:SizeDist}).  Although, such a modification is briefly discussed in \cite{Skorupski2014} it is not a priory clear that an aggregate can comply with one of the laws without violating the other. {\bf In this section we briefly explore where the aggregate may not strictly scale as demanded by the monomer size distribution and given fractal dimension.}\\
In Fig.~\ref{fig:DistP} we present the monomer size distribution $p\left(10\ \mathrm{nm}\right)$ for grains with a size of $a_{\mathrm{eff}}=400\ \mathrm{nm}$. For a fractal dimension of $D_{\mathrm{f}} > 2.0$ the monomer size distribution is no longer followed for the largest values of $a_{\mathrm{mon}}$. This is because for more roundish grains with $a_{\mathrm{eff}}\geq 400\ \mathrm{nm}$ it becomes increasingly unlikely to find a proper position without any overlap. In our implementation of the algorithm presented in \cite{Skorupski2014} after a few thousand attempts to connect a large monomer to an aggregate's surface a kill counter kicks in and a new monomer size is sampled in order to ensure that each aggregate is constructed within an acceptable time frame. Consequently, smaller grains with $a_{\mathrm{eff}}\leq200 \ \mathrm{nm}$ suffer less from this limitation where even grains with $D_{\mathrm{f}} = 2.6$ follow strictly the size distribution.\\
The fractal dimension $D_{\mathrm{f}}$ of a dust aggregate can be estimated by the correlation function \citep[see e.g.][]{Skorupski2014}
\begin{equation}
	C(r)= \frac{n(r)}{4\pi r^2 l N_{\mathrm{mon}}} \propto r^{D_{\mathrm{f}}-3}\,,
\end{equation}
where $r$ is the distance from the center of mass, $l$ is a length with $l\ll a_{\mathrm{eff}}$, and $n(r)$ is the number density of connected monomers within the shell ${[r-l/2;r+l/2]}$.\\
In Fig.~\ref{fig:DistC} we present the resulting fractal dimensions for the ensemble of grains with $a_{\mathrm{eff}} = 400 \ \mathrm{nm}$. Here, the function $C(r)$ is consistent with the demanded fractal dimension. Exceptions are at the very surface of the grains as well as  towards the center. The largest deviation is for grains with $D_{\mathrm{f}}\leq 1.8$ where the correlation function $C(r)$ already starts to break down at $r=a_{\mathrm{eff}} $. This trend is similar for all grains sizes but smaller grains have larger variation in $C(r)$. However, for the most elongated grains $D_{\mathrm{f}}\leq 1.8$ the radius $a_{\mathrm{eff}}$ represents only the inner most region considering the extension of the total aggregate i.e. $a_{\mathrm{eff}} \ll a_{\mathrm{out}}$ (see Fig.~\ref{fig:Grains}
). Hence, we estimate the deviations from the scaling law, the size distribution of monomers on the grains surface and the inner core to influence our results only marginally.

\begin{figure}[ht!]
	\begin{center}
	\includegraphics[width=0.5\textwidth]{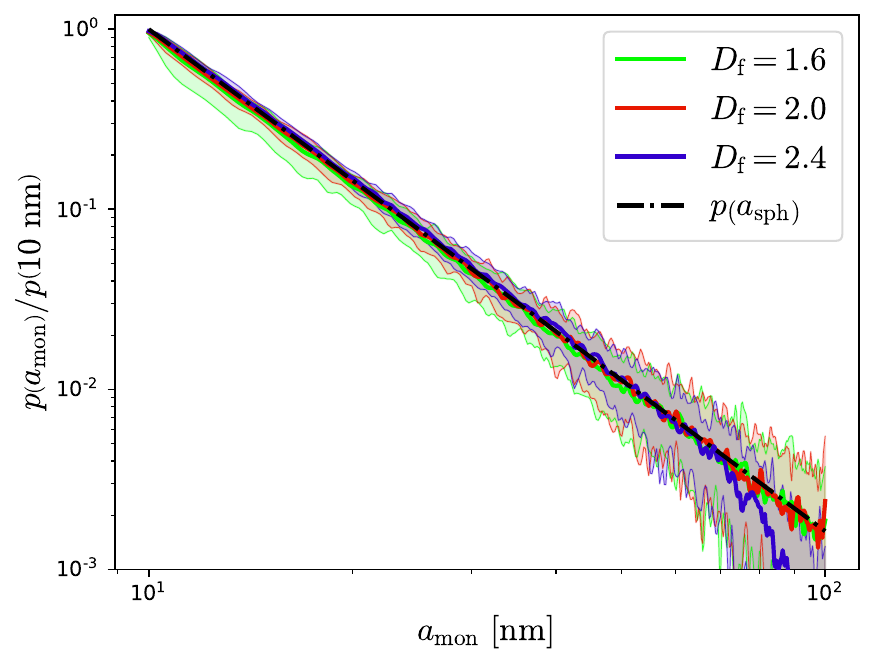}
	\end{center}
\caption{Distribution of the monomer size $a_{\mathrm{mon}}$ averaged over the grain ensemble with $a_{\mathrm{eff}}=400\ \mathrm{nm}$ for the fractal dimensions of $D_{\mathrm{f}}=1.6$ (solid green), $D_{\mathrm{f}}=2.0$ (solid red), $D_{\mathrm{f}}=2.4$ (solid blue), respectively, in comparison with the distribution function $p\left(a_{\mathrm{mon}}\right)$ (dash dotted black). For comparison, all distributions are normalized by $p\left(10\ \mathrm{nm}\right)$.}
\label{fig:DistP}
\end{figure}
\begin{figure}[ht!]
	\begin{center}
	\includegraphics[width=0.5\textwidth]{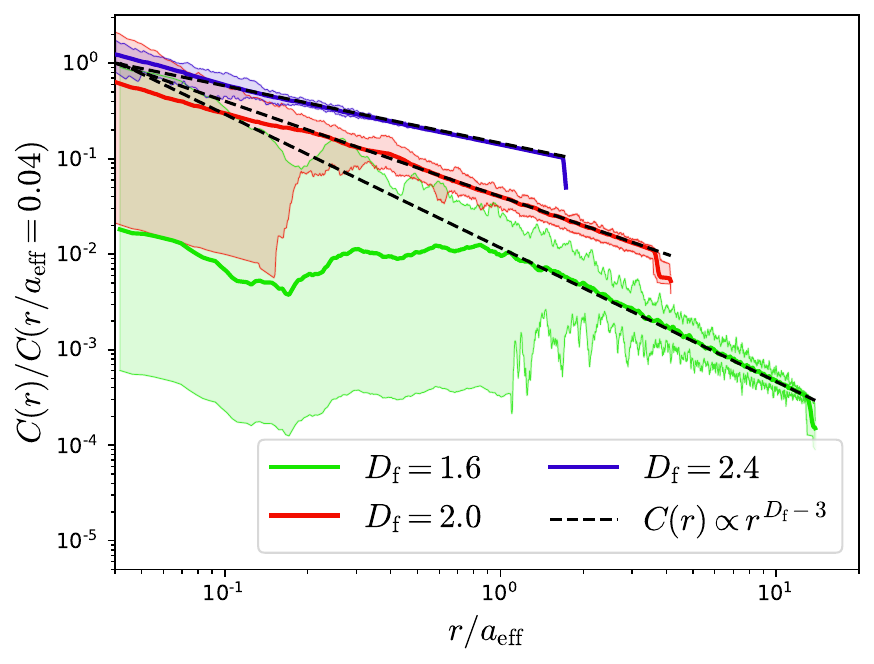}
	\end{center}
\caption{The same as Fig.~\ref{fig:DistP} but for the correlation function $C(r)$ over the distance $r/a_{\mathrm{eff}}$ from the center of mass.}
\label{fig:DistC}
\end{figure}






\section{List of applied physical quantities}
In Table \ref{tab:Quantities} we briefly list all physical quantities utilized in this paper.
\begin{table*}
\centering
\begin{tabularx}{0.98\textwidth} { 
  | p{1.4cm} 
  | >{\centering\arraybackslash}X  
  | p{1.4cm} 
  | >{\centering\arraybackslash}X  |}
 \hline
quantity & description & quantity & description \\
\hline
\hline
$\{ \hat{a}_{\mathrm{1}}, \hat{a}_{\mathrm{2}}, \hat{a}_{\mathrm{3}} \}$ & basis of the aggregate (target-frame) & $N_{\mathrm{des}}$ & number of desorbing gas particles\\ \hline
$a_{\mathrm{eff}}$ & effective radius of of the aggregate & $N_{\mathrm{l}}$ & number of gas particles leaving the surface\\ \hline
$a_{\mathrm{mean}}$ & mean monomer radius & $N_{\mathrm{mon}}$ & total number of monomers\\ \hline
$a_{\mathrm{mon,i}}$ & i-th monomer radius & $\mathcal{P}$ & porosity of the aggregate\\ \hline
$a_{\mathrm{out}}$ & outer radius of of the aggregate & $P_{\Delta s}(\vartheta)$ & angular distribution of gas trajectories\\ \hline
$B$ & magnetic field strength & $\varphi$ & azimuthal angle of the lab-frame\\ \hline
$B_{\lambda}\left( T_{\mathrm{d}} \right)$ & Planck's law & $\phi$ & precession angle around $B$\\ \hline
$\beta$ & grain rotation angle around $\hat{a}_{\mathrm{1}}$ & $\Phi$ & precession angle around $\hat{a}_{\mathrm{1}}$\\ \hline
$\chi$ & magnetic susceptibility & $\psi$ & alignment angle between $\Delta s$ and $B$\\ \hline
$D_{\mathrm{f}}$ & fractal dimension & $Q_{\mathrm{abs}}(\lambda)$ & absorption efficiency\\ \hline
$\mathrm{d}\vec{A}$ & infinitesimal surface element & $\vec{Q}_{\mathrm{coll}}$ & gas collision torque efficiency\\ \hline
$\delta_{\mathrm{m}}$ & ratio of gas drag time to paramagnetic time & $\vec{Q}_{\mathrm{des}}$ & gas desorption torque efficiency\\ \hline
$\Delta s$ & dimensionless gas-dust drift velocity & $\vec{Q}_{\mathrm{drag}}$ & total drag torque efficiency\\ \hline
$\Delta t_{\mathrm{int}}$ & intersection time scale & $\vec{Q}_{\mathrm{gas}}$ & gas drag torque efficiency\\ \hline
$\Delta v$ & gas-dust drift velocity & $\vec{Q}_{\mathrm{mech}}$ & total MET efficiency\\ \hline
$\mathrm{d}\Omega$ & infinitesimal element of the solid angle & $\vec{Q}_{\mathrm{sca}}$ & gas scattering torque efficiency\\ \hline
$\{ \hat{e}_{\mathrm{1}}, \hat{e}_{\mathrm{2}}, \hat{e}_{\mathrm{3}} \}$ & basis of the gas properties (lab-frame) & $r$ & distance from the center of mass\\ \hline
$E_{\mathrm{b}}$ & binding energy between monomers & $\hat{r}_{\mathrm{i}}$ & normalized surface gas impact position\\ \hline
$f_{\mathrm{IR}}( T_{\mathrm{d}} )$ & IR damping factor & $\rho_{\mathrm{dust}}$ & mass density of the dust material\\ \hline
$f_{\rm{vel}}(s,\Delta s)$ & gas velocity distribution & $R$ & Rayleigh Reduction Factor\\ \hline
$\vec{\Gamma}_{\mathrm{coll}}$ & gas collision torque & $R_{\mathrm{c}}$ & critical aggregate radius\\ \hline
$\vec{\Gamma}_{\mathrm{des}}$ & gas desorption torque & $R_{\mathrm{gyr}}$ & radius of gyration\\ \hline
$\vec{\Gamma}_{\mathrm{drag}}$ & total drag torque & $\mathcal{R}_{\mathrm{int}}$ & rate if interaction\\ \hline
$\vec{\Gamma}_{\mathrm{DG}}$ & paramagnetic torque & $\sigma_{\mathrm{mon}}$ & standard deviation of the monomer size\\ \hline
$\vec{\Gamma}_{\mathrm{gas}}$ & gas drag torque & $\mathcal{S}_{\mathrm{max}}$ & maximal tensile strength of the aggregate\\ \hline
$\vec{\Gamma}_{\mathrm{IR}}$ & IR drag torque & $\tau_{\mathrm{DG}}$ & paramagnetic dissipation timescale\\ \hline
$\vec{\Gamma}_{\mathrm{mech}}$ & total MET & $\tau_{\mathrm{drag}}$ & total drag timescale\\ \hline
$\vec{\Gamma}_{\mathrm{sca}}$ & gas scattering torque & $\tau_{\mathrm{gas}}$ & gas drag timescale\\ \hline
$h$ & overlap between monomers & $\tau_{\mathrm{IR}}$ & IR drag timescale\\ \hline
$I_{\mathrm{a_1}}, I_{\mathrm{a_2}}, I_{\mathrm{a_3}}$ & moments of inertia along the target-frame & $T$ & total simulation time\\ \hline
$\vec{J}$ & angular momentum of the aggregate & $T_{\mathrm{d}}$ & dust temperature\\ \hline
$\vec{J}_{\mathrm{disr}}$ & critical angular momentum of disruption & $T_{\mathrm{g}}$ & gas temperature\\ \hline
$\vec{J}_{\mathrm{th}}$ & thermal angular momentum & $\Theta$ & alignmentangle between $\hat{e}_{\mathrm{1}}$ and $\hat{a}_{\mathrm{1}}$\\ \hline
$k_{\mathrm{B}}$ & Boltzmann constant & $\vartheta$ & polar angle of the lab-frame\\ \hline
$k_{\mathrm{f}}$ & scaling factor of the aggregate & $V_{\mathrm{agg}}$ & total volume of the dust aggregate\\ \hline
$K$ & scaling factor of the susceptibility & $\varv_{\rm d}$ & arbitrary dust velocity\\ \hline
$\lambda$ & wavelength & $\left< v_{\mathrm{des}} \right>$ & gas desorption velocity\\ \hline
$m_{\mathrm{g}}$ & gas particle mass & $\varv_{\rm g}$ & arbitrary dust velocity\\ \hline
$\mu_{\mathrm{0}}$ & vacuum permeability & $\varv_{\rm th}$ & peak gas velocity\\ \hline
$n_{\mathrm{g}}$ & gas number density & $\left< \varv_{\rm th} \right>$ & average gas velocity\\ \hline
$N_{\mathrm{att}}$ & number of attractor points & $\varv_{\rm g}$ & arbitrary gas velocity\\ \hline
$\left< N_{\mathrm{con}} \right>$ & average number of monomer connections & $w_{\mathrm{i}}$ & weight of an attractor point\\ \hline
$\hat{N}_{\mathrm{i}}$ & surface normal at the gas impact position & $\xi$ & angle between $\hat{a}_{\mathrm{1}}$ and $B$\\ \hline
$N_{\mathrm{coll}}$ & number of colliding gas particles & $\vec{X}_{\mathrm{i}}$ & monomer position\\ \hline
\end{tabularx}
\caption{Table of the physical quantities and abbreviations utilized in this paper. }
\label{tab:Quantities}
\end{table*}

\begin{acknowledgements}
The authors thank the anonymous reviewer for the concise and
constructive referee report. Special thanks go to Cornelis Dullemond and Sebastian Wolf for numerous enlightening discussions about MC codes and gas-dust dynamics. We also thank Ugo Lebreuilly and Valentin Le Gouellec for providing helpful feedback. S.R., P.M., and R.S.K. acknowledge financial support from the Heidelberg cluster of excellence (EXC 2181 - 390900948) “{\em STRUCTURES}: A unifying approach to emergent phenomena in the physical world, mathematics, and complex data”, specifically via the exploratory project EP 4.4. S.R., P.M., and R.S.K. also thank for support from Deutsche Forschungsgemeinschaft (DFG) via the Collaborative Research Center (SFB 881, Project-ID 138713538) 'The Milky Way System' (subprojects A01, A06, B01, B02, and B08). And we thanks for funding form the European Research Council in the ERC synergy grant “{\em ECOGAL} – Understanding our Galactic ecosystem: From the disk of the Milky Way to the formation sites of stars and planets” (project ID 855130). The project made use of computing resources provided by {\em The L\"{a}nd} through bwHPC and by DFG through grant INST 35/1134-1 FUGG. Data are in part stored at SDS@hd supported by the Ministry of Science, Research and the Arts and by DFG through grant INST 35/1314-1 FUGG.
\end{acknowledgements}



\bibliographystyle{aa}
\bibliography{./bibtex}
\end{document}